\documentclass[a4paper,UKenglish,numberwithinsect]{lipics-v2021}

\usepackage[T2A,T1]{fontenc}
\usepackage[utf8]{inputenc}

\usepackage{tikz}
\usetikzlibrary{arrows.meta,backgrounds,decorations.pathmorphing,calc}

\title{On the Satisfaction~Probabilities of~\textit{k}-CNF~Formulas}

\author{Till Tantau}{
  Institute for Theoretical Computer Science,
  Universit\"at zu L\"ubeck,
  L\"ubeck, Germany
}{tantau@tcs.uni-luebeck.de}{}{}

\authorrunning{T. Tantau} 

\Copyright{Till Tantau}



\relatedversion{}

\EventEditors{}
\EventNoEds{1}
\EventLongTitle{}
\EventShortTitle{}
\EventAcronym{}
\EventYear{}
\EventDate{}
\EventLocation{}
\EventLogo{}
\SeriesVolume{}
\ArticleNo{}

\hideLIPIcs
\expandafter\def\expandafter\UrlBreaks\expandafter{\UrlBreaks\do\-\do\)}

\newcommand\Class[1]{%
  \mathchoice%
  {\text{\normalfont\small$\mathrm{#1}$}}%
  {\text{\normalfont\small$\mathrm{#1}$}}%
  {\text{\normalfont$\mathrm{#1}$}}%
  {\text{\normalfont$\mathrm{#1}$}}%
}

\newcommand{\CNFSSigmaSpectrum}{\Lang{cnfs-pr-spectrum}}
\newcommand{\Lang}[1]{\text{\upshape\normalfont\textsc{#1}}}
\newcommand{\Algo}{\Lang}

\newbox\myTBox\setbox\myTBox=\hbox{\sffamily T}
\newcommand{\TOperator}{\begin{tikzpicture}[baseline]\draw[cap=round,line
      width=.4pt](0,\ht\myTBox-.2pt) -- (.9\wd\myTBox,\ht\myTBox-.2pt)
    (.45\wd\myTBox,\ht\myTBox-.2pt) -- (.45\wd\myTBox,0.2pt);\end{tikzpicture}}
\newcommand{\ThrSO}[1]{\TOperator_{\!\!\scriptscriptstyle#1}\kern-.4pt}
\newcommand{\MajSO}{\ThrSO{\ge\kern-1pt\frac{1}{2}}}

\newcommand\crossout[1]{%
  \tikz[baseline]{
    \node[inner sep=0pt,anchor=base](n){#1};
    \draw[overlay,thick,black!50,line cap=round]([shift={(-1.5pt,-1.5pt)}]n.base west)
    -- ([shift={(1.5pt,1.5pt+1.5ex)}]n.base east);
  }%
}
\newcommand\crossinspec[1]{
  \scoped[shift={#1}]{\draw[line width=.55pt] (-1pt,-1pt) -- (1pt,1pt) (1pt,-1pt)--(-1pt,1pt);}
}

\newcommand\Restrict[2]{#1|_{#2}}
\newcommand\RestrictAdd[2]{#1|^+_{#2}}
\newcommand\Prob[1]{\Pr[#1]}
\newcommand\ProbBig[1]{\Pr\bigl[#1\bigr]}

\newcommand\ProbOr[1]{\Prob{{\textstyle\bigvee}#1}}
\newcommand\FormImpl{\mathrel {\tikz\draw[line cap=round](0,0)--(0,1ex);\kern.25ex}\joinrel \Relbar}

\theoremstyle{plain}
\newtheorem{fact}[theorem]{Fact}
\newenvironment{reclaim*}[1][]{\begin{claim*}[#1]\itshape\ignorespaces}{\end{claim*}}
\newtheorem*{corollary*}{Corollary}
\newtheorem*{observation*}{Observation}

\newtheorem*{notation*}{Notation}
\newtheorem*{definition*}{Definition}
\newtheorem{reduction rule}[theorem]{Rule}

\tikzset{graph/.style = {
    node/.style = {circle, minimum size=5mm, inner sep=1pt, semithick, draw, font=\small},
    thin,
    > = {Stealth[round,sep]}
  },
  figure/.style={font=\small}
}

\usepackage{listings}
\lstdefinelanguage{pseudocode}{
  morekeywords={
    algorithm,method,new,and,or,
    if,then,else,while,do,repeat,until,seq,
    seqdo,return,call,
    for,pardo,in,foreach,print,output,input,exit,
    break,loop,end,begin,goto,par,global,local,
    read,write,to,stop,idle,procedure,function,
    throw,catch
  },
  sensitive=true,
  morecomment=[l]{//},
  morestring=[b]",
  morestring=[s]{``}{''},
}
\lstdefinestyle{pseudocode}{
  language=pseudocode,
  basicstyle=\small\rmfamily,
  commentstyle=\upshape\color{black!50},
  keywordstyle=\bfseries\itshape,
  identifierstyle=\itshape,
  stringstyle=\rmfamily,
  columns=fullflexible,
  mathescape,
  literate={<-}{{$\gets$\ }}2,
  numbers=left,
  numberstyle=\scriptsize\sffamily,
}

\newcommand\commented[1]{\textcolor{black!50}{#1}}
\def\pictureexplain#1{
  \draw [draw=black!25]
  (-1,2.5) node[below right,align=left,text=black!65] {#1}
  rectangle (3,-3.7);
}

\def\pictureliteralsbase{
  \pic (a) at (0.5,1)   [pic text = y] {literal};
  \pic (a') at (0.5,0)  [pic text = \neg y] {literal, height=9mm};
  \pic (b) at (1,0)     [pic text = z] {literal, height=9mm};
  \pic (b') at (1,-2)   [pic text = \neg z] {literal};
  \pic (f') at (2,-3)   [pic text = \neg l] {literal};
  \pic (g') at (2,-1)   [pic text = \neg k] {literal};
  \pic (h) at (-.5,0)   [pic text = f] {literal};
  \pic (i) at (0.5,-2)  [pic text = e] {literal};
  \pic (l) at (0,0)     [pic text = x] {literal, height=9mm};
  \pic (m) at (0,-2)    [pic text = \neg d] {literal};
}
\def\pictureliteralsoutside{
  \pic (f) at (2,-2)    [pic text = l] {literal};
  \pic (c) at (2,2)     [pic text = g] {literal};
  \pic (d') at (2,1)    [pic text = \neg h] {literal};
  \pic (e) at (2,0)     [pic text = j] {literal};
  \pic (j) at (2.5,1)   [pic text = i] {literal};
  \pic (k) at (2.5,-2)  [pic text = m] {literal};
  \pic (i') at (2.5,-3) [pic text = \neg e] {literal};
}
\def\pictureliterals{
  \pictureliteralsbase
  \pictureliteralsoutside
}
\def\pictureliteralsprime{
  \pictureliteralsbase
  \pic (p1) at (2,2)    [red, pic text = v_1] {literal};
  \pic (p2) at (2.5,2)  [red, pic text = v_2] {literal};
  \pic (p3) at (2,1)    [red, pic text = \neg v_1] {literal};
  \pic (p4) at (2.5,1)  [red, pic text = v_3] {literal};
  \coordinate (p5-mid) at (2,0);
  \coordinate (p6-mid) at (2.5,0);
  \node at (2.25,0.3) {$\vdots$};
}

\def\picturejoinstart#1#2{
  [thick,cap = round]
    ([xshift=-.5em,
      yshift=-.25em,
      yshift=-#2mm
    ]#1-mid)
    -- ++ (0.5em,0em)
}
\def\picturejoinon#1#2{
   {[rounded corners=0.95mm]
     -- ++ (.8em,0em)
     --
    ([xshift=-.8em,
      yshift=-.25em,
      yshift=-#2mm
    ]#1-mid) -- ++ (.8em,0)}
}
\def\picturejoinend{-- ++ (.5em,0pt)}
\def\picturekernel{
  \pic [] {join 4 = h on 1 with l on 1 with a  on 1 with b  on 1};
  \pic [] {join 4 = m on 2 with i  on 2 with b' on 2 with f' on 2};
  \pic [] {join 4 = l on 9 with a' on 9 with b' on 1 with g' on 2};
}

\tikzset{
  sunflower figure/.style={figure, x=1.5cm,  y=8mm},
  literal/.pic={
    \draw [thick, black!20, rounded corners, pic actions]
    (-.775em,.75em) rectangle
    (.775em,-.75em-\pgfkeysvalueof{/tikz/pics/height});
    \node [anchor=mid] at (0,0) {$\tikzpictext$};
    \coordinate (-mid) at (0,0);
  },
  join/.pic={
    \draw 
      \picturejoinstart{\pgfkeysvalueof{/tikz/pics/from}}{\pgfkeysvalueof{/tikz/pics/from lane}}
      \picturejoinon{\pgfkeysvalueof{/tikz/pics/to}}{\pgfkeysvalueof{/tikz/pics/to lane}}
      \picturejoinend;      
  },
  pics/height/.initial={1mm},
  pics/from/.initial={},
  pics/from lane/.initial={0},
  pics/from shift/.initial={0pt},
  pics/to/.initial={},
  pics/to lane/.initial={0},
  pics/to shift/.initial={0pt},
  pics/join 2/.style args={#1 on #2 with #3 on #4}{
    code = {
      \draw
        \picturejoinstart{#1}{#2}
        \picturejoinon{#3}{#4}
        \picturejoinend;
    }
  },
  pics/join 3/.style args={#1 on #2 with #3 on #4 with #5 on #6}{
    code = {
      \draw
        \picturejoinstart{#1}{#2}
        \picturejoinon{#3}{#4}
        \picturejoinon{#5}{#6}
        \picturejoinend;
    }
  },
  pics/join 3'/.style args={#1 on #2 with #3 on #4 with #5 on #6}{
    code = {
      \draw
        \picturejoinstart{l}{#2}
        \picturejoinon{#1}{#2}
        \picturejoinon{#3}{#4}
        \picturejoinon{#5}{#6}
        \picturejoinend;
    }
  },
  pics/join 4/.style args={#1 on #2 with #3 on #4 with #5 on #6 with #7 on #8}{
    code = {
      \draw
        \picturejoinstart{#1}{#2}
        \picturejoinon{#3}{#4}
        \picturejoinon{#5}{#6}
        \picturejoinon{#7}{#8}
        \picturejoinend;
    }
  },
  pics/join 4'/.style args={#1 on #2 with #3 on #4 with #5 on #6 with #7 on #8}{
    code = {
      \draw
        \picturejoinstart{l}{#2}
        \picturejoinon{#1}{#2}
        \picturejoinon{#3}{#4}
        \picturejoinon{#5}{#6}
        \picturejoinon{#7}{#8}
        \picturejoinend;
    }
  }
}

\begin{document}

\maketitle

\begin{abstract}
  The \emph{satisfaction probability $\Prob{\phi} :=
  \Pr_{\beta:\operatorname{vars}(\phi) \to \{0,1\}}[\beta\models
    \phi]$} of a propositional formula~$\phi$ is the likelihood that a
  random assignment $\beta$ makes the formula true. We study the
  complexity of the problem $k\Lang{sat-pr}_{>p} = \{ \phi$ is
  a $k\Lang{cnf}$ formula $\mid \Prob{\phi} > p\}$ for fixed
  $k \in \mathbb N$ and $p \in [0,1]$. While $\Lang{3sat-pr}_{>0} =
  \Lang{3sat}$ is $\Class{NP}$-complete and $\Lang{sat-pr}_{>1/2}$
  is $\Class{PP}$-complete, Akmal and Williams recently showed that
  $\Lang{3sat-pr}_{>1/2}$ lies in~$\Class P$ and that
  $4\Lang{sat-pr}_{>1/2}$ is $\Class{NP\text{-}complete}$; but the methods used to prove
  these striking results stay
  silent about, say, $\Lang{4sat-pr}_{>3/4}$, leaving the computational
  complexity of $k\Lang{sat-pr}_{>p}$ open for most $k$
  and~$p$. In the present paper we give a    
  complete characterization in the form of a trichotomy:
  $k\Lang{sat-pr}_{>p}$ lies in $\Class{AC}^0$, is 
  $\Class{NL}$-complete, or is $\Class{NP}$-complete. The proof of the 
  trichotomy hinges on a new order-theoretic insight: Every set of
  $k\Lang{cnf}$ formulas contains a formula of maximum satisfaction
  probability. This deceptively simple statement
  allows us to (1)~kernelize $k\Lang{sat-pr}_{\ge p}$ for the joint
  parameters $k$ and~$p$, (2)~show that the variables of the kernel form a backdoor set
  when the trichotomy states membership in $\Class{AC}^0$
  or~$\Class{NL}$, and (3)~prove locality properties for $k\Lang{cnf}$
  formulas~$\phi$, by which $\Prob{\phi} < p$ implies that 
  $\Prob{\psi} < p$ holds already for a subset~$\psi$ of
  $\phi$'s clauses whose size depends only on~$k$
  and~$p$, and $\Prob{\phi} = p$ implies $\phi \equiv \psi$ for some
  $k\Lang{cnf}$ formula~$\psi$  whose size once more depends only on~$k$ and~$p$. 
\end{abstract}

\section{Introduction}

For a propositional formula~$\phi$ 
like $(x \lor \neg y \lor a) \land (\neg x \lor b \lor \neg c) \land
(y) \land (d)$ it is, in
general, a very hard problem to obtain much information about the
number $\#(\phi)$ of assignments that satisfy~$\phi$ or, equivalently, about
the \emph{satisfaction probability} $\Prob{\phi}$ defined as
\begin{align*}
  \Pr\nolimits_{\beta:\operatorname{vars}(\phi) \to
    \{0,1\}}\bigl[\beta\models \phi\bigr] \;=\; \#(\phi) \bigm/ 2^n
\end{align*}
where $n =
\left|\operatorname{vars}(\phi)\right|$ is the number of variables
in~$\phi$. By the Cook--Levin Theorem \cite{Cook1971,Levin1973} it is
already $\Class{NP}$-complete to determine whether $\Prob{\phi} > 0$
holds; and to determine whether $\Prob{\phi} > 1/2$ holds is complete
for $\Class{PP}$. Indeed, the function $\#(\cdot)$ itself is complete
for $\Class{\#P}$, a counting class high up in the complexity
hierarchies. Writing $\Lang{sat-pr}_{> p}$ for $\{ \phi \mid
\Prob{\phi} > p \}$, we can rephrase these results as
``$\Lang{sat-pr}_{>0}$ is $\Class{NP}$-complete'' (the Cook--Levin
Theorem) and  ``$\Lang{sat-pr}_{>1/2}$ is $\Class{PP}$-complete'' (and
so is $\Lang{sat-pr}_{\ge 1/2}$, see for instance
\cite[Theorem~4.1]{Simon1975}).

The Cook--Levin result on the complexity of $\Lang{sat-pr}_{>0} =\Lang{sat}$
is remarkably robust regarding the kinds of formulas 
one can consider: The problem stays $\Class{NP}$-complete for formulas in
$\Lang{cnfs}$, the set of formulas in conjunctive normal form, so
$\Lang{cnf-sat-pr}_{>0} = \Lang{cnf-sat}$ is 
$\Class{NP}$-complete, and even for formulas 
$\phi \in \Lang{3cnfs}$, that is, when all clauses of~$\phi$ have at most three literals, so
$\Lang{3sat-pr}_{>0} = \Lang{3sat}$ is $\Class{NP}$-complete. Similarly,
$\Lang{cnf-sat-pr}_{>1/2}$ has the same complexity as
$\Lang{sat-pr}_{>1/2}$, see \cite[lemma on page 80]{Simon1975}, and
$\#(\cdot)$ is still $\Class{\#P}$-hard for formulas in 
$\Lang{3cnfs}$ and even in $\Lang{2cnfs}$,
see~\cite{Valiant79}.

In sharp contrast to these well-established hardness results, Akmal
and Williams~\cite{AkmalW2022} recently showed that
$\Lang{3sat-pr}_{> 1/2}$ can be solved in polynomial time -- in
fact, they show this time bound for $3\Lang{sat-pr}_{> p} = \{\phi
\in \Lang{3cnfs} \mid \Prob{\phi} > p\}$ for all rational $p >0$
(intriguingly, their argument does not apply to non-rational~$p$ and
left open the complexity of, say, $3\Lang{sat-pr}_{>\pi/4}$). Yet again in
contrast, they also show that  $k\Lang{sat-pr}_{> 1/2}$ is
$\Class{NP}$-complete for every $k \ge 4$. To complicate things
even further, the $\Class{NP}$-completeness result for
$4\Lang{sat-pr}_{> 1/2}$ can easily be extended to
$4\Lang{sat-pr}_{> 1/4}$, to $4\Lang{sat-pr}_{> 1/8}$, to
$4\Lang{sat-pr}_{> 1/16}$ and so on, and with very little extra work
to more exotic values of~$p$ like $p = 15/32$ -- but apparently
\emph{not} to certain other values like $p = 3/4$ or $p =
63/128$. Indeed, $4\Lang{sat-pr}_{>15/16}$ is a \emph{trivial}
problem as \emph{every} nontrivial, nonempty formula in $4\Lang{cnfs}$
has  a satisfaction  probability of at most $15/16$ (a single
non-tautological clause already rules out $1/16$th of all
assignments). In other words, even for a fixed~$k$ the complexity of
$k\Lang{sat-pr}_{>   p}$ might fluctuate wildly for changing~$p$
(and in fact does as Figure~\ref{fig-spec} on page~\pageref{fig-spec}
illustrates quite clearly) and it is unclear how the methods
introduced   in~\cite{AkmalW2022} could be used to show that, say,
$4\Lang{sat-pr}_{> 63/128}$ is $\Class{NL}$-complete while
$3\Lang{sat-pr}_{>\pi/4}$ lies in~$\Class{AC}^0$.

The main purpose of the present paper is to 
find explanations \emph{why} bounding the satisfaction
probabilities for $k\Lang{cnf}$ formulas is sometimes easy and
sometimes hard. The main insight is that there is a deep 
connection between the order-theoretic structure of the \emph{spectra} of
values that satisfaction probabilities of $k\Lang{cnf}$ formulas can
have and the computational complexity of deciding the question
``$\Prob{\phi} > p$?'' Along the way we will also encounter results
that are more concrete and somewhat ``cute,'' like this one:
\emph{Determining if a $\Lang{3cnf}$ formula is satisfied by exactly
half of all possible assignments, is complete for nondeterministic
logarithmic space.}  

\subparagraph*{Notations and Terminology.} Before proceeding, it will be
useful to fix some perhaps not-quite-so-standard notations. As mentioned already,
$\Lang{cnfs}$ denotes the  
set of propositional formulas in conjunctive normal form, and
$k\Lang{cnfs}$ the restriction to formulas with \emph{at most} $k$
literals per clause. Following Knuth~\cite{Knuth2016}, we consider
the elements of $\Lang{cnfs}$ to be finite \emph{sets} of clauses,
which are finite \emph{sets} of literals, which are variables~$v$ or negated
variables~$\neg v$; so the formula from the paper's first line is
actually $\phi = \bigl\{ \{x,\neg y,a\},\penalty0 \{\neg x,b,\neg
c\},\penalty0 \{y\},\penalty50 \{d\}\bigr\} \in \Lang{3cnfs}$. As
another example, $(x) \land (\neg x)$ is actually
$\bigl\{\{x\},\{\neg x\}\bigr\} \in 1\Lang{cnfs}$. 
A bit nonstandard,  we \emph{syntactically forbid tautological
clauses,} so $\bigl\{\{x,\neg x\}\bigr\}
\notin\Lang{cnfs}$, but allow the \emph{tautological
  formula~$\emptyset$,} which has no clauses, and also the 
(unsatisfiable) empty clause inside formulas, so $\{ \emptyset \}$
and~$\emptyset$ are both members of $\Lang{cnfs}$ (and of all $k\Lang{cnfs}$
for that matter). For a literal~$l$, let $\operatorname{var}(l)$
denote its underlying variable, 
so $\operatorname{var}(v) = \operatorname{var}(\neg v) = v$; for a
clause $c$ let $\operatorname{vars}(c)$ denote
$\{\operatorname{var}(l) \mid l\in c\}$; and for a formula $\phi \in
\Lang{cnfs}$ let $\operatorname{vars}(\phi)$ denote $\bigcup_{c \in
  \phi} \operatorname{vars}(c)$. We use $\mathbb B = \{0,1\}$ to
denote the two possible truth values 0 (corresponding to \emph{false}) and 1
(corresponding to \emph{true}).
Also following Knuth, for an assignment $\beta \colon X \to\mathbb B$,
where $X$ is \emph{not} necessarily a superset of
$\operatorname{vars}(\phi)$, let
$\Restrict{\phi}{\beta}$ result from~$\phi$ by removing all 
clauses~$c$ containing a literal $l\in c$ made true by~$\beta$ (meaning
$v = \operatorname{var}(l) \in X$ with $\beta(v) =1$ for $v = l$ or
$\beta(v) = 0$ for $\neg v = l$) and then removing all remaining
occurrences of literals~$l$ with $\operatorname{var}(l) \in 
X$. For instance, for $\phi$ from above, $X = \{x,y\}$, and $\beta(x)
= 1$ and $\beta(y) = 0$, we have $\Restrict{\phi}{\beta} = 
\bigl\{ \crossout{$\{x,\neg y,a\}$}, \penalty0 \{\neg\crossout{$
  x$},b,\neg c\},\penalty0  \{\,\crossout{$
  y$}\,\},\penalty0  \{d\}\bigr\} =  \bigl\{\{b, \neg
c\}, \emptyset, \{d\}\bigr\}$. For $X \supseteq
\operatorname{vars}(\phi)$, we say that \emph{$\beta$ satisfies
$\phi$} (written $\beta \models \phi$) if $\phi|_\beta = \emptyset$,
that is, if every clause of $\phi$ contains a literal made true
by~$\beta$. Given formulas $\phi,\psi \in \Lang{cnfs}$, let us write
$\phi \FormImpl \psi$ to indicate that every satisfying assignment
of~$\phi$ is also a satisfying assignment~of~$\psi$; and we write
$\phi \equiv \psi$ if $\phi \FormImpl \psi$ and $\psi \FormImpl
\phi$. 

We assume that the classes $\Class{NL}$ or~$\Class{NP}$ have their
standard definitions, see for instance~\cite{Papadimitriou1994},
and any subtleties concerning coding issues will not be relevant for
this paper. The class $\Class{AC}^0$ is perhaps less well-known
and refers to ``functions computable by \textsc{dlogtime}-uniform 
families of circuits of constant depth, polynomial size, and unbounded
fan-in,'' see \cite{Immerman1998} for an introduction; but the details
will not be important and readers unfamiliar with them can just think of
$\Class{AC}^0$ as the class of ``very simple functions'' or of ``functions
computable in constant parallel time.'' A bit imprecisely, we also
consider $\Class{AC}^0$ to contain a language if 
it actually contains its characteristic function. The reductions we
use (and with respect to which all completeness results are meant and
under which all considered classes are closed) are
many-to-one reduction functions in $\Class{AC}^0$, though they are
better known as \emph{first-order many-to-one reductions,} see
\cite{Immerman1998} once more for an introduction. Issues relating to 
circuit uniformity will not be further addressed in the following, but
readers familiar with this notion will find that all circuit families
presented in this paper can be made \textsc{dlogtime}-uniform.

\subsection{Contributions of This Paper}

We continue the investigation initiated by Akmal and Williams of the
complexity of threshold problems for the satisfaction probability of
$k\Lang{cnf}$ formulas. We will look at this complexity from 
three different angles in the three main sections of this paper:
From the \emph{order-theoretic} angle in Section~\ref{section-order}, from the
\emph{algorithmic} angle in Section~\ref{section-algorithmics}, and
from the \emph{structural complexity} angle in
Section~\ref{section-complexity}. 

The \emph{first main contribution} is an analysis of the
\emph{order-theoretic} properties 
of the sets $k\CNFSSigmaSpectrum := \{\Prob{\phi}
\mid \phi \in k\Lang{cnfs}\} \subseteq [0,1]$. The analysis will
uncover that for each~$k$ this spectrum is
\emph{well-ordered by~$>$}. A set $X\subseteq [0,1]$ has this 
property if there is no  infinite strictly \emph{increasing} sequence
of elements of~$X$ or, equivalently, if every subset of~$X$ contains a
maximum. We will show that for all~$k$:
\begin{theorem}[Spectral Well-Ordering Theorem]\label{thm-order}
  $k\CNFSSigmaSpectrum$ is well-ordered by~$>$.
\end{theorem}
Another way of phrasing the theorem is in terms of \emph{spectral gaps,}
which can be ``seen'' in Figure~\ref{fig-spec} on page~\pageref{fig-spec}: For any probability $p
\in [0,1]$ (not necessarily a member of the spectrum),
the spectral gap ``stretches left till the next cross (member of the
spectrum).'' The key observation is that there always \emph{is} a
``stretch to the left'' and the below corollary is an equivalent way of stating the Spectral
Well-Ordering Theorem.
\begin{definition}\label{def-spectral-gap}
  Let $\operatorname{spectral-gap}_{k\Lang{cnfs}}(p) := \sup 
  \{\epsilon \mid (p -\epsilon,p) \cap k\CNFSSigmaSpectrum
  = \emptyset\}$.
\end{definition}

\begin{corollary}\label{cor-spec}
  For all $k$ and $p \in [0,1]$, we have 
  $\operatorname{spectral-gap}_{k\Lang{cnfs}}(p) > 0$.
\end{corollary}

The \emph{second main contribution} is the insight that the
deceptively simple above statement has far-reaching \emph{algorithmic}
consequences: It forms the basis of 
several new algorithms for showing $k\Lang{sat-pr}_{\ge p} \in
\Class{AC}^0$ for $p \in [0,1]$ (note the ``$_{\ge  p}$'' rather than  
``$_{>  p}$'' subscript). 

In their ground-breaking work, Akmal and Williams~\cite{AkmalW2022}
already presented a sophisticated and complex algorithm for
showing $k\Lang{sat-pr}_{\ge p}  \in \Class{P}$ for $p \in
[0,1]\,\cap\,\mathbb Q$. As we will see, because of the Spectral
Well-Ordering Theorem, a simple folklore 
algorithm for computing $\Prob{\phi}$ approximately (just sample
assignments randomly and measure the satisfying fraction) already places
$k\Lang{sat-pr}_{\ge p}$ in~$\Class{BPP}$ and a simple linear-time
derandomization due to Trevisan~\cite{Trevisan2004} places
$k\Lang{sat-pr}_{\ge p}$ even in~$\Class{LINTIME}$, yielding the
Akmal--Williams result for all $p \in [0,1]$. Unfortunately, even
though these algorithms are conceptually extremely \emph{simple}, they
are by no means \emph{practical} as they involve prohibitively large
hidden constants (namely the reciprocals of spectral gap sizes,
which we do not even know how to compute exactly and can only lower-bound). 
Moving beyond the simple approximation approach, we have a look at
conceptually different algorithms that give arguably deeper 
insights, starting with two \emph{kernel algorithms} in the sense of
fixed-parameter tractability (\textsc{fpt}) 
theory.
The properties of the kernel computed by the first algorithm will be
crucial in the proofs of the main trichotomy theorem of this paper and
of the following locality result:
\begin{theorem}\label{thm-constant-influence}
  For every $k$ and~$p$ there is
  a number~$S$ so that for every $\phi \in k\Lang{cnfs}$ with 
  $\Prob{\phi} = p$ there is a $\psi \in k\Lang{cnfs}$ of
  size $|\psi| \le S$ with $\phi \equiv \psi$.
\end{theorem}
The second kernel algorithm is instrumental in the proof a second
locality result:
\begin{theorem}[Threshold Locality Theorem]\label{thm-local}
  For every $k$ and $p$ there is a size~$S$ so that for
  every $\phi \in k\Lang{cnfs}$ we have $\Prob{\phi} \ge p$, iff
  $\Prob{\phi'} \ge p$ holds for every $\phi' \subseteq \phi$ with $|\phi'|
  \le S$.
\end{theorem}
By the theorem, the question of whether $\Prob{\phi} \ge p$ holds is
determined by whether $\Prob{\phi'} \ge p$ holds 
\emph{locally} for all constant-size clause subsets. In sharp contrast, $\phi
\in 2\Lang{cnfs}$ can be a contradiction while (\emph{``globally''}) every 
$\phi' \subsetneq \phi$ is satisfiable (an example would be any~$\phi$
expressing that an odd cycle is bipartite). The theorem could thus be summarized as
``satisfaction thresholds are local, satisfiability is global''; but
besides justifying a slogan, it forms the basis of a proof,
given in Corollary~\ref{cor-maj-maj}, of the conjecture by  Akmal and
Williams~\cite{AkmalW2022} that $\Lang{maj-maj-}k\Lang{sat} \in \Class
P$ holds for all~$k$, where 
\begin{align*}
  \Lang{maj-maj-}k\Lang{sat} = \bigl\{ (\phi,X_1,X_2) \bigm|{}
&\phi \in k\Lang{cnfs},\penalty0\,\operatorname{vars}(\phi) = X_1
  \mathbin{\dot\cup} X_2,\\
  &\textstyle\Pr_{\beta: X_1\to \mathbb B}[\Restrict{\phi}{\beta} \in
    k\Lang{sat-pr}_{\ge 1/2}]\ge 1/2 \bigr\}.
\end{align*}
As a final algorithmic contribution, we present and 
analyze a conceptually very different algorithm, namely a
\emph{gap size oblivious} algorithm, meaning that unlike all other known 
algorithms (including that of Akmal and Williams and also the
earlier-mentioned approximation algorithms) one can run the algorithm
without knowing the size of the spectral gap. This makes it especially
appealing for practical implementations. 

The \emph{third main contribution} is to apply the developed
theory to~$k\Lang{sat-pr}_{> p}$;
a problem whose complexity is somewhat more, well, complex than that of
$k\Lang{sat-pr}_{\ge p}$. We establish a complete
classification of the complexity for all $k$ and~$p$ in the
form of a trichotomy: 

\begin{theorem}[Spectral Trichotomy Theorem]\label{thm-main-simple}
  Let $k \ge 1$ and $p \in [0,1]$ be a real number. Then
  $k\Lang{sat-pr}_{> p}$ is
  $\Class{NP}$-complete or
  $\Class{NL}$-complete or
  lies in $\Class{AC}^0$.    
\end{theorem}

\begin{figure}[htpb]
  \begin{tikzpicture}[
      spectrum/.style={black!50},
      nl/.style={green!50!black!75},
      np/.style={red!75},
      figure
    ]
    \def\explanation#1{
      \node [left=1mm] at (0,0) {$k=#1$};
    }
    \scoped[x=12cm,font=\small] {
      
      \node [anchor=mid west,spectrum] at (0,-28mm)    {$p \in k\CNFSSigmaSpectrum$};
      \node [anchor=mid east]          at (0.75,-28mm) {$k\Lang{sat-pr}_{>p}$};
      \node [anchor=mid west,spectrum] at (0.75,-24mm) {$\in{} $AC$^0$};
      \node [anchor=mid west,green!50!black!75]       at (0.75,-28mm) {$\in{} $NL-complete};
      \node [anchor=mid west,red!75]       at (0.75,-32mm) {$ \in{} $NP-complete};
      
      \scoped[yshift=-4mm] {
        \explanation{1}

        \scoped[spectrum] {
          \foreach \pos in {
            0, 0.0005, 0.001, 0.001953125, 0.00390625, 0.0078125, 0.015625, 0.03125, 0.0625, 0.125, 0.25, 0.5, 1
          }
          {
            \crossinspec{(\pos,0)}
          }
        }
      }

      \scoped[yshift=-8mm] {
        \explanation{2}

        \scoped[spectrum] {
          \foreach \pos in {
            1, 0.75, 0.625, 0.5625, 0.53125, 0.515625, 0.5078125,
            0.50390625, 0.5, 0.46875, 0.421875, 0.3984375, 0.390625,
            0.38671875, 0.380859375, 0.3779296875, 0.375, 0.3515625,
            0.33203125, 0.322265625, 0.3173828125, 0.31640625,
            0.31494140625, 0.3125, 0.298828125, 0.29296875,
            0.2900390625, 0.28564453125, 0.283447265625, 0.2822265625,
            0.28125, 0.27392578125, 0.269775390625, 0.2677001953125,
            0.265869140625, 0.265625, 0.263671875, 0.2618408203125,
            0.25982666015625, 0.25787353515625, 0.2578125,
            0.25588989257812, 0.25392150878906, 0.25390625, 0.251953125,
            0.25,0
          }
          {
            \crossinspec{(\pos,0)}
          }
          \draw [line width=2.4pt] (0,0) -- (.25,0);
          \fill[xshift=.375pt] (-1.2pt,1.2pt) -- (0,1.2pt) -- (0,-1.2pt) --
          (-1.2pt,-1.2pt) -- (0,0) -- cycle;
        }

        \scoped[nl] {
          \fill (0,0) -- ++(60:1.5mm) -- ++(180:1.5mm) -- cycle;
        }
      }

      \scoped[yshift=-12mm] {
        \explanation{3}

        \scoped[spectrum] {
          \foreach \pos in {
            1, 0.875, 0.8125, 0.78125, 0.765625, 0.7578125, 0.75390625, 0.751953125, 0.7509765625, 0.75, 0.734375, 0.7109375, 0.69921875, 0.6953125, 0.693359375, 0.6904296875, 0.68896484375, 0.688232421875, 0.6875, 0.67578125, 0.666015625, 0.6611328125, 0.65869140625, 0.658203125, 0.657470703125, 0.6568603515625, 0.65625, 0.6494140625, 0.646484375, 0.64501953125, 0.642822265625, 0.6417236328125, 0.64117431640625, 0.64111328125, 0.640625, 0.636962890625, 0.6348876953125, 0.63385009765625, 0.63333129882812, 0.6329345703125, 0.6328125, 0.6318359375, 0.63092041015625, 0.62991333007812, 0.62940979003906, 0.62893676757812, 0.62890625, 0.62794494628906, 0.62744903564453,0
          }
          {
            \crossinspec{(\pos,0)}
          }
          \draw [line width=2.4pt] (0,0) -- (.627,0);
          \fill[xshift=.375pt] (-1.2pt,1.2pt) -- (0,1.2pt) -- (0,-1.2pt) --
          (-1.2pt,-1.2pt) -- (0,0) -- cycle;
        }

        \scoped[nl] {
          \foreach \pos in {
            0.5, 0.375, 0.3125, 0.28125, 0.265625, 0.2578125, 0.25390625, 0.251953125, 0.2509765625, 0.25, 0.234375, 0.2109375, 0.19921875, 0.1953125, 0.193359375, 0.1904296875, 0.18896484375, 0.188232421875, 0.1875, 0.17578125, 0.166015625, 0.1611328125, 0.15869140625, 0.158203125, 0.157470703125, 0.1568603515625, 0.15625, 0.1494140625, 0.146484375, 0.14501953125, 0.142822265625, 0.1417236328125, 0.14117431640625, 0.14111328125, 0.140625, 0.136962890625, 0.1348876953125, 0.13385009765625, 0.13333129882812, 0.1329345703125, 0.1328125, 0.1318359375, 0.13092041015625, 0.12991333007812, 0.12940979003906, 0.12893676757812, 0.12890625, 0.12794494628906, 0.12744903564453, 0.12696075439453, 0.126953125, 0.12646865844727, 0.12597846984863, 0.1259765625, 0.12548828125, 0.125, 0.12451171875, 0.1220703125, 0.120849609375, 0.1190185546875, 0.11865234375, 0.11810302734375, 0.11764526367188, 0.1171875, 0.112060546875, 0.10986328125, 0.1087646484375, 0.10711669921875, 0.10629272460938, 0.10588073730469, 0.1058349609375, 0.10546875, 0.103759765625, 0.10272216796875, 0.10116577148438, 0.1007080078125, 0.10038757324219,0.0001
          } 
          {
            \fill (\pos,0) -- ++(60:1.5mm) -- ++(180:1.5mm) -- cycle;
          }
          \draw [line width=0.86602540*1.5mm] (0.0001,0.86602540*.75mm) -- ++(.1,0);
        }

        \scoped[np] {
          \fill (0,0) -- ++(-60:1.5mm) -- ++(180:1.5mm) -- cycle;
        }
      }

      \scoped[yshift=-16mm] {

        \explanation{4}

        \scoped[spectrum] {
          \foreach \pos in {
            1, 0.9375, 0.90625, 0.890625, 0.8828125, 0.87890625,
            0.876953125, 0.8759765625, 0.87548828125, 0.875,
            0.8671875, 0.85546875, 0.849609375, 0.84765625,
            0.8466796875, 0.84521484375, 0.844482421875,
            0.8441162109375, 0.84375, 0.837890625, 0.8330078125,
            0.83056640625, 0.829345703125, 0.8291015625,
            0.8287353515625, 0.82843017578125, 0.828125,
            0.82470703125, 0.8232421875, 0.822509765625,
            0.8214111328125, 0.82086181640625, 0.82058715820312,
            0.820556640625, 0.8203125, 0.8184814453125,
            0.81744384765625, 0.81692504882812, 0.81666564941406,
            0.81646728515625, 0.81640625, 0.81591796875,
            0.81546020507812, 0.81495666503906, 0.81470489501953,
            0.81446838378906, 0.814453125, 0.81397247314453,
            0.81372451782227, 0.81348037719727, 0.8134765625,
            0.81323432922363, 0.81298923492432, 0.81298828125,
            0.812744140625, 0.8125, 0.812255859375, 0.81103515625,
            0.8104248046875, 0
            } {
            \crossinspec{(\pos,0)}
          }
          \draw [line width=2.4pt] (0,0) -- (.81,0);
          \fill[xshift=.375pt] (-1.2pt,1.2pt) -- (0,1.2pt) -- (0,-1.2pt) --
          (-1.2pt,-1.2pt) -- (0,0) -- cycle;
        }

        \scoped[nl] {
          \foreach \pos in {
            0.75, 0.703125, 0.6796875, 0.66796875, 0.662109375, 0.6591796875, 0.65771484375, 0.656982421875, 0.6566162109375, 0.65625, 0.650390625, 0.6416015625, 0.63720703125, 0.6357421875, 0.635009765625, 0.6339111328125, 0.63336181640625, 0.63308715820312, 0.6328125, 0.62841796875, 0.624755859375, 0.6229248046875, 0.62200927734375, 0.621826171875, 0.62155151367188, 0.62132263183594, 0.62109375, 0.6185302734375, 0.617431640625, 0.61688232421875, 0.61605834960938, 0.61564636230469, 0.61544036865234, 0.61541748046875, 0.615234375, 0.61386108398438, 0.61308288574219, 0.61269378662109, 0.61249923706055, 0.61235046386719, 0.6123046875, 0.6119384765625, 0.61159515380859, 0.6112174987793, 0.61102867126465, 0.6108512878418, 0.61083984375, 0.6104793548584, 0.6102933883667, 0.61011028289795, 0.610107421875, 0.60992574691772, 0.60974192619324, 0.6097412109375, 0.60955810546875, 0.609375, 0.60919189453125, 0.6082763671875, 0.60781860351562, 0.60713195800781, 0.60699462890625, 0.60678863525391, 0.60661697387695, 0.6064453125, 0.60452270507812, 0.60369873046875, 0.60328674316406, 0.60266876220703, 0.60235977172852, 0.60220527648926, 0.60218811035156, 0.60205078125, 0.60140991210938, 0.60102081298828, 0.60043716430664, 0.60026550292969, 0.60014533996582, 0.59999942779541, 0.59988784790039, 0.599853515625, 0.59969329833984, 0.59957885742188, 0.59940719604492, 0.59932136535645, 0.59926414489746, 0.59912109375, 0.59903812408447, 0.59889650344849, 0.59876346588135, 0.5987548828125, 0.5984845161438, 0.59834504127502, 0.59820771217346, 0.59820556640625, 0.59806931018829, 0.59793144464493, 0.59793090820312, 0.59779357910156, 0.59765625, 0.59751892089844, 0.59683227539062,0.001
          } 
          {
            \fill (\pos,0) -- ++(60:1.5mm) -- ++(180:1.5mm) -- cycle;
          }
          \draw [line width=0.86602540*1.5mm] (0.001,0.86602540*.75mm) -- ++(.5968,0);

        }

        \scoped[np] {
          \foreach \pos in {
            0.5, 0.46875, 0.453125, 0.4453125, 0.44140625, 0.439453125, 0.4384765625, 0.43798828125, 0.437744140625, 0.4375, 0.43359375, 0.427734375, 0.4248046875, 0.423828125, 0.42333984375, 0.422607421875, 0.4222412109375, 0.42205810546875, 0.421875, 0.4189453125, 0.41650390625, 0.415283203125, 0.4146728515625, 0.41455078125, 0.41436767578125, 0.41421508789062, 0.4140625, 0.412353515625, 0.41162109375, 0.4112548828125, 0.41070556640625, 0.41043090820312, 0.41029357910156, 0.4102783203125, 0.41015625, 0.40924072265625, 0.40872192382812, 0.40846252441406, 0.40833282470703, 0.40823364257812, 0.408203125, 0.407958984375, 0.40773010253906, 0.40747833251953, 0.40735244750977, 0.40723419189453, 0.4072265625, 0.40698623657227, 0.40686225891113, 0.40674018859863, 0.40673828125, 0.40661716461182, 0.40649461746216, 0.406494140625, 0.4063720703125, 0.40625, 0.4061279296875, 0.405517578125, 0.40521240234375, 0.40475463867188, 0.4046630859375, 0.40452575683594, 0.40441131591797, 0.404296875, 0.40301513671875, 0.4024658203125, 0.40219116210938, 0.40177917480469, 0.40157318115234, 0.40147018432617, 0.40145874023438, 0.4013671875, 0.40093994140625, 0.40068054199219, 0.40029144287109, 0.40017700195312, 0.40009689331055, 0.39999961853027, 0.39992523193359, 0.39990234375, 0.39979553222656, 0.39971923828125, 0.39960479736328, 0.3995475769043, 0.39950942993164, 0.3994140625, 0.39935874938965, 0.39926433563232, 0.3991756439209, 0.399169921875, 0.3989896774292, 0.39889669418335, 0.39880514144897, 0.3988037109375, 0.39871287345886, 0.39862096309662, 0.39862060546875, 0.39852905273438, 0.3984375, 0.39834594726562, 0.39788818359375,0
          } 
          {
            \fill (\pos,0) -- ++(-60:1.5mm) -- ++(180:1.5mm) -- cycle;
          }
          \draw [line width=0.86602540*1.5mm] (0,-0.86602540*.75mm) -- ++(.398,0);
        }
      }
        
      \scoped[thick] {
        \node [xshift=-1mm,anchor=mid east] at (0,4mm) {$p=$};
        \foreach \pos/\val/\r in {
          1/1/0,
          .875/{\frac{7}{8}}/0,
          .75/{\frac{3}{4}}/0,
          .5/{\,\frac{1}{2}}/0,
          .25/{\frac{1}{4}}/0,
          .125/{\frac{1}{8}}/0,
          .0625/{\frac{1}{16}}/0,
          .03125/{\frac{1}{32}}/0,
          0.703125/{\frac{3\cdot 15}{4\cdot 16}}/0,
          0.4921875/{\llap{$\frac{3\cdot 3\cdot 7}{4\cdot 4 \cdot 8}{=}$}\frac{63}{128}}/1,
          0.46875/{\llap{$\frac{1\cdot 15}{2\cdot 16} {=}$}\frac{15}{32}\;}/0,
          0/0/0
        } {
          \draw [help lines]  (\pos,1mm+6*\r mm) coordinate(x) -- (\pos,-18mm);
          \node [anchor=mid] at (\pos,4mm+6*\r mm) {$\val$};
        }
      }
    }
  \end{tikzpicture}
  \caption{
    Visualization of the complexity of $k\Lang{sat-pr}_{>p}$
    for $k=1$, $k=2$, 
    $k=3$, and $k=4$. Each green triangle represents a value of~$p$
    for which the characterization from
    Theorem~\ref{thm-main} states $\Class{NL}$-completeness, while
    for each red triangle it states $\Class{NP}$-completeness (there are
    no green triangles directly above red triangles, but this is not
    possible to visualize as these are highly intertwined). Each gray
    cross is an element of $k\CNFSSigmaSpectrum$. For them,
    $k\Lang{sat-pr}_{>p}$ lies in $\Class{AC}^0$ according to
    Theorem~\ref{thm-main}. For ``white'' values of $p$, which
    lie outside the spectra, $k\Lang{sat-pr}_{>p}$ also lies in
    $\Class{AC}^0$.  Note that while the visualization may suggest
    that the spectra become dense close to~$0$, they are in fact
    nowhere dense (by Theorem~\ref{thm-order}, the Spectral
    Well-Ordering Theorem). For a discussion of
    the marked specific values like $p = 63/128$, please see the
    conclusion. 
  }\label{fig-spec}
\end{figure}

In the following, (just) the \emph{ideas} underlying the above
contributions are elaborated.  

\subparagraph*{Overview of the Order-Theoretic Results.}
In a sense, the ``reason'' why the functions $\Prob{\cdot}$ and
$\#(\cdot)$ are so hard to compute, lies in the fact that
the spectrum
$\CNFSSigmaSpectrum := 
\{\Prob{\phi} \mid \phi \in \Lang{cnfs}\} = \bigcup_k
k\CNFSSigmaSpectrum$ is just the set $\mathbb D$ of dyadic rationals
(numbers of the 
form $m/2^e$ for integers $m$ and~$e$) between $0$ and~$1$ (see
Lemma~\ref{lemma-dyadic}). In particular, it is  a dense 
subset of~$[0,1]$ and in order to conclusively decide
whether, say, $\Prob{\phi} \ge 1/3$ holds for an arbitrary $\phi \in \Lang{cnfs}$, we may need to
determine all of the first $n$ bits of~$\Prob{\phi}$.

A key insight of Akmal and Williams is that for fixed~$k$, the spectra
$k\CNFSSigmaSpectrum$ behave differently, at least near to~$1$: There
are ``holes'' like $3\CNFSSigmaSpectrum \cap (7/8,1) = \emptyset$
since for a $\Lang{3cnf}$ formula~$\phi$ we cannot have $7/8 <
\Prob{\phi} < 1$ (a single size-$3$ clause already lowers the satisfaction
probability to at most $7/8$). This implies immediately that, say,
$3\Lang{sat-pr}_{>9/10}$ is actually a quite trivial problem: The
\emph{only} formula in $\Lang{3cnfs}$ having a satisfaction probability
larger than $9/10$ has probability~$1$ and is the trivial-to-detect
tautology $\phi = \emptyset$. In general, for all~$k$ we have 
$k\CNFSSigmaSpectrum \cap (1-2^{-k},1) = \emptyset$.

Of course, $3\CNFSSigmaSpectrum$ does not have ``holes above every
number~$p$'' as we can get arbitrarily close to, say, $p =
3/4$: Just consider the sequence of $\Lang{3cnf}$ formulas
$\phi_1=\bigl\{\{a,b, x_1\}\bigr\}$, $\phi_2=\bigl\{\{a,b,x_1\},\{a,b, 
x_2\}\bigr\}$, $\phi_3=\bigl\{\{a, b,x_1\},\{a, b,x_2\},\{a,b, x_3\}\bigr\}$, and so
on with $\Prob{\phi_i} = 3/4 + 2^{-i-2}$ and
$\lim_{i\to\infty}\Prob{\phi_i} = 3/4$. Nevertheless, Akmal and Williams point out that their algorithm is
in some sense based on the intuition that there are ``lots of holes'' in
$k\CNFSSigmaSpectrum$. The new Spectral Well-Ordering Theorem,
Theorem~\ref{thm-order} above, turns this intuition into a formal statement.

\emph{Well-orderings} are a standard notion of order theory; we will
just need the special case that we are given a set $X$ of non-negative
reals and consider the total order~$>$ on it. Then \emph{$X$ is
well-ordered (by~$>$)} if there is no infinite strictly \emph{increasing}
sequence $x_0 < x_1 < x_2 < \cdots$ of numbers $x_i \in
X$ or, equivalently, if $X$ is bounded and for every~$x \in \mathbb
R^{\ge 0}$ there is an $\epsilon > 0$ 
such that $(x -\epsilon, x) \cap X = \emptyset$ or, again
equivalently, if every subset of $X$ contains a maximum. 
In particular, Theorem~\ref{thm-order} tells us that every 
$\Phi \subseteq k\Lang{cnfs}$ contains a formula $\phi \in \Phi$ 
of maximum satisfaction probability, that is, $\Prob{\phi} \ge
\Prob{\phi'}$ for all $\phi' \in \Phi$.  (Observe that this is
certainly no longer true  when  we replace ``maximum'' by ``minimum''
as the set $\Phi = \bigl\{\bigl\{\{x_1\},\dots,\{x_n\}\bigr\} \mid n
\in \mathbb N\bigr\} \subseteq \Lang{1cnfs}$ shows.)

While $k\CNFSSigmaSpectrum$ is not well-ordered by~$<$ (only
by~$>$), with a small amount of additional work we will be able to show
that it is at least \emph{topologically closed,} 
see Corollary~\ref{corollary-closed}. A succinct way of stating both this and
Theorem~\ref{thm-order} is that for every set
$X \subseteq k\CNFSSigmaSpectrum$ of numbers we have $\sup X \in X$ and $\inf X
\in k\CNFSSigmaSpectrum$.

The proof of Theorem~\ref{thm-order} will need only basic
properties of well-ordered sets of reals, like their 
being closed under finite sums and unions, and a simple
relationship (which also underlies Akmal and
Williams' analysis~\cite{AkmalW2022})  
between the satisfaction probability of a formula~$\phi$ and the size of 
\emph{packings} $\pi \subseteq \phi$, which are just sets of
pairwise variable-disjoint clauses:

\begin{lemma}[Packing Probability Lemma]\label{lemma-pr}
  Let $\phi \in k\Lang{cnfs}$ and let $\pi \subseteq \phi$ be a
  packing. Then $\Prob{\phi} \le (1-2^{-k})^{|\pi|}$ and,
  equivalently, $\log_{1-2^{-k}}(\Prob{\phi}) \ge |\pi|$.
\end{lemma}
\begin{proof}
  We have $\Prob{\phi} \le \Prob{\pi} = \prod_{c \in \pi}
  (1-2^{-|c|}) \le (1-2^{-k})^{|\pi|}$
  as all clauses of~$\pi$ are variable-disjoint and, hence, their
  satisfaction probabilities are pairwise independent.  
\end{proof}

A simple consequence of the Packing Probability Lemma will be that for
every $\phi 
\in k\Lang{cnfs}$, we can write $\Prob{\phi}$ as a sum $\sum_{i=1}^s 
\Prob{\phi_i}$ with $\phi_i \in (k-1)\Lang{cnfs}$ in such a way that
$s$ depends only on~$\Prob{\phi}$, which will almost immediately
yield Theorem~\ref{thm-order}.

\subparagraph*{Overview of the Algorithmic Results.} A first,
surprisingly simple application of the \emph{existence of (spectral)
gaps} will be in the form of decision procedures for $k\Lang{sat-pr}_{\ge
  p}$ (and, ultimately, also for the  ``$_{>  p}$'' version). Using the
existence of gaps is actually common in algorithmic theory, though the 
objective is usually to prove that certain algorithms do \emph{not} 
exist (for instance, in \cite[Theorem 13.4]{Papadimitriou1994}
gaps in the spectrum of possible $\Lang{tsp}$-tour lengths are
used to show that $\Lang{tsp}$ cannot be approximated unless $\Class P
= \Class{NP}$).
The satisfaction probabilities of $k\Lang{cnf}$ formulas \emph{can} be
approximated efficiently and, because of the existence of spectral
gaps, we \emph{are} able to solve $k\Lang{sat-pr}_{\ge p}$ 
just as efficiently. Indeed, there is a randomized folklore
algorithm for approximating~$\Prob{\phi}$ with
small \emph{additive} error $\epsilon > 0$: \emph{Randomly sample assignments
$\beta_1,\dots,\beta_r \colon \operatorname{vars}(\phi) 
\to \mathbb B$ for  $r = O(1/\epsilon^2)$ and output the fraction $f :=
\bigl|\bigl\{ i \mid \beta_i \models \phi\bigr\}\bigr| \bigm/ r$  of assignments satisfied in the sample.} By the
Chernoff bound, this fraction differs from~$\Prob{\phi}$ by less
than~$\epsilon$ with high probability. Setting $\epsilon$ to half the
size of the spectral gap below some~$p$, see also  
Figure~\ref{fig-gap-only}, with high probability $f > p - \epsilon$
implies $\Prob{\phi} \ge p$ and $f < p - \epsilon$ implies
$\Prob{\phi} < p$. All told, we get $k\Lang{sat-pr}_{\ge p} \in
\Class{BPP}$ via an algorithm whose runtime is linear in
$|\phi|/\bigl(\operatorname{spectral-gap}_{k\Lang{cnfs}}(p)\smash{\bigr){}^2}$. Even
better, 
Trevisan~\cite{Trevisan2004} showed that there is a simple
derandomization of the sketched approximation algorithm that runs in
linear time for fixed~$\epsilon$. 
Trevisan's algorithm and the Spectral Well-Ordering Theorem are all
that is needed to prove Akmal and Williams' striking result
$k\Lang{sat-pr}_{\ge p} \in \Class{LINTIME}$.

\begin{figure}[htpb]
  \hfil\begin{tikzpicture}[x=28.9cm,figure] 

    \scoped {

      \draw [black!15,line width=.5mm] (0.46875,0) -- (0.421875,0);

      \node [anchor=mid,font=\small] at (0.315,0) {$0$};
      \node [anchor=mid,font=\small] at (0.785,0) {$1$};
      \draw [red!15,line width=1mm] (0.34,0) -- (0.421875,0);
      \draw [green!50!black!20,line width=1mm] (0.46875,0) -- (0.76,0);

      \draw [green!50!black,line width=1mm] (0.485-0.0234375,-2mm) -- ++(0.046875,0);
      \draw [red!75!black,line width=1mm] (0.415-0.0234375,-2mm) -- ++(0.046875,0);

      \foreach \pos/\col in {
        0.76/green!50!black!20,
        0.76+0.005/green!50!black!20,
        0.76+0.010/green!50!black!20,
        0.76+0.015/green!50!black!20,
        0.34/red!15,
        0.34-0.005/red!15,
        0.34-0.010/red!15,
        0.34-0.015/red!15%
      } { \fill[\col] (\pos,0) circle[radius=.5mm]; }
    }

    \scoped[thick] {
      \draw [green!50!black] (.46875,-.5mm) -- ++(0,2.5mm) node [above] {$p$};

      \draw [black!25] (0.4453125,0) -- ++(0,2mm) node [above,black!50] {$p-\epsilon$};

      \draw [green!50!black] (.485,-2mm) -- ++(0,7mm) node [above] {$\Prob{\psi'}$};

      \draw [red!75!black] (.415,-2mm) -- ++(0,7mm) node [above] {$\Prob{\psi}$};

      \node[below,black!75] at (0.4453125,-4mm) {$\underbrace{\kern1.3cm}_{\operatorname{spectral-gap}_{k\Lang{cnfs}}(p)\rlap{$\scriptscriptstyle=2\epsilon$}}$};

      \node[below,green!50!black] at (.485,-2mm) {$I_{\psi'}$};
      \node[below,red!75!black] at (.415,-2mm) {$I_{\psi}$};

    }
    
  \end{tikzpicture}
  \caption{ {
      The long line visualizes the set $[0,1]$ of possible
      satisfaction probabilities with a gray \emph{spectral gap} below
      some value~$p$, meaning that $\Prob{\phi}$ cannot lie in this
      gap for any $\phi \in k\Lang{cnfs}$. In particular, for two
      formulas $\psi,\psi' \in k\Lang{cnfs}$ the values $\Prob{\psi}$
      and $\Prob{\psi'}$ must either lie in the light red part (as is the
      case for $\Prob{\psi} \le p -
      \operatorname{spectral-gap}_{k\Lang{cnfs}}(p) < p$) or in the
      light green part (as is the case for $\Prob{\psi'} \ge p$). The open
      interval $I_{\psi}$ is centered on $\Prob{\psi}$ and protrudes to the
      left and to the right by $\epsilon = 
      \operatorname{spectral-gap}_{k\Lang{cnfs}}(p)/2$; likewise for
      $I_{\psi'}$. Suppose that for an arbitrary value $f \in [0,1]$
      we (just) know that $f$ must lie in (some) interval $I_{\phi}$, centered
      on (an unknown) $\Prob{\phi}$ and having the size of the
      spectral gap. Then $f < p - \epsilon$ implies 
      that $\Prob{\phi}$ must ``lie in the red part'' and
      $\Prob{\phi}<p$ holds, whereas $f > p - \epsilon$ implies that
      $\Prob{\phi}$ ``lies in the green part'' and $\Prob{\phi}
      \ge p$ holds.
  } }
  \label{fig-gap-only}
\end{figure}

While the above insights demonstrate the importance of the existence
of spectral gaps from an algorithm point of view, they are not enough
to reach our ultimate goal, the classification of the
complexity of $k\Lang{sat-pr}_{>p}$. We will need additional
results on the properties of the formulas~$\phi$ with
$\Prob{\phi} > p$. We will gain the necessary insights 
\emph{by looking at $k\Lang{sat-pr}_{\ge p}$ through the lens of
{\slshape\textsc{fpt}} theory.} Specifically, we will \emph{kernelize} this
problem (for the joint parameters $k$ and~$p$) using two different
algorithms, each of which just applies a simple reduction rule
exhaustively.   
Kernels are one of the core tools of \textsc{fpt} theory: Given an  
instance (a formula $\phi \in k\Lang{cnfs}$ in our case) for a problem
($k\Lang{sat-pr}_{\ge p}$ in our case), a \emph{kernel} is a
membership-equivalent instance $\phi^* \in k\Lang{cnfs}$ (meaning
$\Prob{\phi} \ge p$ iff $ \Prob{\phi^*} \ge p$ in our case) whose
\emph{size} can be bounded purely in terms of the parameters (in terms
of $k$ and~$p$ in our case). Given a formula $\phi\in k\Lang{cnfs}$, a
\emph{reduction rule} may be \emph{applicable to~$\phi$} and, if so,
yields a simpler formula $\phi'\in k\Lang{cnfs}$ (\emph{simpler}
meaning $|\phi'| < |\phi|$ in our case). The rule is \emph{safe} if
$\phi'$ and $\phi$ are always membership-equivalent (meaning
$\Prob{\phi} \ge p$ iff $\Prob{\phi'}\ge p$ in~our~case). 

The key property of the two reduction rules analyzed in the present
paper (Rules~\ref{rule-pluck} and~\ref{rule-prune}) is that
they both compute a~$\phi'$ 
whose satisfaction probability is nearly the same as than
of~$\phi$. In fact, the formulas will be \emph{gap-close:}
\begin{definition}[Gap-Close Formulas]\label{def-gap-close}
  Formulas $\phi,\phi' \in k\Lang{cnfs}$ are \emph{gap-close (for $k$
  and~$p$)} if $\bigl|\Prob{\phi}- \Prob{\phi'}\bigr| <
  \operatorname{spectral-gap}_{k\Lang{cnfs}}(p)$. 
\end{definition}
For such formulas a curious thing happens: \emph{Both}
of $\Prob{\phi}$ and $\Prob{\phi'}$ must be
at least~$p$ or \emph{both} must be below~$p$ -- it is not possible
that the probability ``tunnels through the gap'':
\begin{lemma}[No Tunneling Lemma]\label{lem-tunnel}
  Let $\phi,\phi' \in k\Lang{cnfs}$ be gap-close for
  $k$ and~$p$. Then $\Prob{\phi} \ge p$ iff $\Prob{\phi'} \ge p$.  
\end{lemma}
\begin{proof}
  Without loss of generality, consider the case $\Prob{\phi} \ge
  \Prob{\phi'}$. Then $\Prob{\phi'} \ge p$ clearly implies
  $\Prob{\phi} \ge p$. If $\Prob{\phi} \ge p$, then
  $\Prob{\phi'} > p - 
  \operatorname{spectral-gap}_{k\Lang{cnfs}}(p)$. As
  $\Prob{\phi'}$ cannot lie in the spectral gap, 
  $\Prob{\phi'} \ge p$.
\end{proof}
In other words, any reduction rule for which $\phi'$ is always
gap-close to~$\phi$ is \emph{safe} with respect to
$k\Lang{sat-pr}_{\ge p}$ (see Figure~\ref{fig-gap} for a visualization
of the effect of this safety property when such a rule is applied
repeatedly).

\begin{figure}[htpb]
  \hfil\begin{tikzpicture}[x=28.9cm,figure] 
    
    \def\skips#1#2{
      \draw[shorten <=1pt,arrows=-{>[sep]}] (#1,2pt) to[bend right] (#2,2pt);
    }
    \def\skipsprime#1#2{
      \draw[shorten <=1pt,arrows=-{>[sep]}] (#1,2pt) to[bend left] (#2,2pt);
    }

    \def\spectrum{
      \scoped {
        \node [anchor=mid,font=\small] at (0.315,0) {$0$};
        \node [anchor=mid,font=\small] at (0.785,0) {$1$};
        \draw [black!15,line width=.5mm] (0.46875,0) -- (0.421875,0);
        \draw [cap=round, red!15,line width=1mm] (0.34,0) -- (0.421875,0);
        \draw [cap=round, green!50!black!20,line width=1mm] (0.46875,0) -- (0.76,0);
        \foreach \pos/\col in {
          0.76+0.005/green!50!black!20,
          0.76+0.010/green!50!black!20,
          0.76+0.015/green!50!black!20,
          0.34-0.005/red!15,
          0.34-0.010/red!15,
          0.34-0.015/red!15
        } { \fill[\col] (\pos,0) circle[radius=.5mm]; }
          
        \foreach \pos in {
          0.75, 0.625, 0.5625, 0.53125, 0.515625, 0.5078125,
          0.50390625, 0.5, 0.46875, 0.421875, 0.3984375, 0.390625,
          0.38671875, 0.380859375, 0.3779296875, 0.375, 0.3515625
        } {
          \crossinspec{(\pos,0)}
        }
      }
    }
    
    \scoped[yshift=1.8cm]{
      \scoped[thick] {
        \foreach \pos/\val/\r/\style in {
          0.46875/{p\rlap{\color{black}${}=\Prob{\phi^*} = \cdots =\Prob{\phi_4}$}}/7/green!50!black,
          0.5625/{\kern2mm\Prob{\phi_1}}/3/very thin,
          0.53125/{\Prob{\phi_2}}/3/very thin,
          0.5/{\Prob{\phi_3}\kern2mm}/3/very thin%
        } {
          \draw [\style]  (\pos,\r mm) coordinate(x) -- (\pos,0mm);
          \node [anchor=mid,\style] at (\pos,\r mm+2mm) {$\val$};
        }
        \node[below] at (0.4453125,0) {$\underbrace{\kern1.3cm}_{\operatorname{spectral-gap}_{k\Lang{cnfs}}(p)}$};

        \node[below] at (0.546875,0)
             {$\underbrace{\kern.8cm}_{\left|\Prob{\phi_i} - \Prob{\phi_{i+1}}\right|\rlap{$\scriptstyle{}<\operatorname{spectral-gap}_{k\Lang{cnfs}}(p)$}}$};
      }
      
      \spectrum
      \skips{0.5625}{0.53125}
      \skips{0.53125}{0.5}
      \skips{0.5}{0.46875}
      \draw [->>,shorten <=1pt,shorten >=1mm] (.46875,4pt) .. controls
      +(35:5mm) and +(65:5mm) .. +(0,0);
      \scoped[black!50,densely dashed]{\skips{0.46875}{0.44}}
    }

    \draw [help lines] (0.34,.8) -- (0.76,.8);
    
    \scoped[yshift=-6mm]{
      \scoped[thick] {
        \foreach \pos/\val/\r/\style in {
          0.46875/{p}/3/green!50!black,
          0.421875/{\Prob{\phi_4}\rlap{${}=\cdots=\Prob{\phi^*}$}}/8.5/very thin,%
          0.38671875/{\Prob{\phi_3}}/4.5/very thin,%
          0.3515625/{\Prob{\phi_1}=\Prob{\phi_2}}/8.5/very thin%
        } {
          \draw [\style]  (\pos,\r mm) coordinate(x) -- (\pos,0mm);
          \node [anchor=mid,\style] at (\pos,\r mm+2mm) {$\val$};
        }
      }
      
      \spectrum
      \skipsprime{0.38671875}{0.421875}
      \draw [->,shorten <=1pt,shorten >=1mm] (0.3515625,4pt) .. controls
      +(35:5mm) and +(65:5mm) .. +(0,0);
      \draw [->>,shorten <=1pt,shorten >=1mm] (0.421875,4pt) .. controls
      +(35:5mm) and +(65:5mm) .. +(0,0);
      \skipsprime{0.3515625}{0.38671875}
      \scoped[black!50,densely dashed]{\skipsprime{0.421875}{0.45}}
    }

    \draw [help lines] (0.34,-.9) -- (0.76,-.9);
    
    \scoped[yshift=-2cm] {
      \scoped[thick] {
        \foreach \pos/\val/\r/\style in {
          0.46875/{p}/3/green!50!black,
          0.75/{\llap{$\Prob{\phi^*} = \cdots =\Prob{\phi_2}={}$}\Prob{\phi_1}}/4/very thin%
        } {
          \draw [\style]  (\pos,\r mm) coordinate(x) -- (\pos,0mm);
          \node [anchor=mid,\style] at (\pos,\r mm+2mm) {$\val$};
        }
      }
      
      \spectrum
      \draw [->>,shorten <=1pt,shorten >=1mm] (.75,2pt) .. controls
      +(35:5mm) and +(65:5mm) .. +(0,0);
      \scoped[black!50,densely dashed]{\skips{0.75}{0.72}}
    }
    
  \end{tikzpicture}
  \caption{ {
      Different ways how $\Prob{\phi_i}$ can change in a sequence
      $(\phi_1,\phi_2,\dots,\phi_q)$, ending with some $\phi^* =
      \phi_q$, when $\phi_i$ is  always gap-close to $\phi_{i-1}$. In each row, the crosses in the
      red lines are elements of $k\CNFSSigmaSpectrum$ smaller
      than~$p$, while the crosses on the green lines are at
      least~$p$. In the first line, $\Prob{\phi_1} \ge p$ holds 
      and each $\phi_{i+1}$ has smaller or equal satisfaction
      probability, but the gap-closeness ensures that
      $\Prob{\phi^*}$ gets ``stuck'' at $p$ as it cannot ``tunnel
      through'' the spectral gap by Lemma~\ref{lem-tunnel} and the
      dashed arrow is an impossible change in the satisfaction 
      probability. In the second line, the satisfaction probabilities
      increase, but also get stuck, only now at the lower
      end of the spectral gap. In the third line, $\Prob{\phi_i}$ is
      stuck at a much larger value than~$p$.
  } }
  \label{fig-gap}
\end{figure}

It is surprisingly easy to come up with a rule that is safe because of
the No Tunneling Lemma: \emph{Find a large sunflower in $\phi$ and 
pluck its petals} (Rule~\ref{rule-pluck} later on). The rule
also underlies the algorithm of Akmal and Williams~\cite{AkmalW2022},
but see the Section~\ref{sec-related} on related work for a discussion of the
differences. The central concept, \emph{sunflowers,} are generalizations of
packings (packings are sunflowers with an empty core): 
\begin{definition}\label{def-sunflower}
  A \emph{sunflower with core~$c$} is a formula $\psi \in \Lang{cnfs}$
  such that $c \subseteq e$ 
  holds for all $e \in \psi$ and such that for any two different $e,e' \in
  \psi$ we have $\operatorname{vars}(e) \cap \operatorname{vars}(e') =
  \operatorname{vars}(c)$.
\end{definition}
\begin{figure}[htbp]
  \centering
  \begin{tikzpicture}[sunflower  figure]

    \scoped{
      \pictureexplain{$\phi$}
      \pictureliterals
      \scoped[dashed]{\picturekernel}
      
      \scoped[red!75!black] {
        \pic {join 3' = a' on 2 with b on 2 with c on 1};
        \pic {join 4' = a' on 3 with b on 3 with d' on 1 with j on 1};
        \pic {join 3' = a' on 4 with b on 4 with e on 1};
        \pic {join 3' = a' on 5 with b on 5 with g' on 1};
        \pic {join 4' = a' on 6 with b on 6 with f on 1 with k on 1};
        
        \pic [dotted] {join 4' = a' on 7 with b on 7 with f on 2 with i' on 1};
        \pic [dotted] {join 3' = a' on 8 with b on 8 with f' on 1};
      }

      \scoped[xshift=7cm] {
        \pictureexplain{$\phi' = (\phi \setminus \psi) \cup \{c\}$}
        \pictureliterals
        \scoped[dashed]{\picturekernel}
        
        \pic [dotted] {join 4' = a' on 7 with b on 7 with f on 2 with i' on 1};
        \pic [dotted] {join 3' = a' on 8 with b on 8 with f' on 1};
        
        \scoped[green!50!black] {
          \pic {join 3 = l on 4 with a' on 4 with b on 4};
        }
      }
    }
  \end{tikzpicture}
  \caption{Left, a formula $\phi \in \Lang{5cnfs}$ is
    visualized by drawing, for each clause in~$\phi$, a line that
    ``touches'' exactly the clause's literals; so the upper dashed
    line represents the clause $\{f,x,y,z\}$. The solid(-line) clauses
    form a sunflower $\psi \subseteq \phi$ with core 
    $c = \{x,\neg y,z\}$. Although the dotted clauses also contain the
    core, they are not part  of the sunflower: The upper dotted clause 
    $\{x,\neg y,z,l,\neg e\}$ shares the literal ``$l$'' with the petal
    $\{x,\neg y, z,l,m\}$ of the sunflower, while the second dotted
    clause shares the variable ``$l$'' (though not the literal) with
    this petal. The dashed clauses are not part of the sunflower as
    they do not contain all of the literals of the 
    core (containing the variables is not enough). A key property of
    a sunflower is that it is ``unlikely that an assignment makes the
    sunflower true, but not its core'': For $\phi$,
    this happens only when $g$, $j$, and $\neg k$ are all set to
    true as well as at least one of $\neg h$ or $i$, and one of $l$
    or~$m$. The probability that all of this happens is just
    $\frac{1}{2}\cdot\frac{1}{2}\cdot\frac{1}{2} \cdot
    \frac{3}{4}\cdot\frac{3}{4} = \frac{9}{128}$. In particular, for
    $\phi' = (\phi \setminus \psi) \cup \{c\}$ shown right, we have $\Prob{\phi} -
    \Prob{\phi'} \le \frac{9}{128}$.
  }
  \label{fig-sunflower-only}
\end{figure}

The clauses of a sunflower ``agree on the
literals in~$c$, but are variable-disjoint otherwise,'' see the
clauses represented by solid lines in Figure~\ref{fig-sunflower-only} for an
example. The sunflower clauses will be referred to
as \emph{petals} in the following (but note that in the literature
this term may also refer to the clauses \emph{without} the shared
core~$c$). For any sunflower~$\psi$ with some core~$c$, we have  
\begin{align}
  \Prob{\psi} =  \Prob{\{c\}} +
  2^{-|c|}\ProbBig{\{e \setminus c \mid e \in \psi\}} \label{eq-pr-sunflower}
\end{align}
and as $\{e \setminus c \mid e \in \psi\}$ is clearly a packing, the Packing
Probability Lemma implies $\Prob{\psi} - \Prob{\{c\}} \le 
2^{-|c|}(1-2^{-k})^{|\psi|}$. Since the right-hand side decreases
exponentially as the size~$|\psi|$ of the sunflower
increases, \emph{large enough sunflowers are gap-close
to their cores} and replacing such a sunflower $\psi \subseteq \phi$
by its core inside a larger formula~$\phi$ yields a new formula $\phi'
= (\phi \setminus \psi) \cup \{c\}$ that is gap-close to~$\phi$. Thus,
the reduction rule \emph{``find a large sunflower and pluck its petals''}
is safe (Lemma~\ref{lemma-safety}). 
Of course, the rule is no longer applicable when there are
no large sunflowers left, but, then, the  Erdős--Rado Sunflower
Lemma~\cite{ErdosR60} kicks in and states that \emph{the
formula has constant size} and we have thus \emph{computed a kernel!} 
All told, with very little effort, we can augment the standard
\emph{Sunflower Kernel Algorithm} from \textsc{fpt} theory
(``as long as possible, find a sunflower and pluck its petals'') so
that it decides $k\Lang{sat-pr}_{\ge p}$ in linear time (or in
$\Class{AC}^0$ as one can parallelize this kernel
algorithm~\cite{BannachT18}).

Instead of replacing a sunflower by its core, one can also
\emph{prune} the sunflower, meaning that one replaces it by a
fixed-size subset of its clauses. The resulting 
Rule~\ref{rule-prune} also yields gap-close formulas, is hence
also safe (Lemma~\ref{lemma-safety}), and also results in a
kernel. However, the kernel now has the desirable property that it is
a subset of the original~$\phi$, which is needed to prove the
Threshold Locality Theorem (Theorem~\ref{thm-local}).

To summarize, the
findings on the kernelization of satisfaction thresholds are as follows:

\begin{theorem}[Kernel Theorem for \textit{k}{\footnotesize CNFS} Satisfaction
    Thresholds]\label{thm-kernel}
  For each $k$ and $p$, on input $\phi \in k\Lang{cnfs}$ we can
  compute formulas $\phi^*_{\mathrm{pluck}}, \phi^*_{\mathrm{prune}}
  \in k\Lang{cnfs}$ such that:
  \begin{enumerate}
  \item The computation takes linear time or is done by
    $\Class{AC}^0$ circuits. 
  \item $|\phi^*_{\mathrm{pluck}}| \le S_{k,p}$ and
    $|\phi^*_{\mathrm{prune}}| \le S_{k,p}$ for a constant~$S_{k,p}$
    depending only on $k$ and~$p$. 
  \item $\phi^*_{\mathrm{prune}} \subseteq \phi$.
  \item $\phi^*_{\mathrm{pluck}}(\phi) \FormImpl \phi \FormImpl
    \phi^*_{\mathrm{prune}}(\phi)$ and hence $\Prob{\phi^*_{\mathrm{pluck}}} \le \Prob{\phi} \le \Prob{\phi^*_{\mathrm{prune}}}$.
  \item $\Prob{\phi} \ge p$ iff $\Prob{\phi^*_{\mathrm{pluck}}} \ge p$ iff $\Prob{\phi^*_{\mathrm{prune}}} \ge p$.
  \end{enumerate}
\end{theorem}
(By item~4, applying the plucking rule repeatedly can only decrease
the probability, while applying the pruning rule can only increase
it. Since we will use this fact quite often in the following and since
``plucking'' and ``pruning'' sound a bit similar, here is an easy
mnemonic: p\emph{l}ucking only \emph{l}owers probabilities,
p\emph{r}uning only \emph{r}aises probabilities.)

We also study a conceptually quite different algorithm for deciding
$k\Lang{sat-pr}_{\ge p}$. Its key property will be that it is
\emph{gap size oblivious,} meaning that we can run the 
algorithm without knowledge of  
(even just a bound on) the size of the spectral gap below 
some~$p$. Instead, there will be an easy-to-check termination property
which we show to always hold after a constant number of steps
(with the constant depending on the size of the spectral gap, but we
do not need to know it beforehand). The algorithm might be of
practical interest: While it is well-established in \textsc{fpt} theory 
that  being able to compute kernels for a parameterized problem
is in some sense the best one can hope for from a theoretical point of
view, ``in practice'' it is of high interest
how we can actually decide whether $\Prob{\phi^*} \ge p$ holds for a
kernel~$\phi^*$. Of course, as the kernel size is fixed, this 
can be decided in constant time by brute-forcing all assignments,
but a practical algorithm will need to use different ideas -- such as
those of the gap size oblivious algorithm.

\subparagraph*{Overview of the Structural Complexity Results.}
The proof of Theorem~\ref{thm-main-simple}, which states that 
$k\Lang{sat-pr}_{>p}$ is always $\Class{NP}$-complete,
$\Class{NL}$-complete, or lies in $\Class{AC}^0$, will be
based on a characterization of which case applies for which values of
$k$ and~$p$ in terms of ``formulas that have room for
$t\Lang{sat}$'' (think of~$t$ as ``two'' or ``three,'' which will be
the values we are mostly interested~in). We need three simple
definitions:
\begin{definition}\label{def-tight}
  A formula $\omega \in \Lang{cnfs}$ is \emph{irredundant} if there is no
  $\omega' \subsetneq \omega$ with $\omega' \equiv \omega$.
\end{definition}
In an irredundant formula~$\omega$ there are no redundant clauses in
the sense that no clause is already implied by the other clauses. In
particular, for each clause $c_* \in \omega$ some assignment $\beta_*
\colon \operatorname{vars}(\omega) \to \mathbb B$ 
``witnesses $c_*$'s irredundancy in~$\omega$,''
meaning $\beta_* \models \omega\setminus\{c_*\}$ but $\beta_*
\not\models \{c_*\}$.  
\begin{definition}\label{def-room1}
  For $k$ and $t$, a formula $\omega \in k\Lang{cnfs}$ \emph{has
  room for $t\Lang{sat}$} if it is irredundant and contains a clause~$c_*$
  of size $|c_*| \le k-t$. 
\end{definition}
\begin{definition}\label{def-open}
  For $k$, $t$, and $p$, \emph{the $k\Lang{cnfs}$ have room
  for $t\Lang{sat}$ at~$p$} if some $\omega \in k\Lang{cnfs}$
  with $\Prob{\omega} = p$  has room for $t\Lang{sat}$.
\end{definition}
For example, the $3\Lang{cnf}$ formulas have room for $2\Lang{sat}$ at $p =
7/32$. To see this, consider  $\omega = 
\bigl\{\{a\},\penalty50 \{b\},\penalty0 \{c_1,c_2,c_3\}\bigr\} \in \Lang{3cnfs}$. It is
clearly irredundant,  contains a clause
$c_* = \{a\}$ of size $3-2 = 1$ (of course, $\{b\}$ is another
possible choice for~$c_*$), and $\Pr[\omega] = \frac{1}{2} \cdot
\frac{1}{2} \cdot\frac{7}{8} = 7/32$. The curious phrase
``$\omega$ has room for $t\Lang{sat}$'' is motivated by a simple
observation: The very existence of~$\omega$ implies that we can reduce 
$\Lang{2sat}$ to $3\Lang{sat-pr}_{>7/32}$ by mapping each input formula
$\psi \in \Lang{2cnfs}$ (with fresh variables, that is,
$\operatorname{vars}(\psi) \cap \operatorname{vars}(\omega) = \emptyset$) to
\begin{align*}
  \rho = \underbrace{\bigl\{\overbrace{c_*
  \cup d}^{\text{size$\le$3}} \mid d \in
    \overbrace{\psi}^{\kern-4ex\in\Lang{2cnfs}\kern-4ex}\bigr\}}_{\in\Lang{3cnfs}}
  \cup
  \underbrace{\bigl\{\{b\}, \{c_1,c_2,c_3\}\bigr\}}_{\in\Lang{3cnfs}} \in \Lang{3cnfs},
\end{align*}
that is, by ``adding the clauses of $\psi$ to the small clause $c_* =
\{a\}$.'' The important observation is that all satisfying assignments
of $\omega$ are also satisfying assignments of~$\rho$ (so $\Prob{\rho}
\ge \Prob{\omega}$); but when $\psi$ is satisfiable via some $\alpha
\colon \operatorname{vars}(\psi) \to \mathbb B$, merging $\alpha$ with any
witness~$\beta_*$ of $c_*$'s irredundancy in~$\omega$
will \emph{result in an additional satisfying assignment of~$\rho$.} In
other words, $\psi$ is satisfiable iff $\Pr[\rho] > \Prob{\omega} = p$. A
visualization of this reduction idea (but now from $2\Lang{sat}$ to
$5\Lang{sat-pr}_{>p'}$ for another $\omega'$ and $p' = \Pr[\omega']$)
is shown in Figure~\ref{fig-collapsed}.

\begin{figure}[htpb]
  \centering
  \begin{tikzpicture}[sunflower figure]
    
      \scoped[xshift=0cm] {
        \pictureexplain{$\omega'$}
        \pictureliteralsbase
        \scoped{\picturekernel}
        
        \scoped[green!50!black] {
          \pic {join 3 = l on 4 with a' on 4 with b on 4};
        }
      }
      
    \scoped[xshift=7cm] {
      \pictureexplain{$(\omega' \setminus
        \{\textcolor{green!50!black}{c_*}\}) \cup{}$\\ $ \{\textcolor{green!50!black}{c_*} \cup \textcolor{red}{d} \mid \textcolor{red}{d} \in \textcolor{red}{\psi}\}$}
      \pictureliteralsprime
      \scoped{\picturekernel}

      \scoped[green!50!black] {
        \pic[] {join 4' = a' on 3 with b on 3 with p1 on 1 with p2 on 1};
        \pic[] {join 4' = a' on 4 with b on 4 with p3 on 1 with p2 on 2};
        \pic[] {join 4' = a' on 5 with b on 5 with p3 on 2 with p4 on 2};
        \pic[] {join 4' = a' on 6 with b on 6 with p5 on 1 with p6 on 1};
      }
    }
  \end{tikzpicture}
  \caption{
    Left, an irredundant formula $\omega' \in 5\Lang{cnfs}$ is shown that has
    room for $2\Lang{sat}$ since it contains the green clause $c_* =
    \{x,\neg y, z\} \in \omega'$, which is ``small,'' meaning of size
    $5-2 = 3$ and implying that it ``has room'' for adding up to two
    literals in a reduction from $\Lang{2sat}$ to
    $5\Lang{sat-pr}_{>p'}$ for $p' = 
    \Prob{\omega'}$. How the reduction  works is  shown right: The
    clauses of a formula $\psi \in 
    \Lang{2cnfs}$ with ``fresh'' variables~$v_i$ are added to~$c_*$, which will
    \emph{result in a 5\textsl{\textsc{cnf}} formula since there is always
    enough room to add only two literals to a size-3 clause.} The
    resulting formula $\rho$ has a satisfaction probability strictly
    larger than that of~$\omega'$ iff $\psi$ is satisfiable.
  }
  \label{fig-collapsed}
\end{figure}

\begin{theorem}[Spectral Trichotomy Theorem, Detailed Version]\label{thm-main}
  For each $k$ and $p$:
  \begin{enumerate}
  \item If $k\Lang{cnfs}$ have room for $3\Lang{sat}$ at~$p$, then
    $k\Lang{sat-pr}_{> p}$ is $\Class{NP}$-complete.
  \item If $k\Lang{cnfs}$ have room for $2\Lang{sat}$ at~$p$, but not for
    $3\Lang{sat}$, then $k\Lang{sat-pr}_{> 
    p}$  is $\Class{NL}$-complete. 
  \item In all other cases, $k\Lang{sat-pr}_{> p}$ lies in
    $\Class{AC}^0$.
  \end{enumerate}
\end{theorem}

We already saw how the definition of ``$k\Lang{cnfs}$ have room for
$2\Lang{sat}$ (or $3\Lang{sat}$) at probability~$p$'' is tailored
towards making it easy to show that $\Lang{2sat}$ (or $\Lang{3sat}$)
can be reduced to $k\Lang{sat-pr}_{> p}$ -- and this will make it
easy to prove the hardness results implicit in the claim of the
theorem.
The tricky part are the upper bounds for the last two items of
Theorem~\ref{thm-main}. To prove 
them, we proceed as follows: For fixed $k$ and~$p$, we introduce two
conditions that a number~$t$ may or may not satisfy, dubbed the
\emph{hardness condition} (which is just ``$k\Lang{cnfs}$ have room
for $t\Lang{sat}$ at~$p$'') and the \emph{membership condition} (which has a
complex definition). We then proceed to prove three lemmas
(Lemmas~\ref{lem-lower}, \ref{lem-upper}, and \ref{lem-link}), which 
state: 
\begin{enumerate}
\item If $t$ meets the hardness condition, $t\Lang{sat}$ reduces to
  $k\Lang{sat-pr}_{>p}$. 
\item If $t$ meets the membership condition, $k\Lang{sat-pr}_{>p}$
  reduces to $t\Lang{sat}$.
\item $t+1$ meets the hardness condition or $t$ meets the membership condition.
\end{enumerate}
Even without knowing the exact definitions of the conditions, it will
not be hard to derive the Spectral Trichotomy Theorem just from these
three items. It is worth mentioning one key idea behind the lemmas:
We introduce a new notion of \emph{weak backdoor sets for a
threshold.} Backdoor sets are an important tool in \textsc{fpt}
theory, commonly used to decide satisfiability, that is, to decide
whether $\Prob{\phi} > 0$ holds. The trick is that while $\phi$ may be
a difficult formula, for a cleverly chosen small set~$X$ the formulas
$\phi|_\beta$ for $\beta \colon X\to\mathbb B$ might be syntactically
simple: If $\phi|_\beta \in \Lang{2cnfs}$ holds (let alone
$\phi|_\beta \in \Lang{1cnfs}$), deciding 
satisfiability is easy. By modifying this approach appropriately, we
get ``backdoors for thresholds,'' meaning that they will allow us to decide
whether $\Prob{\phi} > p$ holds for fixed~$p$. The core insight will
be that \emph{the variables $\operatorname{vars}(\phi^*_{\mathrm{pluck}})$ of the
plucking kernel from the Kernel Theorem} (Theorem~\ref{thm-kernel}) are a good candidate  
backdoor set: Larger clauses in $\phi|_\beta$ for $\beta \colon
\operatorname{vars}(\phi^*_{\mathrm{pluck}}) \to \mathbb B$ mean smaller clauses
in $\phi^*_{\mathrm{pluck}}$ and thus clauses with more room, possibly implying that
$k\Lang{cnfs}$ have room for $\Lang{2cnfs}$ (or $\Lang{1cnfs}$) at~$p =
\Prob{\phi^*_{\mathrm{pluck}}}$. While linking all this together is   
technically challenging, it will allow us to show membership in
$\Class{AC}^0$~or~$\Class{NL}$.   

Curiously, it remains an open problem how we can decide
algorithmically on input of $k$ and~$p$ (encoded appropriately) which
case applies in Theorem~\ref{thm-main}, see the conclusion for some
ideas towards resolving this problem.

\subsection{Related Work}
\label{sec-related}

The history of determining the complexity of the many different
variants of the satisfiability problem for propositional formulas
dates back all the way to Cook's original $\Class{NP}$-completeness
proof~\cite{Cook1971} from 1971. In parallel and unaware of Cook's
work, Levin studied \emph{perebor} (``brute-force'') algorithms in the
\textsc{ussr} and in 
1973 also pointed out~\cite[Theorem~1]{Levin1973} that, in
modern parlance, $\Lang{sat}$ as well as five other problems are
$\Class{NP}$-complete. Unfortunately, unlike Cook, Levin gives no
proof in his paper (Trakhtenbrot~\cite{Trakhtenbrot1984}
rightfully calls the paper ``absolutely crisp'' but also ``laconic'')
and  the graph isomorphism problem is one of the five other
problems mentioned (a problem that is 
typically \emph{not} believed to be $\Class{NP}$-complete), making the
missing proof in Levin's paper a bit of a sore point.  

Since these early times, it has become textbook knowledge that
$k\Lang{sat}$ is in $\Class{AC}^0$ for $k=1$, is $\Class{NL}$-complete
for $k=2$, and is $\Class{NP}$-complete for $k \ge 3$.
Determining whether the number of satisfying assignments of a formula
is not just positive, but whether ``a lot'' of assignments are
satisfying, is a quite different problem, though: Determining whether a
majority of assignments are satisfying is a canonical 
$\Class{PP}$-complete problem~\cite{Gill1974,Simon1975}; and it does not matter whether one
considers ``strictly more than $1/2$'' ($\Lang{sat-pr}_{>1/2}$)
or ``more than or equal to $1/2$''
($\Lang{sat-pr}_{\ge1/2}$). Indeed, any fixed value different from
$1/2$ can also be used and it does not matter  whether ``$>$'' or ``$\ge$''
is used \cite[Theorem 4.1]{Simon1975}. Because of the indifference of 
the complexity to the exact problem definition, it is often a bit
vague how the problem ``$\Lang{majority-sat}$'' is defined, exactly,
in a paper (indeed, the common meaning of ``majority'' in voting
suggests that ``strictly more than one half'' is perhaps the natural
interpretation). 

Given that the tipping point between ``easy'' and ``hard''
satisfaction problems is exactly from $k=2$ to $k=3$, it seemed 
natural to assume that $k\Lang{sat-pr}_{\ge 1/2}$ and
$k\Lang{sat-pr}_{>1/2}$ are also both $\Class{PP}$-complete for $k
\ge 3$. Indeed, given that computing $\#(\phi)$ for $\phi
\in 2\Lang{cnfs}$ is known to be $\#\Class P$-complete \cite{Valiant79}, even
$2\Lang{sat-pr}_{\ge   1/2}$ being $\Class{PP}$-complete seemed
possible and even natural. 
It was thus (extremely) surprising that Akmal and Williams~\cite{AkmalW2022} were
recently able to show that $k\Lang{sat-pr}_{\ge p} \in \Class
{LINTIME}$ holds for \emph{all}~$k$ and~$p \in \mathbb Q$. As pointed out by Akmal
and Williams, not only has the opposite generally been believed to hold,
this has also been claimed repeatedly (page~1 of \cite{AkmalW2021}
lists no less than 15 different papers from the last 20 years that
conjecture or even claim $\Class{PP}$-hardness of $3\Lang{sat-pr}_{\ge
  1/2}$). Similarly, when the participants of the Computational
Complexity Conference 2022 were asked to guess the complexity of
$3\Lang{sat-pr}_{=\frac{1}{2}}$, the leading experts unanimously
voiced the proposition ``$\Class{C_=P}$-complete.'' This was certainly
a highly educated guess since a natural way of defining this class
(which is extremely powerful as it happens to
equal~\cite{10.1098/rspa.1999.0485} a quantum 
version of $\Class{coNP}$), is 
as the reduction closure of $\Lang{sat-pr}_{=\frac{1}{2}}$, so it
seemed natural to assume that when $\Lang{3sat-pr}_{=0}$ and
$\Lang{sat-pr}_{=0}$ have the same complexity (namely, being
$\Class{coNP}$-complete), so should $\Lang{3sat-pr}_{=\frac{1}{2}}$
  and $\Lang{sat-pr}_{=\frac{1}{2}}$. In fact, by the results of 
the present paper, $\Lang{3sat-pr}_{=\frac{1}{2}}$
is $\Class{NL}$-complete. Just as surprising was  
the result of Akmal and Williams that while $4\Lang{sat-pr}_{\ge
  1/2}$ lies in $\Class P$, the seemingly almost identical problem
$4\Lang{sat-pr}_{> 1/2}$ is $\Class{NP}$-complete. This has led
Akmal and Williams to insist on a precise notation
in~\cite{AkmalW2022}: They differentiate clearly between
$\Lang{majority-sat}$ and $\Lang{gt-majority-sat}$ and consider these
to be special cases of the threshold problems
$\Lang{thr}_p\Lang{-sat}$ and
$\Lang{gt-thr}_p\Lang{-sat}$ -- and all of these problems can
arise in a ``-$k\Lang{sat}$'' version. The notations
$k\Lang{sat-pr}_{\ge p}$ and $k\Lang{sat-pr}_{> p}$ from
the present paper are a proposal to further simplify, unify, and
clarify the notation, no new problems are introduced.

We will use tools from \textsc{fpt} theory, namely
kernels algorithms in Section~\ref{sec-kernels} and backdoor sets in
Section~\ref{section-complexity}. As computing (especially hitting 
set) kernels is very well-understood from a
complexity-theoretic point of view (see~\cite{vanBevern2014} for
the algorithmic state of the art and \cite{BannachT20} for
upper bounds on the parallel parameterized complexity), we can base
proofs on this for $k\Lang{sat-pr}_{\ge p} \in 
\Class{AC}^0$ for all $k$ 
and~$p$. Of course, different parameterized versions of
$\Lang{sat}$ are studied a lot in \textsc{fpt} theory, see~\cite{FlumG06} for a
starting point, but considering the satisfaction probability as a
parameter (as we do in the present paper) is presumably new. 

While the proofs of the locality theorems
(Theorems~\ref{thm-constant-influence} and~\ref{thm-local}) are
largely based on the just-mentioned kernel algorithm, their
statements are purely model-theoretic and not algorithmic. They
concern the structure of the set of satisfying assignments of
$k\Lang{cnf}$ formulas which has, of course, been studied a lot in the
literature, 
for instance in the form of the \emph{influence of variables}
\cite{10.1109/SFCS.1988.21923} or \emph{average sensitivity,} see
\cite{BOPPANA1997257,10.1007/978-3-642-40328-6_47}
for starting points. A typical result from this line of research~\cite{Amano2011TightBO} is
that the average sensitivity of a
$k\Lang{cnf}$ formula is at most~$k$, meaning that, on average, for
any assignment only $k$ variables have the property that flipping
their value flips whether the assignment is
satisfying. In comparison, Theorem~\ref{thm-constant-influence}
implies that when we consider any $\phi\in k\Lang{cnfs}$ with
$\Pr[\phi] = p$ for a fixed $p\in[0,1]$, there is a constant number of
variables (namely the at most $k\cdot S_{k,p}$ variables in~$\psi$
from the claim of the theorem) so that \emph{all} assignments are
sensitive \emph{only} to these variables. 

The algorithm of Akmal and Williams in \cite{AkmalW2022} was the main
inspiration for the results of the present paper and 
it shares a number of characteristics with the kernel algorithm based
on the plucking rule: Both   
algorithms search for and then pluck sunflowers. However, without the 
Spectral Well-Ordering Theorem, 
one faces the problem that plucking large sunflowers
\emph{repeatedly} could conceivably lower $\Prob{\phi}$
past~$p$. To show that this does not happen (without using the
Spectral Well-Ordering Theorem) means that one has to redo all the
arguments used in the proof of the Spectral  Well-Ordering Theorem,
but now with explicit parameters and 
constants and one has to intertwine the algorithmic and the
underlying order-theoretic arguments in rather complex ways (just the
analysis of the algorithm in~\cite{AkmalW2021} takes eleven pages plus
two pages in the appendix). The fact that these many parameters are
hardwired into the algorithm (arguably) also means that the Akmal--Williams
algorithm is not gap size oblivious: While the spectral gap is never
explicitly mentioned in the algorithm, the hardwired sizes of the
sunflowers that are identified and then plucked are actually lower
bounded in a similar way as the quantitative bounds on the spectral
gaps established in Section~\ref{sec-bounds} of the present paper.

Approximating $\Prob{\phi}$ for $\phi \in \Lang{cnfs}$ and also for
$\phi \in \Lang{dnfs}$ is an active research field. From a practical
point of view, finding \emph{multiplicative} approximations is of high
interest given that $\Prob{\phi}$ is typically an exponentially small
value in practical settings -- and a lot of energy and clever
algorithms are directed towards addressing this problem, starting with
the Karp--Luby algorithm~\cite{KarpL1983,KarpLM1989}, and see for
instance~\cite{ChakrabortyMV2016} for some recent results. For our
purposes, because ``spectral gaps are everywhere,'' approximating
$\Prob{\phi}$ by an 
\emph{additive} error is all that is needed to decide $\Prob{\phi} \ge
p$. We already saw that the ``obvious'' randomized sampling algorithm
yields such an approximation with high
probability. Trevisan~\cite{Trevisan2004} noted that for fixed~$k$,
for $k\Lang{cnfs}$ one can ``derandomize'' this algorithm, meaning
that for each $\epsilon > 0$ and~$k$ one can design an algorithm
(not really related to the randomized one, conceptually) that on input
$\phi \in k\Lang{cnfs}$ outputs an interval $I \subseteq [0,1]$ of
size at most~$\epsilon$ with $\Prob{\phi} \in I$. Note that when one
is interested in additive errors, it makes no difference whether one
considers formulas in $k\Lang{cnfs}$ or in $k\Lang{dnfs}$ as (in
not-so-slight abuse of notation and terminology) for $\phi \in
\Lang{cnfs}$ we have by duality ``$\neg \phi \in \Lang{dnfs}$ and $\Prob{\phi} = 1 -
\Prob{\neg\phi}$.''  

The majority-of-majority problem for arbitrary $\Lang{cnf}$ formulas,
called $\Lang{maj-maj-sat}$ in~\cite{AkmalW2022}, is known to be
complete for $\Class{PP}^{\Class{PP}}$ and of importance in ``robust''
satisfaction probability
estimations~\cite{ChoiXD2012,OztokCD2016}. The arguments
from~\cite{AkmalW2022} on $k\Lang{sat-pr}_{\ge 1/2}$ do 
not generalize in any obvious way to $\Lang{maj-maj-}k\Lang{sat}$: The
difficulty lies in the ``mixed'' clauses that contain both  
$X_1$- and $X_2$-variables. In a clever argument, Akmal and
Williams were able to show that for $k=2$ one \emph{can} ``separate''
the necessary satisfaction probability estimations for the $X_1$- and
$X_2$-variables in polynomial time (so $\Lang{maj-maj-2sat} \in \Class
P$). This feat was considerably facilitated by the fact that a mixed size-2
clause must contain exactly one $X_1$-literal and one
$X_2$-literal. They conjectured that $\Lang{maj-maj-}k\Lang{sat} \in
\Class P$ holds for all~$k$ (which is indeed the case by
Corollary~\ref{cor-maj-maj}), but point out that it is unclear how (or whether) 
their algorithm can be extended to larger~$k$. The approach taken in
the present paper (via locality arguments) seems quite different and
not directly comparable.

\subsection{Structure of This Paper}

As already mentioned, this paper consists of three main sections, each
of which addresses a different angle from which to look at the
satisfaction probability of $k\Lang{cnf}$ formulas:
From the \emph{order-theoretic} angle in Section~\ref{section-order},
from the \emph{algorithmic} angle in
Section~\ref{section-algorithmics}, and from the
\emph{structural complexity} angle in
Section~\ref{section-complexity}. The conclusion spells out the
complexity of $k\Lang{sat-pr}_{>p}$ for some
concrete values of $k$ and~$p$, and it contains
an outlook on possible applications and extensions of the presented
methods to new problem variants and versions.

\section{Order-Theoretic Results}

\label{section-order}

There is a sharp contrast between the structural properties of the
``full'' spectrum of satisfaction probabilities of arbitrary
propositional formulas and the spectrum of values $k\Lang{cnf}$
formulas can have. The full spectrum $\CNFSSigmaSpectrum = \{
\Prob{\phi} \mid \phi \in \Lang{cnfs}\}$ is, well, ``full'' as it
is the set of all dyadic rationals between 0 and~1 (recall $\mathbb D = \{m/2^e
\mid m,e \in \mathbb Z\}$):
\begin{lemma}\label{lemma-dyadic}
  $\CNFSSigmaSpectrum = \mathbb D \cap [0,1]$.
\end{lemma}
\begin{proof}
  For any $\phi \in \Lang{cnfs}$ we have, by definition, $\Prob{\phi} =
  m/2^e$ for $m = \#(\phi)$ and $e =
  \left|\operatorname{vars}(\phi)\right|$. For the other direction,
  let $e \in \mathbb N$ and $m \in
  \{0,\dots,2^e\}$. Consider the truth table over the variables
  $X=\{x_1,\dots,x_e\}$ in which the first $m$ lines are set to~$1$
  (satisfying assignments) and the rest are set to~$0$ (non-satisfying
  assignments). Then every
  $\Lang{cnf}$ formula~$\phi$ with $\operatorname{vars}(\phi) = X$
  having this truth table has exactly $m$ satisfying assignments and, hence,
  $\Prob{\phi} = m/2^e$.
\end{proof}
By the lemma, $\CNFSSigmaSpectrum$ is a dense subset of the real
interval $[0,1]$, it is not closed topologically, and it is
order-isomorphic to $\mathbb Q \cup \{-\infty,\infty\}$ with respect
to both~$<$ and~$>$. We will soon see that the properties
of each $k\CNFSSigmaSpectrum$ could hardly be more different: They
are nowhere-dense, they are closed, and they are well-ordered. Of
these properties, the well-orderedness is the most important one both
for algorithms in later sections and because the other properties
follow from the well-orderedness rather easily. 

To get a better intuition about the spectra, let us have a closer look
at the first few of them. As a slightly pathological case,
$0\CNFSSigmaSpectrum = \{1,0\}$ as $\Lang{0cnfs} =
\bigl\{\emptyset,\{\emptyset\}\bigr\}$ just contains the trivial
tautology $\emptyset$ and the trivial contradiction
$\{\emptyset\}$. The spectrum $1\CNFSSigmaSpectrum$ is more
interesting, but still simple:
\begin{align}
  \textstyle 1\CNFSSigmaSpectrum = \{1,\frac{1}{2},\penalty0
  \frac{1}{4},\penalty0 \frac{1}{8},\penalty0 \frac{1}{16}, \dots\}
  \cup \{0\}
  \label{eq-1spec}
\end{align}
as a $1\Lang{cnf}$ formula (a conjunction of literals) has a
satisfaction probability of the 
form $2^{-e}$ or is~$0$. Readers familiar with order theory will
notice immediately that $1\CNFSSigmaSpectrum$ is order-isomorphic to
the ordinal $\omega+1$ with respect to~$>$. The spectrum
$2\CNFSSigmaSpectrum$ is already much more complex:
\begin{align}
  \textstyle
  2\CNFSSigmaSpectrum = \{1, \frac{3}{4}, \frac{5}{8}, \frac{9}{16},
  \frac{17}{32}, \frac{33}{64}, \dots \} 
  \cup \{\frac{1}{2}, \frac{15}{32}\} \cup \{\dots\} \label{eq-2spec}
\end{align}
where $\{\dots\}$ 
contains only numbers less than $\frac{15}{32}$. To see that this is, indeed,
the case, observe that the formulas $\bigl\{\{a, x_1\}\bigr\}$, $\bigl\{\{a, x_1\},
\{a, x_2\}\bigr\}$, $\bigl\{\{a, x_1\},\{a, x_2\},\{a,x_3\}\bigr\}$, $\dots$\ show that
every number of the form $1/2 + 2^{-e}$ is in the 
spectrum. Furthermore, there are no other numbers larger than $1/2$ in
the spectrum as the first formula with two variable-disjoint clauses
has a satisfaction probability of $\Prob{\{\{a, b\},\{c,d\}\}} =
\frac{9}{16}$ which we happen to have already had; and adding any 
additional clause makes the probability drop to at most
$\ProbBig{\bigl\{\{a,b\}, \{c,d\},\{c,e\}\bigr\}} = \frac{3}{4} \cdot \frac{5}{8} =
\frac{15}{32}$. Below this, the exact structure of $2\CNFSSigmaSpectrum$
becomes ever more complex as we get nearer to~$0$ and it is unclear
what the order-type of $2\CNFSSigmaSpectrum$ with respect to $>$
actually is (an educated guess is $\omega^\omega + 1$).

Our aim in the rest of this section is to prove the Spectral
Well-Ordering Theorem, Theorem~\ref{thm-order}, by which all
$k\CNFSSigmaSpectrum$ are well-ordered by~$>$. The surprisingly short
proof, presented in Section~\ref{section-proof-spec}, will combine
results from the following Section~\ref{section-spec-simple} on some
simple properties of well-orderings with some simple properties
of $\Prob{\phi}$ for $k\Lang{cnf}$ formulas~$\phi$.  The theorem
implies the \emph{existence} of spectral gaps below each~$p$ in the
spectra, but the proof does not provide us with any quantitative
information about the \emph{sizes} of these gaps. This is remedied in
Section~\ref{sec-bounds}, where we derive bounds on the sizes of
spectral gaps. \emph{Note that the somewhat technical
Section~\ref{sec-bounds} can safely be skipped upon a first reading.}

\subsection{Well-Orderings and Their Properties}
\label{section-spec-simple}
Well-orderings are a basic tool of set theory, but for our
purposes only a very specific type of orderings will be of interest
(namely only sets of non-negative reals with the 
strictly-greater-than relation as the only ordering relation). For this reason,
we reserve the term ``well-ordering'' only for the following kind of
orderings, where a \emph{strictly increasing sequence in~$X$} is a
sequence $(x_i)_{i \in \mathbb N}$ with $x_i \in X$ for all~$i$ and
$x_0 < x_1 < x_2 < \cdots$:  

\begin{definition}
  A set $X \subseteq \mathbb R^{\ge 0}$ is \emph{well-ordered
  (by~$>$)} if there is \emph{no} strictly increasing sequence
  in~$X$. Let
  $\Class{WO}$ denote the set of all (such) well-ordered sets.
\end{definition}
There is extensive literature on the properties of well-orderings in
the context of 
classical set theory, see for instance~\cite{Jech2003} as a starting
point. We will need only those properties stated in the following lemma, where the
first items are standard, while the last are specific to the present
paper. For $X,Y \subseteq \mathbb R^{\ge 0}$ let $X + Y$ denote $\{x+y
\mid x \in X, y \in Y\}$ and $n \cdot X = X + \cdots + X$ (of course,
$n$ times).

\begin{lemma}\label{lemma-wo}
  \hfil
  \begin{enumerate}
  \item
    Let $X \in \Class{WO}$. Then $X$ contains a largest element.
  \item
    Let $Y \subseteq X \in \Class{WO}$. Then $Y \in \Class{WO}$.\label{lemma-item-subset}
  \item
    Let $X, Y \in \Class{WO}$. Then $X \cup Y \in \Class{WO}$. Thus, $\Class{WO}$ is closed under finite unions.
  \item
    Let $(X_n)_{n \in \mathbb N}$  with $X_n \in
    \Class{WO}$ for all~$n$ and $\lim_{n\to \infty} \max X_n =
    0$. Then $\bigcup_{n \in \mathbb N} X_n \in \Class{WO}$.\label{lemma-item-infinite}
  \item
    Let $X \in \Class{WO}$ and let $(x_i)_{i \in \mathbb N}$ be an
    arbitrary sequence of $x_i \in X$. Then there is an infinite $I
    \subseteq \mathbb N$ such that $(x_i)_{i \in I}$ is 
    decreasing (that is, $x_i \ge x_j$ for $i<j$ and $i,j \in I$).
  \item
    Let $X, Y \in \Class{WO}$. Then $X + Y \in \Class{WO}$. Thus,
    $\Class{WO}$ is closed under finite sums.
  \item
    Let $X \in \Class{WO}$ and $n \in \mathbb N$. Then $n\cdot X \in
    \Class{WO}$.
  \end{enumerate}
\end{lemma}

\begin{proof}\hfil
  \begin{enumerate}
  \item There would otherwise be an infinite
    strictly increasing sequence in~$X$.
  \item Any strictly increasing sequence in~$Y$
    would be a strictly increasing sequence in~$X$.
  \item Any strictly increasing sequence in $X \cup Y$ would
    contain a subsequence fully in~$X$~or~$Y$.
  \item
    Suppose there is a strictly increasing sequence $(x_i)_{i \in
      \mathbb N}$ in $\bigcup_{n \in \mathbb N} X_n$ and
    w.\,l.\,o.\,g.\ assume $x_0 > 0$. 
    As the
    maxima of the $X_n$ tend towards~$0$, there is some~$m$ such that
    $x_0 > \max X_n$ holds for all $n \ge m$. In particular, all
    $x_i$ lie in
    $\bigcup_{n < m} X_n$. By the previous item, this is well-ordered,
    contradicting that it contains an infinite strictly increasing
    sequence.
  \item The set $\{x_i \mid i > 0\}$ is a subset of~$X$
    and must hence contain a maximum~$x_{i_0}$ by the first
    item. Then $\{x_i \mid i > i_0\}\subseteq X$ must contain a
    maximum~$x_{i_1}$ for some $i_1 > i_0$. Next, consider  $\{x_i 
    \mid i > i_1\} \subseteq X$ and let $x_{i_2}$ for some $i_2>i_1$ be
    a maximum. In this way, for $I = \{i_0,i_1,\dots\}$ we get
    an infinite subsequence $(x_i)_{i\in I}$ that is clearly
    (not necessarily strictly) decreasing as each chosen element was
    the maximum of all following elements.
  \item
    Suppose there is a sequence $z_0 < z_1 <
    z_2 < \cdots$ of numbers $z_i \in X+Y$. Then for each~$i$ there
    must exist $x_i \in X$ and $y_i \in Y$ with $z_i = x_i + y_i$. By
    the previous item there is a  decreasing subsequence
    $(x_i)_{i \in I}$ for some infinite~$I$. Then $(y_i)_{i\in I}$ is
    an infinite strictly increasing sequence in~$Y$ as for any $i,
    j\in I$ with $i < j$ we have $y_i = z_i - x_i \le  z_i - x_j < z_j
    - x_j =  y_j$. This contradicts $Y \in \Class{WO}$.
  \item This follows immediately from the previous item.\qedhere
  \end{enumerate}
\end{proof}

\subsection{Proof of the Spectral Well-Ordering Theorem}
\label{section-proof-spec}

The proof of the Spectral Well-Ordering Theorem, by which
$k\CNFSSigmaSpectrum$ is well-ordered by~$>$, is by induction
on~$k$. For the inductive step, we need one more observation:
Every $x \in k\CNFSSigmaSpectrum$ equals the sum of ``a few'' elements
of $(k-1)$\textsc{cnfs-pr-spectrum}, where ``few'' means ``some function of
$1/x$.'' To formulate this statement rigorously, a notation will be
useful: Let $\RestrictAdd{\phi}{\beta} := \Restrict{\phi}{\beta} \cup 
\bigl\{\{v\} \mid v \in\penalty50 X,\penalty0 \beta(v) = 1\bigr\} \cup
\bigl\{\{\neg v\} \mid v \in\penalty50 X,\penalty0 
\beta(v) = 0\bigr\}$ where $\beta \colon X \to \mathbb B$ for some
finite~$X$. Basically, $\RestrictAdd{\phi}{\beta}$ is
``$\Restrict{\phi}{\beta}$ with unit clauses added that ensure that
$\beta$ is the only model on the $X$-variables'' and note that 
$\Prob{\RestrictAdd{\phi}{\beta}} = \Prob{\Restrict{\phi}{\beta}}/2^{|X|}$.

\begin{lemma}\label{lemma-simple-obs}
  Let $\phi \in \Lang{cnfs}$ and let $X$ be a finite set. Then
  the set of satisfying assignments of~$\phi$ over $X\cup \operatorname{vars}(\phi)$
  is exactly the disjoint union, taken over all $\beta \colon X \to
  \mathbb B$, of the sets of satisfying assignments of
  $\RestrictAdd{\phi}{\beta}$ over $X \cup  \operatorname{vars}(\phi)$.
\end{lemma}

\begin{proof}
  Each satisfying assignment~$\alpha \colon X \cup  \operatorname{vars}(\phi) \to \mathbb B$ of $\phi$ satisfies  
  $\RestrictAdd{\phi}{\beta}$ for the assignment $\beta\colon X \to
  \mathbb B$ that agrees with $\alpha$ on~$X$, but $\alpha$ satisfies no $\RestrictAdd{\phi}{\beta'}$ for
  $\beta' \neq \beta$.
\end{proof}
\begin{corollary}\label{col-bounds}
  Let $\phi \in \Lang{cnfs}$ and $X$ be a finite set. Then
  $\Prob{\phi} = \sum_{\beta:X\to\mathbb B} \Prob{\RestrictAdd{\phi}{\beta}}$. 
\end{corollary}
\begin{lemma}\label{lemma-res}
  Let $\phi \in k\Lang{cnfs}$ for $k\ge2$ and let $\pi \subseteq \phi$
  be a maximal packing. Then $\RestrictAdd{\phi}{\beta} \in (k-1)\Lang{cnfs}$ for
  all $\beta\colon \operatorname{vars}(\pi) \to \mathbb B$. 
\end{lemma}
\begin{proof}
  In $\RestrictAdd{\phi}{\beta}$, we either remove a clause or remove at
  least one literal from it as $\operatorname{vars}(\pi)$ intersects
  $\operatorname{vars}(c)$ for all $c \in \phi$ (as $\pi$ would not be
  maximal, otherwise).
\end{proof}

\begin{proof}[Proof of Theorem~\ref{thm-order}, the Spectral Well-Ordering Theorem]\label{proof-order}
  By induction on~$k$. The base case is $k=1$ where
  $1\CNFSSigmaSpectrum = \{1,
  \frac{1}{2},\frac{1}{4},\frac{1}{8},\dots\} \cup \{0\} =
  \{2^{-i}\mid i \in \mathbb N \cup \{\infty\}\}$, which is clearly
  well-ordered (with order type $\omega+1$).
  For the inductive step from $k-1$ to~$k$, we show that
  \begin{align*}
    k\CNFSSigmaSpectrum \subseteq \bigcup_{r
      \in \mathbb N} \bigl(2^{k\cdot r} \cdot (k-1)\CNFSSigmaSpectrum\bigr) \cap
    \bigl[0,(1-2^{-k})^r\bigr]. 
  \end{align*}
  This will prove $k\CNFSSigmaSpectrum \in \Class{WO}$ as
  $(k-1)\CNFSSigmaSpectrum \in \Class{WO}$ by the induction
  hypothesis, and thus $2^{k\cdot r} \cdot (k-1)\CNFSSigmaSpectrum
  \in\Class{WO}$ as a finite sum of well-orderings. By
  intersecting this with the ever-smaller intervals
  $\bigl[0,(1-2^{-k})^r\bigr]$, we still get elements of $\Class{WO}$ by
  item~\ref{lemma-item-subset} of Lemma~\ref{lemma-wo} and can
  then apply item~\ref{lemma-item-infinite} to get
  that the union lies in~$\Class{WO}$. 

  It remains to prove the inclusion. Let $p \in k\CNFSSigmaSpectrum$
  be witnessed by $\phi \in k\Lang{cnfs}$, that is, $p =\Prob{\phi}$.
  Let $\pi$ be a maximal
  packing $\pi \subseteq \phi$ and let $r = |\pi|$. By the Packing
  Probability Lemma, $\Prob{\phi} \le (1-2^{-k})^r$. By Corollary~\ref{col-bounds}, we have
  $\Prob{\phi} = \sum_{\beta:\operatorname{vars}(\pi)\to\mathbb B}
  \Prob{\RestrictAdd{\phi}{\beta}}$ and by Lemma~\ref{lemma-res}, each $\RestrictAdd{\phi}{\beta}$ is
  a $(k-1)\Lang{cnf}$ formula. Thus, $\Prob{\phi}$ is the sum
  of at most $2^{\left|\operatorname{vars}(\pi)\right|} \le 2^{k\cdot r}$ values from $(k-1)\CNFSSigmaSpectrum$, proving
  that we have $p \in \bigl(2^{k\cdot r} \cdot
  (k-1)\CNFSSigmaSpectrum\bigr) 
  \cap \bigl[0, (1-2^{-k})^r\bigr]$. 
\end{proof}

\subsection{Bounding Spectral Gaps}
\label{sec-bounds}

The Spectral Well-Ordering Theorem (or Corollary~\ref{cor-spec} to be
precise) tells us that there exist 
spectral gaps below all reals $p \in [0,1]$ and this will be
``all we need'' to establish the complexity-theoretic and most of the algorithmic
results in the rest of this paper. Indeed, basing the proofs of these results
just on the existence of spectral gaps allows for particularly simple
arguments. However, both in practical settings and for future theoretical
work we would like to
have bounds on the size of the spectral gaps and the proof given
earlier does not provide any. This is fixed in the following
in the form of an alternative proof of Corollary~\ref{cor-spec} (and,
thereby, of the Spectral Well-Ordering Theorem) that provides us with
lower bounds for spectral gaps (we are 
interested in \emph{lower} bounds as smaller gaps will mean larger
running times). Since the analysis is somewhat technical, readers may
wish to skip at least the proof details in this section upon first
reading. 

\subparagraph*{Spectral Gaps and the Distribution of 1-Bits.}
To get some intuition on how spectral gaps behave, observe that
the sizes of spectral gaps fluctuate 
wildly with~$p$: Recall the spectrum $2\CNFSSigmaSpectrum$ from
equation~\eqref{eq-2spec} and observe that the gap below $\frac{1}{2}
+ \frac{1}{16}$ is 
$\frac{1}{32}$, the gap below $\frac{1}{2} + \frac{1}{2^{1000}}$ is
$\frac{1}{2^{1001}}$, but the gap below $\frac{1}{2}$ is  $\frac{1}{32}$ once more. 
As we will see in a moment, the deeper reason for this erratic
behaviour lies in the \emph{distribution of the 1-bits in the binary
representations of the values in the spectrum}. To get a better
handle on these, a lemma and some notations will be useful (we
restrict attention to the nontrivial values $p \in (0,1)$
in this section, although some of the definitions also make
sense for $p = 1$ or for $p = 0$):

\begin{lemma}\label{lemma-unique}
  Let $p \in (0,1)$. Then there is a unique infinite sequence
  $j_1 < j_2 < j_3 < \cdots$ of positive integers with $p =
  \sum_{i=1}^\infty 2^{-j_i}$. 
\end{lemma}

\begin{proof}
  Write $p$ as a binary fraction, that is, as $0.b_1b_2b_3\cdots_2$
  in base~$2$ with $b_i \in \{0,1\}$, so $p = \sum_{i=1}^\infty
  b_i2^{-i}$. When $p$ is a dyadic rational (an 
  element of $\mathbb D = \{m/2^e \mid m,e\in\mathbb Z\}$), there are
  two possible ways of writing $p$, namely the standard way with
  infinitely many 0-bits at the end, but also alternatively with
  infinitely many 1-bits at the end (for instance, $1/4 = 0.01_2 =
  0.001111\cdots_2$) -- and we pick the second way in this case. This
  ensures that for every $p \in (0,1)$ we have infinitely many
  $i$ with $b_i = 1$. Let $j_i$ be the position of the $i$th 1-bit in
  the sequence, that is, $j_i = \min \{n \mid \sum_{l=1}^n b_l =
  i\}$. Clearly, we then have $j_1 < j_2 < j_3 <
  \cdots$ and $p = \sum_{i=1}^\infty 2^{-j_i}$.
\end{proof}

\begin{definition}
  For $p \in (0,1)$ and $n \ge 1$, let $p_{=n} := 2^{-j_n}$, let
  $p_{\le n} := \sum_{i=1}^n p_{=i}$, and $p_{\ge n}:=
  \sum_{i=n}^\infty p_{=i}$. Here, $j_n$ is the number from    
  Lemma~\ref{lemma-unique}.
\end{definition}
Observe $p =  \sum_{i=1}^\infty p_{=i} = \lim_{i\to \infty}
p_{\le i}$ and $p - p_{\le i} = p_{\ge i+1}$. As
an example, $p = 1/3$ can be written as $0.\overline{01}_2$ in
binary representation (with the line indicating the infinitely
repeated repetend) and we have $p_{\le 1} = 0.01_2 =
\frac{1}{4}$, $p_{\le 2}= 0.0101_2 = \frac{5}{16}$, and
$p_{\le 3} = 0.010101_2 = \frac{21}{64}$, and so on. As 
another example, $p = 1/2 = 0.1_2$ can be written (only) as
$0.0\overline{1}_2$ with infinitely many 1-bits in the fractional
part. Then $p_{\le 1} = 0.01_2 = \frac{1}{4}$, $p_{\le 2} =
0.011_2 = \frac{3}{8}$, $p_{\le 3} = 0.0111_2 = \frac{7}{16}$, and so on.

\subparagraph*{Expansion Sequences: The Intuition.}

We now introduce our tool for bounding the sizes of spectral gaps:
\emph{expansion sequences.} Before introducing the formal definition,
let us try to first develop some intuition. For this, in turn, we need
some simple notations: First, for a formula $\psi \in \Lang{cnfs}$ let  
$\operatorname{pack}(\psi) \subseteq \psi$ denote a packing of maximum
size (obtained, for instance, greedily and deterministically, so
$\operatorname{pack}(\cdot)$ is an 
easy-to-compute ordinary function). Second, a set $\Psi \subseteq \Lang{cnfs}$
is \emph{pairwise model-disjoint} if for any two different $\psi,\psi'
\in \Psi$ there is no assignment that satisfies both $\psi$
and~$\psi'$. Third, for a pairwise model-disjoint
$\Psi\subseteq\Lang{cnfs}$, let $\ProbOr{\Psi} := \sum_{\psi\in\Psi}
\Prob{\psi}$ denote the probability that an assignment satisfies some 
formula in~$\Psi$. 

The objective of expansion sequences is to show that for every
$\phi \in k\Lang{cnfs}$ for which $\Prob{\phi}$ is 
below~$p$, it is actually ``well below''~$p$. If this is
always the case, we know that there is a spectral gap below~$p$
whose size depends on the ``quantitative meaning of `well below'.''
Towards proving ``below $p$ implies well below 
$p$,'' we keep track of a changing set~$\Phi_i$ of formulas (these
sets will be the elements of the expansion sequence) that starts with
$\Phi_1 = \{\phi\}$. There is a simple update rule to generate the
next set $\Phi_{i+1}$ of formulas from~$\Phi_i$: If $\Phi_i \subseteq
1\Lang{cnfs}$, the sequence ends and is \emph{exhausted}. Otherwise
pick a formula $\psi \in \Phi_i \setminus \Lang{1cnfs}$ for which
$\Prob{\operatorname{pack}(\psi)}$ is maximum, remove $\psi$, and add 
$\bigl\{\RestrictAdd{\psi}{\beta} \bigm| \beta \colon
\operatorname{vars}(\operatorname{pack}(\psi)) \to \mathbb B\bigr\}$ in 
its stead. Note that this update will not change the
sum of satisfaction probabilities by Corollary~\ref{col-bounds}, so
$\ProbOr{\Phi_i} = \ProbOr{\Phi_{i+1}}$. In particular, each
$\ProbOr{\Phi_i}$ will always be equal to the original $\Prob{\phi}$
as we started with $\Phi_1 = \{\phi\}$ (this is just the claim of
the first of four lemmas on the properties of expansion sequences). 

Up to now, we have a sequence of sets $\Phi_i$ of formulas that,
sooner or later, becomes exhausted and ends at some set~$\Phi_r$
(namely when $\Phi_r \subseteq \Lang{1cnfs}$) -- the link to~$p$
and spectral gaps is still missing. This is remedied by introducing a
condition on the~$\Phi_i$ that, when met, tells us
that we have reached a point where we are sure that $\Prob{\phi}$ is
``well below''~$p$:
\begin{definition}[Technical Condition]\label{tc-a}
  A set $\Psi \subseteq \Lang{cnfs}$ meets the \emph{technical
  condition} if 
  \begin{align}
    \sum_{\psi \in \Psi} \Prob{\operatorname{pack}(\psi)} \le p_{\le
      |\Psi|+1}. \label{eq-tc-a}
  \end{align}
\end{definition}
This condition needs to be so, well, technical, since it must
serve two purposes: First, we want to be able to check it 
efficiently, though this will only be important later, in
Section~\ref{sec-oblivious}. What we need in the present section is,  
second, that it implies $\ProbOr{\Phi_i} \le p_{\le
  |\Phi_i|+1}$ (which it does as $\ProbOr{\Phi_i} = \sum_{\psi \in
  \Phi_i} \Prob{\psi} \le 
\sum_{\psi \in \Phi_i} \Prob{\operatorname{pack}(\psi)}$). Thus, when 
$\Phi_i$ meets the technical condition,
$\Prob{\phi} = \ProbOr{\Phi_i} \le 
p_{\le |\Phi_i|+1} < p - p_{=|\Phi_i|+2}$, meaning that $\Prob{\phi}$ is
``below $p$ by at least $p_{=|\Phi_i|+2}$'' and this will count as
``well below~$p$'' (even though $p_{=|\Phi_i|+2}$ is typically a
\emph{very} small number). The second lemma of the four lemmas
formalizes these observations.  

\begin{figure}[htpb]
  \centering
  \begin{tikzpicture}

    \def\ShowLine#1#2#3{
      \scoped [shift={(0,-#1)}]  {
        \node [anchor=mid east] {
          #2
        };
        \ShowString{#3}
      }
    }
    
    \def\ShowString#1{
      \foreach \v[count=\i] in {#1} {
        \ifnum \i<11 \tikzset{h/.style={red!75!black}} \fi
        \node at (\i-1.5,0) [anchor=mid,h/.try] {\rlap{\v}}; 
      }
    }

    \scoped [x=2mm,y=1.1\baselineskip] {

      \ShowLine{0}{$p = 0.$}{
        0,1,0,1,0,1,0,1,0,1,0,1,0,1,0,1,0,1,0,{$\cdots_2 = \frac{1}{3}$}
      }
      \ShowLine{2}{$>p_{\le |\Phi_i|} = 0.$}{
        0,1,0,1,0,1,0,1,0,1,0,0,0,0,0,0,0,0,0,{$\cdots_2
          = \frac{1}{4} + \frac{1}{16} + \frac{1}{64} + \frac{1}{256}
          = \frac{85}{256}$}
      }
      \ShowLine{1}{$> P_1 = 0.$}{?,?,?,?,?,?,?,?,?,?,?,?,?,?,?,?,?,?,?$_2$}

      \draw[semithick,gray] (-3.5cm,-3) -- (10cm,-3);
      
      \ShowLine{4}{$\Prob{\psi_1} = 0.$}{0,0,1\textcolor{black}{$_2$}}
      \ShowLine{5}{$\Prob{\psi_2} = 0.$}{0,0,1\textcolor{black}{$_2$}}
      \ShowLine{6}{$\Prob{\psi_3} = 0.$}{0,0,0,0,0,0,0,0,0,0,0,1$_2$ }

      \draw[help lines] (-2cm,-6.6) -- (2.5cm,-6.6);

      \ShowLine{7.25}{$\sum_{i=1}^3 \Prob{\psi_i}  = 0.$}{0,1,0,0,0,0,0,0,0,0,0,1$_2= P_1$}

    }

  \end{tikzpicture}
  
  \caption{ {
      An example of how 1-bits can be distributed for a set~$\Phi_i$
      for which an expansion sequence still continues. In the example,
      $p = 1/3$ and $\Phi_i$ has size $|\Phi_i| = 5$. The number
      $p_{\le|\Phi_i|} = p_{\le 5} = 85/256$ equals $p$ with all
      1-bits after the fifth 1-bit replaced by 0-bits.
      Consider the number 
      $P_1 = \ProbOr{(\Phi_i \cap \Lang{1cnfs})}$. Suppose $\Phi_i
      \cap \Lang{1cnfs} = \{\psi_1,\psi_2,\psi_3\}$ with the
      probabilities shown. Each formula can contribute at most one
      1-bit to the sum~$P_1$, so $P_1$ can have at most $|\Phi_i|
      \ge |\Phi_i \cap \Lang{1cnfs}|$ many 1-bits in its red part
      (just  one in the example). In Lemma~\ref{lemma-third} we show
      that under the assumption that another number~$P_2$ is very small,
      $P_1$ would have to lie between  $p_{\le|\Phi_i|} =
      p_{\le 5}$ and~$p$, see the upper part, which would imply that the red
      part of $P_1$ would have to be \emph{identical} to that of $p$
      and $p_{\le |\Phi_i|}$ and that the black part would have to contain
      \emph{another} 1-bit. This is clearly impossible when there
      are only $|\Phi_i|$ many 1-bits in total in the binary
      representation of~$P_1$.
  } }
  \label{fig-ones}
\end{figure}

One crucial piece is, of course, still missing: The size of the $\Phi_i$
could conceivably grow arbitrarily and, thus, the notion of ``well
below~$p$'' would just mean ``below~$p$ by arbitrarily small
amounts'' -- exactly what we do not wish to prove. This is where the
third lemma comes in: It gives a bound
on~$|\Phi_{i+1}|$ purely in terms of $|\Phi_i|$ and~$p$; in 
particular, it is \emph{independent of the initial~$\phi$}. The core
idea is to write the term $\sum_{\psi \in \Phi_i}
\Prob{\operatorname{pack}(\psi)}$ from \eqref{eq-tc-a} as two sums
$P_1 + P_2$ with $P_1 = \sum_{\psi \in 
  \Phi_i \cap \Lang{1cnfs}} \Prob{\operatorname{pack}(\psi)}$ and $P_2
= \sum_{\psi \in \Phi_i \setminus \Lang{1cnfs}}
\Prob{\operatorname{pack}(\psi)}$. Whenever we are below~$p$, but not 
yet \emph{well} below~$p$, we have $P_1 < p$ and $P_1 + P_2 > p_{\le
  |\Phi_i| + 1}$. From this, we derive that if $P_2$ were
smaller than $p_{=|\Phi_i| + 1}$, the number $P_1$ would have to lie
between $p_{\le |\Phi_i|}$ and~$p$. In particular, the first
$|\Phi_i|$ many 1-bits of this sum must be at exactly the same
positions as in~$p$ (let us call this the ``front'' part of these
numbers, see the upper part and the red bits in Figure~\ref{fig-ones} for
a concrete example). However, each $\psi \in \Phi_i \cap \Lang{1cnfs}$
can contribute at most one 1-bit to the sum, which will yield a
contradiction, see Figure~\ref{fig-ones} once more. So $P_2 \ge
p_{=|\Phi_i| + 1}$ and then the average satisfaction probability of the
packings of the formulas in $\Phi_i\setminus \Lang{1cnfs}$ is at least 
$p_{=|\Phi_i|+1}/|\Phi_i|$. Consider some $\psi \in 
\Phi_i\setminus \Lang{1cnfs}$ for which
$\Prob{\operatorname{pack}(\psi)}$ is maximum. Then this maximum is at
least the average. This gives us a \emph{lower} bound on the
satisfaction probability of $\operatorname{pack}(\psi)$ for the~$\psi$
picked in the expansion step and thus an \emph{upper} bound on the size of
$\operatorname{pack}(\psi)$ by the Packing Probability Lemma. In
particular, we can bound the size of the next $\Phi_{i+1}$ purely in
terms of $|\Phi_i|$ and $p$. 

The final lemma, Lemma~\ref{lemma-fourth}, gives a recursive,
quantitative bound on how long an expansion sequence can
continue. The key observation is that just like the increase in size
from one $\Phi_i$ to the next, the length of any expansion
sequences starting with $\{\phi\}$ can be bounded purely in terms of
$k$ and~$p$ -- it does not depend on~$\phi$.

Taken together, the lemmas give a quantitative lower bound on the size
of spectral gaps. The details follow.

\subparagraph*{Expansion Sequences: Definition and Properties.}

The formal definition of expansion sequences will be a bit more
general than sketched above as, for the analysis, we will need to consider the
behaviour not only of expansion sequences starting with $\{\phi\}$,
but also of certain subsequences of such sequences -- namely
those, in which only clauses are picked for expansion that have
maximum size~$l$ for some number~$l$ resulting from a recursion. For this reason,
we formally allow expansion sequences to start with any finite set $\Phi
\subseteq \Lang{cnfs}$ that is pairwise model-disjoint. In the
definition, $\operatorname{maxpack}(\Phi)$ denotes the formula $\psi \in \Phi$ for
which $\Prob{\operatorname{pack}(\psi)}$ is maximum (resolve ties in
some deterministic way; and note that the formula returned by
$\operatorname{maxpack}(\Phi)$ is $\psi$, not
$\operatorname{pack}(\psi)$, unless $\psi$ happens to be a
packing). For $\Phi = \emptyset$ let    
$\operatorname{maxpack}(\Phi)$ be the special error symbol $\bot
\notin \Lang{cnfs}$.

\vskip 0pt plus 1cm\penalty-100\vskip 0pt plus -1cm

\begin{definition}[Expansion Sequence]\label{def-exp}
  Let $p \in (0,1)$, let $l \in \mathbb N$, and let $\Phi
  \subseteq \Lang{cnfs}$ be finite and pairwise model-disjoint. The
  \emph{expansion sequence $\operatorname{exp-seq}_{l,p}(\Phi)$} is the following sequence
  $(\Phi_1,\Phi_2,\dots,\Phi_r)$ of sets $\Phi_i \subseteq
  \Lang{cnfs}$: Start with $\Phi_1 = \Phi$. For $i \in \{1,\dots,r\}$ define:
  \begin{enumerate}
  \item \label{item-at-least} If $\ProbOr{(\Phi_i\cap1\Lang{cnfs})} \ge p$, the sequence
    ends (so $r=i$) \emph{at or above~$p$.}
  \item \label{item-bound} Otherwise, if $\sum_{\psi \in \Phi_i}
    \Prob{\operatorname{pack}(\psi)} \le p_{\le |\Phi_i|+1}$, the sequence
    ends \emph{well below~$p$.}
  \item Otherwise, if $\operatorname{maxpack}(\Phi_i \setminus
    \Lang{1cnfs}) \not\in l\Lang{cnfs}$, the sequence ends \emph{exhausted.} 
  \item Otherwise, for $\psi := \operatorname{maxpack}(\Phi_i \setminus
    \Lang{1cnfs}) \in l\Lang{cnfs}$ the sequence \emph{continues} with
    \label{item-expand}
    \begin{align}
      \Phi_{i+1} = \Phi_i \setminus \{\psi\} \cup
      \bigl\{\RestrictAdd{\psi}{\beta} \bigm| \beta \colon
      \operatorname{vars}(\operatorname{pack}(\psi)) \to
      \mathbb B\bigr\}. \label{eq-expand}
    \end{align}
  \end{enumerate}
\end{definition}

Note that item~\ref{item-bound} is exactly the technical
condition from Definition~\ref{tc-a}. Note furthermore that an expansion
sequence could become exhausted because $\Phi_i$ 
contains only $1\Lang{cnf}$ formulas (and, thus, $\Phi_i \setminus
1\Lang{cnfs}$ is empty and $\operatorname{maxpack}(\Phi_i \setminus
1\Lang{cnfs}) = \bot \notin l\Lang{cnfs}$) or because the formula in
$\Phi_i \setminus 1\Lang{cnfs}$ whose packing's satisfaction
probability is maximum does not lie in $l\Lang{cnfs}$ (has a clause of
size $l+1$ or larger). Finally, observe that an expansion sequence does, indeed, 
always end after a finite number of steps: Each time the sequences
continues, all formulas $\RestrictAdd{\psi}{\beta}$ that are added
have smaller clauses than $\psi$ had (see Lemma~\ref{lemma-res}). As
there are only finitely many $\Lang{cnf}$ formulas over the finite set
of variables that the initial~$\Phi$ had, sooner or later the sequence
is exhausted.

The first lemma links expansion sequences to $\Prob{\phi}$.
\begin{lemma}\label{lemma-first}
  Let $p \in (0,1)$, $l \in \mathbb N$, and $\phi \in \Lang{cnfs}$ be
  given. Then for $(\Phi_1,\dots,\Phi_r) := 
  \operatorname{exp-seq}_{l,p}(\{\phi\})$ and $i \in \{1,\dots,r\}$,
  the set $\Phi_i$ is pairwise model-disjoint and $\ProbOr{\Phi_i} = 
  \Prob{\phi}$.
\end{lemma}
\begin{proof}
  By induction on~$i$ sequence. The case $i=1$ is 
  trivial. For the inductive step, $\Phi_{i+1}$ is obtained
  from~$\Phi_i$ according to item~\ref{item-expand} from 
  Definition~\ref{def-exp}, that is, by removing a formula~$\psi$ from
  $\Phi_i$ and adding all formulas $\RestrictAdd{\psi}{\beta}$ in its stead for
  $\beta \colon X \to \mathbb B$ with $X =
  \operatorname{vars}(\operatorname{pack}(\psi))$. Lemma~\ref{lemma-simple-obs}
  tells us that $\Phi_{i+1}$ remains pairwise model-disjoint and 
  Corollary~\ref{col-bounds} tells us that this does not change the
  sum of the satisfaction probabilities.
\end{proof}

The second lemma links expansion sequences to spectral gaps:
\begin{lemma}\label{lemma-second}
  Let $p \in (0,1)$ and $k \in \mathbb N$. Suppose there is a
  number~$s$ such that for all $\phi \in k\Lang{cnfs}$ with
  $\Prob{\phi} <p$ the sequence
  $\operatorname{exp-seq}_{k,p}(\{\phi\})$ ends well below~$p$ and
  ends with a set $\Phi_r$ of size at most~$s$. Then 
  $\operatorname{spectral-gap}_{k\Lang{cnfs}}(p) > p_{=s+2}$.
\end{lemma}
\begin{proof}
  We have 
  \begin{align}
     \Prob{\phi} = \ProbOr{\Phi_r} = \sum_{\psi \in \Phi_r}
    \Prob{\psi} \le  \sum_{\psi \in \Phi_r}
    \Prob{\operatorname{pack}(\psi)} \le p_{\le |\Phi_r|+1} \le p_{\le
      s+1},\label{eq-second}
  \end{align}
  where the first equality holds by Lemma~\ref{lemma-first}, the second by
  definition, the first inequality as $\Prob{\psi} \le
  \Prob{\operatorname{pack}(\psi)}$, the second by the
  definition of ``ending well below~$p$,'' and the last because
  of the assumption $|\Phi_r| \le s$. As \eqref{eq-second} holds for all
  $\phi \in k\Lang{cnfs}$ with $\Prob{\phi} <p$, we conclude 
  that the interval $(p_{\le s+1}, p)$ contains no elements of
  $k\CNFSSigmaSpectrum$. Thus,
  $\operatorname{spectral-gap}_{k\Lang{cnfs}}(p) \ge p -
    p_{\le s+1} = p_{\ge s+2} > p_{=s+2}$.
\end{proof}

The third lemma bounds how quickly the sets in an expansion
sequence can grow.
\begin{lemma}\label{lemma-third}
  Let $p \in (0,1)$ and $l \in \mathbb N$. Define
  \begin{align}
    g_{l,p}(x) := x - 1 +
    2^{l\cdot\log_{1-2^{-l}}((p_{=x+1}) / x)} = 
    x - 1 + (x/p_{=x+1})^{-l/\log_2 (1-2^{-l})}. \label{eq-g}
  \end{align}
  Then for every pairwise model-disjoint $\Phi
  \subseteq \Lang{cnfs}$ we have
  $|\Phi_{i+1}| \le g_{l,p}(|\Phi_i|)$ for all $i < r$, where  
  $(\Phi_1,\dots,\Phi_r) = \operatorname{exp-seq}_{l,p}(\Phi)$.
\end{lemma}
\begin{proof}
  Let $i < r$ be given. Define
  \begin{align*}
     P_1 &= \textstyle\sum_{\psi \in \Phi_i \cap \Lang{1cnfs}}
    \Prob{\operatorname{pack}(\psi)}, \\
     P_2 &= \textstyle\sum_{\psi \in \Phi_i \setminus \Lang{1cnfs}}
     \Prob{\operatorname{pack}(\psi)}     
  \end{align*}
  and observe that $P_1 = \ProbOr{(\Phi_i\cap \Lang{1cnfs})}$, since
  $\Lang{1cnf}$ formulas are already 
  packings, and that $P_1 + P_2 = \sum_{\psi \in \Phi_i}
  \Prob{\operatorname{pack}(\psi)}$, which is the left-hand side
  in inequality~\eqref{eq-tc-a} from the technical condition. Since
  the expansion sequence has not ended, it is neither at or above~$p$
  nor well below~$p$, which means $P_1 < p$ and $P_1 + P_2 >
  p_{\le|\Phi_i|+1}$. 
  
  We claim that  $P_2 \ge p_{=|\Phi_i|+1}$ must hold. Suppose this
  were not the case, so $P_2 < p_{=|\Phi_i|+1}$. Then the two
  inequalities $P_1 < p$ and $P_1 + P_2 > p_{\smash{\le|\Phi_i|+1}}$ imply $p
  > P_1 > p_{\smash{\le|\Phi_i|+1}} - P_2 >  p_{\le|\Phi_i|+1} -  
  p_{=|\Phi_i|+1} = p_{\le|\Phi_i|}$.
  This means that $P_1$ lies strictly between $p$ and
  $p_{\le|\Phi_i|}$. In particular, the first $j_{|\Phi_i|}$ many
  (recall Lemma~\ref{lemma-unique}) leading bits of the fractional
  part of~$P_1$ must be identical to those of $p$
  and~$p_{\le|\Phi_i|}$; and there must be one additional 1-bit in
  $P_1$ following after this leading part (because of the strict
  inequality $P_1 > p_{\le|\Phi_i|}$). As, by definition,  there are
  exactly $|\Phi_i|$ many 1-bits among the first $j_{|\Phi_i|}$ bits
  of~$p$, we conclude that $P_1$ would have to have at least
  $|\Phi_i|+1$ many 1-bits in its binary  representation. However,
  $P_1$ is a sum of at 
  most $|\Phi_i \cap \Lang{1cnfs}| \le |\Phi_i|$ many numbers that
  have exactly one 1-bit in their binary representation. Since the
  number of 1-bits in the sum of two numbers is at most the sum of the
  1-bits in the individual numbers, we conclude that $P_1$ can have at
  most $|\Phi_i|$ many 1-bits it is binary representation -- leading
  to a contradiction.  

  We now know that $P_2 \ge p_{=|\Phi_i|+1}$ holds and point out (for
  later reference) a consequence:
  \begin{align}
    \Phi_i \setminus \Lang{1cnfs}\text{ is not
      empty.}\label{eq-notempty}
  \end{align}
  By definition, $P_2$ is a sum of values, so the maximum of these
  values will be larger 
  than their average. In particular, $\operatorname{pack}(\psi)$ for $\psi :=
  \operatorname{maxpack}(\Phi_i \setminus \Lang{1cnfs})$ will have a
  satisfaction probability that is at least $P_2 / |\Phi_i|\ge p_{=|\Phi_i|+1} /
  |\Phi_i|$. As the sequence is not yet exhausted, we know $\psi \in 
  l\Lang{cnfs}$ by definition. The Packing Probability Lemma now tells
  us $\left|\operatorname{pack}(\psi)\right| \le
  \log_{1-2^{-l}}(p_{=|\Phi_i|+1} / |\Phi_i|)$. This
  implies that 
  \begin{align*}
    |\Phi_{i+1}|=
    \underbrace{|\Phi_i|}_{=:x} -
    \underbrace{|\{\psi\}|}_{=1} +
    \underbrace{
      \bigl|\bigl\{
        \RestrictAdd{\psi}{\beta} \bigm| \beta \colon
        \operatorname{vars}(\operatorname{pack}(\psi)) \to
        \mathbb B\bigr
        \} \bigr|
      }_{=2^{\left|\operatorname{vars}(\operatorname{pack}(\psi))\right|}
      \rlap{$\scriptstyle\le2^{l\cdot
          \left|\operatorname{pack}(\psi)\right|
        }\le2^{l\cdot\log_{1-2^{-l}}((p_{=x+1})
          / x)}$}}. 
  \end{align*}
  Comparing this to the term $g_{l,p}$ from the lemma's claim, we
  see that $|\Phi_{i+1}| \le g_{l,p}(|\Phi_i|)$.   
\end{proof}

Note that if we consider $l\le k$ to be a constant, $g_{l,p}(x) =
(x/p_{=x+1})^{O(1)}$. Furthermore, when $p \in \mathbb Q$
holds, then the distance between consecutive 1-bits in $p$ is
bounded by a constant, so $p_{=x+1} = 2^{-O(x)}$ and
$g_{l,p}(x) = 2^{O(x)}$. In contrast, when $p \in \mathbb R
\setminus \mathbb Q$, the distance between consecutive 1-bits can get
arbitrarily large and $g_{l,p}$ can grow arbitrarily fast (just
consider $p = \sum_{i=s}^\infty 2^{-2^i}$ for any~$s$, then
$g_{l,p}(x) = 2^{2^{O(x)}}$).

The fourth lemma gives a recursive upper bound on the size of the last
set in an expansion sequence. Recall from Lemma~\ref{lemma-second}
that this bound is exactly what we need to lower-bound the spectral
gap. We use the following notation: For a function $f$ and a number
$i$ let $f^i$ denote the $i$-fold application of~$f$, so $f^0(x) =x$
and $f^{i+1}(x) = f(f^i(x))$.

\begin{lemma}\label{lemma-fourth}
  For $p \in (0,1)$ and $l \in \{1,2,3,\dots\}$ define
  recursively
  \begin{align*}
    s_{1,p}(x) &:= x,\\
    s_{l+1,p}(x) &:= s_{l,p}\bigl((g_{l+1,p} \circ
    s_{l,p})^x(x)\bigr)\text{\quad  with $g_{l,p}$ from \eqref{eq-g}}.
  \end{align*}
  Then for every finite, pairwise model-disjoint $\Phi \subseteq \Lang{cnfs}$,
  for $(\Phi_1,\dots,\Phi_r) = \operatorname{exp-seq}_{l,p}(\Phi)$
  we have
  $|\Phi_r| \le s_{l,p}(|\Phi|)$.
\end{lemma}

\begin{proof}
  To simplify the notation, we omit the dependency ``$_{,p}$'' in
  the subscript in the following 
  (the subscript just stresses the fact that both the functions as
  well as the expansion sequence itself depends on~$p$ through 
  the distribution of its 1-bits; but $p$ is fixed in this
  proof). 

  The proof is by induction. For $l=1$ the claim is simple: As it is
  impossible to continue an expansion sequence for $l=1$ (we would
  need to have $\psi \in \Phi_1 \setminus \Lang{1cnfs}$ and also $\psi
  \notin \Lang{1cnfs}$), the sequence is immediately at or above~$p$ or
  well below~$p$ or is exhausted and $|\Phi_r| = |\Phi| = s_1(|\Phi|)$.

  For the inductive step, consider $l+1$ and the sequence
  $(\Phi_1,\dots,\Phi_r) = \operatorname{exp-seq}_{l+1}(\Phi)$. By
  definition, for each $i < r$ there is a $\psi \in \Phi_i \cap
  \bigl((l+1)\Lang{cnfs} \setminus 1\Lang{cnfs}\bigr)$ such that
  $\Phi_{i+1}$ arises from $\Phi_i$ by expanding $\psi$ according
  to~\eqref{eq-expand}. In each such step, $\psi$ is removed from
  $\Phi_i$ and replaced by formulas $\RestrictAdd{\psi}{\beta}$ with smaller
  clauses. In particular, when a formula $\psi \in (l+1)\Lang{cnfs}
  \setminus l\Lang{cnfs}$ is removed, the total number of such
  formulas drops by $1$ and cannot be raised later on in the
  sequence. Let $i_1 < i_2 < \dots < i_y$ be the positions in the 
  sequence such that a formula in $(l+1)\Lang{cnfs} \setminus
  l\Lang{cnfs}$ is expanded at that position. As we cannot expand more
  formulas than there were in the original $\Phi$,
  we see that $y \le |\Phi|$. Most importantly, setting $i_0
  = 1$ and $i_{y+1} = r$, we see that for all~$j$ the expanded formulas in
  steps $i_j+1$, $i_j+2$, $\dots$, $i_{j+1}-1$ lie in
  $l\Lang{cnfs}$. This means that we can apply our induction
  hypothesis to~$\Phi_{i_j+1}$. 

  In detail, consider the expansion sequence
  $\operatorname{exp-seq}_{l}(\Phi_{i_j+1})$ starting at $\Phi_{i_j+1}$
  for $l\Lang{cnfs}$ (rather than $(l+1)\Lang{cnfs}$). This sequence
  will be exactly the subsequence of
  $\operatorname{exp-seq}_{l+1}(\Phi)$ between positions $i_j+1$ and
  $i_{j+1}$ (the formula $\psi =
  \operatorname{maxpack}(\Phi_{i_{j+1}})$ is an $(l+1)\Lang{cnf}$
  formula, causing $\operatorname{exp-seq}_{l}(\Phi_{i_j+1})$ to end
  as it is exhausted). By the induction hypothesis applied to
  $\operatorname{exp-seq}_{l}(\Phi_{i_j+1})$ we get $|\Phi_{i_{j+1}}|
  \le s_l(|\Phi_{i_j+1}|)$. This gives the sequence of size changes
  shown in Figure~\ref{fig-size-changes}. 

  \begin{figure}[htpb]
  \begin{tikzpicture}
    \foreach \t/\p/\col [count=\i] in {
      1 / 1 /black,
      i_1 / 2.25 /red!75!black,
      i_1+1 / 3.25 /black,
      i_2 / 4.75 /red!75!black,
      i_2+1 / 5.75 /black,
      i_3 / 7.1 /red!75!black,
      i_3+1 / 8.1 /black,
      i_{y-1}+1 / 10.1 /black,
      i_y / 11.6 /red!75!black,
      i_y+1 / 12.65 /black,
      r / 14.2 /black!50
    } {
      \node[anchor=mid,inner sep=1pt,outer sep=0pt,color=\col]
        (n\i) at (\p,0) {$\bigl|\Phi_{\t}\bigr|$};
    }

    \foreach \a/\b/\s in {
      1/2/\scriptsize,
      3/4/\scriptsize,
      5/6/\scriptsize,
      7/8/\normalsize,
      8/9/\scriptsize,
      10/11/\scriptsize%
    } {
      \node[anchor=mid,font=\s] at ($ (n\a.east)!.5!(n\b.west) $) {$\cdots$};
    }

    \foreach \s/\t/\what in {
      1/2/s_l,
      2/3/g_{l+1},
      3/4/s_l,
      4/5/g_{l+1},
      5/6/s_l,
      6/7/g_{l+1},
      8/9/s_l,
      9/10/g_{l+1},
      10/11/s_l
    } {
      \draw
      ([xshift=1mm]n\s.south)
      edge[bend right,-{To}] node[below]{$\what$}
      ([xshift=-1mm]n\t.south);
    }
    
    \foreach \s/\t/\what in {
      1/3/g_{l+1} \circ s_l,
      3/5/g_{l+1} \circ s_l,
      5/7/g_{l+1} \circ s_l,
      8/10/g_{l+1} \circ s_l
    } {
      \draw
      ([xshift=1mm,yshift=-1.5em]n\s.south)
      edge[bend right,-{To}] node[below]{$\what$}
      ([xshift=-1mm,yshift=-1.5em]n\t.south);
    }
    
    \draw
    ([xshift=1mm,yshift=-4em]n1.south)
    edge[bend right=5,-{To}] node[below]{$(g_{l+1} \circ s_l)^y$}
    ([xshift=-1mm,yshift=-4em]n10.south);

    \draw
    ([xshift=1mm,yshift=-6em]n1.south)
    edge[bend right=3,-{To}] node[ below]{$s_l \circ (g_{l+1} \circ s_l)^y$}
    ([xshift=-1mm,yshift=-6em]n11.south);
  \end{tikzpicture}
  \caption{ {
    The sequence of possible size changes in the expansion sequence
    $(\Phi_1,\dots,\Phi_r) = \operatorname{exp-seq}_{l+1}(\Phi)$. The
    positions where a formula in $(l+1)\Lang{cnfs}\setminus
    l\Lang{cnfs}$ is expanded are marked red (and $\Phi_r$ is gray as
    no expansion happens there). An arrow from $|\Phi_a|$ 
    to $|\Phi_b|$ with label~$f$ means $|\Phi_b| \le f(|\Phi_a|)$.
  } }
  \label{fig-size-changes}
  \end{figure}
  This shows $|\Phi_r| \le s_l\bigl((g_{l+1}\circ
  s_l)^y(|\Phi_1|)\bigr)$. As $y \le |\Phi| = |\Phi_1|$, we can
  conclude that we have
  $|\Phi_r| \le s_l\bigl((g_{l+1}\circ
  s_l)^{|\Phi|}(|\Phi|)\bigr) = s_{l+1}(|\Phi|)$. This was the claim.
\end{proof}

Jointly, the four lemmas yield (we repeat some of the definitions in
the statement of the theorem to keep it self-contained):
\begin{theorem}[Quantitative Version of
    Corollary~\ref{cor-spec}]\label{thm-spec-quant} 
  Let $p \in (0,1)$ and $k \in \mathbb N$. Then 
  $\operatorname{spectral-gap}_{k\Lang{cnfs}}(p) \ge
  p_{=s_{k,p}(1)+2}$ where
  \begin{enumerate}
  \item $p_{=i} := 2^{-j_i}$ where $j_i$ is the position of the $i$th
    1-bit in $p$'s binary representation,
  \item $s_{l,p}$ is recursively defined as $s_{1,p}(x) :=
    x$ and $s_{l+1,p}(x) := s_{l,p}\bigl((g_{l+1,p} \circ
    s_{l,p})^x(x)\bigr)$ where
  \item $g_{l,p}(x) := 
    \bigl\lceil x - 1 + \smash{(x/p_{=x+1})^{-l/\log_2 (1-2^{-l})}}\bigr\rceil$.
  \end{enumerate}
\end{theorem}
\begin{proof}
  Let any $\phi \in k\Lang{cnfs}$ be given with $\Prob{\phi} <p$. By
  Lemma~\ref{lemma-fourth}, the expansion sequence
  $\operatorname{exp-seq}_{k,p}(\{\phi\})$ ends with a set $\Psi$ of size $|\Psi| 
  \le s_{k,p}(|\{\phi\}|) =
  s_{k,p}(1)$. Lemma~\ref{lemma-first} tells us that
  $\Prob{\phi} = \ProbOr{\Psi}$ holds. The sequence ends well below~$p$
  (and is not just exhausted): Looking at the definition, we see that the
  only way for a sequence starting with $\Phi \subseteq k\Lang{cnfs}$
  to become exhausted for $l=k$ is that we reach a point where $\Phi_i
  \subseteq 1\Lang{cnfs}$ holds (otherwise we could still pick a
  $\psi \in \Phi_i \setminus 1\Lang{cnfs}$ and the requirement ``$\psi
  \in l\Lang{cnfs}$'' is trivially satisfied for $l=k$). However, by
  \eqref{eq-notempty} in Lemma~\ref{lemma-third}, when $\Phi_i$ is not
  yet well below~$p$, there must exist a formula in $\Phi_i \setminus
  1\Lang{cnfs}$. We conclude that for all $\phi \in k\Lang{cnfs}$ with
  $\Prob{\phi}<p$ the sequence
  $\operatorname{exp-seq}_{k,p}(\{\phi\})$ is well below~$p$ and its last 
  set has size at most $s_{k,p}(1)$. Lemma~\ref{lemma-second}
  now yields the claim.
\end{proof}

As was pointed out earlier, for non-rational~$p$ already the term
$p_{=s}$ can get very small very fast even when $s$ is, say, linear
in~$k$. However, even for rational~$p$ where $p_{=i} =
2^{-O(i)}$, the recursion from Lemma~\ref{lemma-fourth} gives us a
spectral gap of at least $2^{-s_k(1)}$ for a recursion roughly of the
form $s_l(x) = (\exp \circ s_{l-1})^{x} (x)$, which yields
hyperexponential terms.

\section{Algorithmic Results}
\label{section-algorithmics}

The Spectral Well-Ordering Theorem, by which the spectra
$k\CNFSSigmaSpectrum$ ``have gaps everywhere,'' has profound
algorithmic consequences: For fixed $k$ and~$p$, if on input of any
$\phi \in k\Lang{cnfs}$ we are able to compute an interval $I
\subseteq [0,1]$ with $|I| <
\operatorname{spectral-gap}_{k\Lang{cnfs}}(p)$ and $\Prob{\phi} \in I$, then
$\Prob{\phi} \ge p$ iff $\max I \ge p$ (recall
Figure~\ref{fig-gap-only}). We saw already in the 
introduction that a randomized algorithm can compute such an interval
\emph{very} easily with high probability (which implies
$k\Lang{sat-pr}_{\ge p} \in \Class{BPP}$); and  Trevisan
showed~\cite{Trevisan2004} that with a bit more effort, we can also
compute $I$ deterministically in linear time, implying
$k\Lang{sat-pr}_{\ge p} \in \Class{LINTIME}$. However, ``just''
being able to compute such an interval $I$ will not be ``enough''
later on: We need more insight into the structure of the
formulas~$\phi$ with $\Prob{\phi} \ge p$. (Of course, we are ultimately
interested in deciding $\Prob{\phi} > p$ rather than $\Prob{\phi} \ge
p$, but we will focus on the latter question in the present section.)

Section~\ref{sec-plucking} uses \emph{kernels} as a way of better
understanding the structure of those $\phi \in k\Lang{cnfs}$ for which
$\Prob{\phi} \ge p$ holds: We analyze two sunflower-based algorithms that
compute kernels for $k\Lang{sat-pr}_{\ge p}$ and differ only in the used
reduction rule (``plucking'' or ``pruning''). The computed kernels,
denoted $\phi^*_{\mathrm{pluck}}$ and $\phi^*_{\mathrm{prune}}$,
respectively, have a number of useful properties. For instance, those of
$\phi^*_{\mathrm{pluck}}$ will allow us to show that the
spectra $k\CNFSSigmaSpectrum$ are \emph{topologically closed,} and
will also be important later in Section~\ref{section-complexity}. For
$\phi^*_{\mathrm{prune}}$ we will always have $\phi^*_{\mathrm{prune}}
\subseteq \phi$, which will imply the Threshold
Locality Theorem (Theorem~\ref{thm-local} in the introduction). Like
the Spectral Well-Ordering Theorem, the Threshold Locality Theorem is
not algorithmic in nature, 
but has a strong algorithmic consequence: We will use it to
reduce certain satisfaction probability threshold problems to model
checking problems. This will allow us to ``chip away'' one
``majority-of-'' in any majority-of-majority-of-$\dots$-majority
problem. In particular, we will get a proof of the Akmal--Williams
conjecture $\Lang{maj-maj-}k\Lang{sat} \in \Class P$. As a final
application, we show that the computed kernels are, in a sense,
``small witnesses of $\Prob{\phi} \ge p$ or $\Prob{\phi} < p$.'' While
this observation is a simple consequence of the previous results, it
will prove useful later on when we 
study the complexity of $k\Lang{sat-pr}_{>p}$. 

In Section~\ref{section-algo-intervals} we develop a new algorithm
for deciding $k\Lang{sat-pr}_{\ge p}$ that is \emph{gap size
oblivious,} meaning that it \emph{does not need to know the size of the spectral gap}
and we do not need to hardwire constants into the algorithm that 
depend on it. Instead, the algorithm will contain a simple-to-check termination
property, which we will show to hold -- at the latest -- after a fixed
number of steps. Ironically, even though the algorithm neither
``knows'' nor ``refers to'' the sizes of spectral gaps, the termination
proof is build on top of the rather technical analysis from
Section~\ref{sec-bounds} on quantitative bounds for spectral~gaps.

\subsection{Kernelizing the Satisfaction Probability Threshold Problem}
\label{sec-plucking}
\label{sec-kernels}

The computation of kernels for $k\Lang{sat-pr}_{\ge p}$  will
use two reduction rules that are based on ``plucking'' or
``pruning'' sunflowers. After having reviewed some basic properties of
sunflowers and having proved the safety of the rules, we will use the
properties of the kernels to show results that 
are of independence interest: We show that the spectra of
$\Prob{\phi}$ for $\phi \in k\Lang{cnfs}$ are topologically closed for
all~$k$, that two locality theorems holds,  
that the  majority-of-majority problem lies in $\Class P$ for $k\Lang{cnfs}$, and that
both $\Prob{\phi} \ge p$ and $\Prob{\phi} < p$ have small witnesses.

(A brief note to readers familiar with \textsc{fpt} theory at this
point: We do, indeed, compute kernels in the sense of \textsc{fpt}
theory, but for the parameterized problem 
$\mathrm p\Lang{-cnf-sat-pr}_{\ge} := \bigl\{\bigl(\phi,(k,p)\bigr) \in
\Lang{cnfs} \times 
\mathbb N \times \mathbb Q \bigm|\penalty-50 \phi \in k\Lang{cnfs},\penalty0
\Prob{\phi} \ge p\bigr\}$, where the pair $(k,p)$ is the parameter. However, since
we will neither need nor explore kernel theory in detail, we will keep
the description intuitive and will -- imprecisely -- refer to
``kernels for $k\Lang{sat-pr}_{\ge p}$.'')

\subparagraph*{The Erd\H os--Rado Sunflower Lemma.}
Recall from Definition~\ref{def-sunflower} that a
\emph{sunflower with core~$c$} is a formula $\psi \in \Lang{cnfs}$ such that $c 
\subseteq e$ holds for all clauses $e \in \psi$ and such that for
any two different \emph{petals} $e,e' \in \psi$ we have $\operatorname{vars}(e) \cap
\operatorname{vars}(e') = \operatorname{vars}(c)$. 
Sunflowers play a key role in the computation of kernels for the
hitting set problem~\cite{FlumG06} and 
related problems: Suppose that for a given formula~$\phi$ we want to
find a size-$h$ set~$V$ of variables such that for each clause $e \in
\phi$ we have $\operatorname{vars}(e) \cap V \neq \emptyset$ (each
clause is ``hit'' by~$V$). Then if there is a sunflower $\psi
\subseteq \phi$ of size $h+1$ in $\phi$ with some core~$c$, any
size-$h$ hitting set~$V$ must hit~$c$ since, otherwise, we would need
$h+1$ variables to hit the ``petals outside the core'' of the
sunflower (we would need to have $V \cap (\operatorname{vars}(e)
\setminus \operatorname{vars}(c)) \neq  \emptyset$ for $h+1$ pairwise
disjoint sets $\operatorname{vars}(e) \setminus
\operatorname{vars}(c)$). This means that $\phi$ has a size-$h$ 
hitting set iff $(\phi \setminus \psi) \cup \{c\}$ has one (indeed, iff
$(\phi \setminus \operatorname{link}_\phi(c)) \cup \{c\}$ has a
size-$h$ hitting set, where
$\operatorname{link}_\phi(c) = \{e \in \phi \mid c \subseteq
e\}$). Most importantly, 
applying this reduction rule ``as often as possible'' leads to a
formula whose size is bounded by a constant depending only on~$h$, not
on the original formula (this is known as a ``kernelization'' in
\textsc{fpt} theory). The reason for this size bound is the
following Sunflower Lemma (rephrased in terms of positive formulas
rather than hypergraphs, as would be standard, where a \emph{positive}
formula is a formula without negations):

\begin{fact}[Sunflower Lemma, \cite{ErdosR60}]
  Every positive $\phi \in k\Lang{cnfs}$ with more than
  $h^{k}\cdot k!$ clauses contains a sunflower of
  size~$h+1$.  
\end{fact}
The ``positive'' in the statement is due to the fact that in
combinatorics sunflowers usually do not care about the ``sign'' of the
variables (whether or not it is negated). In particular, for a formula
$\phi$ let $\phi^+$ be the formula where all negations are simply
removed. Then the Sunflower Lemma tells us that if $\phi^+$ is
sufficiently large, then it has a large sunflower $\psi \subseteq
\phi^+$ with core~$c$. This large sunflower does not
necessarily become a large sunflower of the original~$\phi$ if we just
reinsert the negations: While this makes no difference for the petals
outside the core, there may now suddenly be up to $2^{|c|}$ different
versions of the core. However, for the version of this core that is
present in the maximum number of petals, the number of these petals is
at least a fraction of $1/2^{|c|} \ge 1/2^k$ of the size of the
``unsigned'' sunflower. This yields  the following corollary:

\begin{corollary}\label{cor-sunflower-lemma}
  Every $\phi \in k\Lang{cnfs}$ with more than
  $(2h)^{k}\cdot k!$ clauses contains a sunflower of
  size~$h+1$.
\end{corollary}
For our purposes, the contraposition of the corollary will be of
particular interest: When a formula contains \emph{no} (longer a)
sunflower of size~$h+1$, the formula has size at most
$(2h)^{k}\cdot k!$ (and a recent breakthrough by Alweiss,
Lovett, Wu, and Zhang \cite{10.4007/annals.2021.194.3.5} shows that we
can even improve on this size bound, see \cite{10.19086/da.11887} for
a simplified proof and as a starting point for further
improvements). The important point for us will be that this size is a
\emph{constant} when $h$ and~$k$ are constants.

\subparagraph*{The Plucking and Pruning Rules.}
As just pointed out, sunflowers are of interest in the context of
computing hitting set kernels since a small hitting set of a large
sunflower ``must hit the core,'' which allows us to replace large sunflowers
by their cores (which can be envisioned as ``plucking the petals'' so
that only the core remains). In our context, plucking petals from a sunflower \emph{does}
cause a change in the satisfaction probability -- but only a small one
for large sunflowers. Because of the No Tunneling Theorem, this
will mean that it is \emph{safe} to do such a replacement for large
enough sunflowers. Instead of completely replacing a sunflower 
by its core, it is also possible to just ``prune'' the sunflower,
meaning that one completely removes as many clauses as needed so that
only a certain number remain. If the remaining clauses are still
numerous enough, then the pruning will also cause only a small change
in the satisfaction probability. The details follow.

\begin{definition}
  A reduction rule is \emph{safe for $k\Lang{sat-prop}_{\ge p}$} if
  for every $\phi \in k\Lang{cnfs}$ to which it is \emph{applicable,} it
  yields a $\phi' \in k\Lang{cnfs}$ with $\Prob{\phi} \ge p$ iff
  $\Prob{\phi'} \ge p$.
\end{definition}

To formally state the two rules, fix $k$ and~$p$. Both rules will (only)
be applicable to a given $\phi \in k\Lang{cnfs}$ if it contains a 
large enough sunflower $\psi \subseteq \phi$, which will mean that
the number of petals in $\psi$ is at least 
\begin{align}
  h_{k,p} := 1+
  \log_{1-2^{-k}}\bigl(\operatorname{spectral-gap}_{k\Lang{cnfs}}(p)\bigr). \label{eq-h} 
\end{align}
Both rules then basically wish to remove all petals of $\psi$
from~$\phi$ and to then add either the core~$c$ of~$\psi$ (plucking rule)
or a size-$h_{k,p}$ subset of the petals (pruning rule). However, as hinted at
earlier, when removing the petals of~$\psi$ from~$\phi$, we can
actually remove \emph{all clauses that contain~$c$ for free} (meaning
that the resulting formula will still be gap-close). For this reason,
the following rules remove from~$\phi$ all clauses in $\operatorname{link}_\phi(c)$,
which was defined as $\{e \in \phi \mid
c\subseteq e\}$ and for which $\operatorname{link}_\phi(c) \supseteq
\psi$ holds. 

\begin{reduction rule}[Plucking Rule]\label{rule-pluck}
  Let $\phi \in k\Lang{cnfs}$. The \emph{plucking rule} is
  \emph{applicable to~$\phi$} if there
  exists a sunflower $\psi \subseteq \phi$ 
  with $|\psi| \ge h_{k,p}$. In this case, let $c$ be the core of
  $\psi$. The rule \emph{yields} 
  \begin{align*}
    \operatorname{pluck}_{\psi}(\phi) := (\phi
    \setminus \operatorname{link}_\phi(c)) \cup \{c\}.
  \end{align*}
\end{reduction rule}

\begin{reduction rule}[Pruning Rule]\label{rule-prune}
  Let $\phi \in k\Lang{cnfs}$. The \emph{pruning rule} is
  \emph{applicable to~$\phi$} if there
  exists a sunflower $\psi \subseteq \phi$ 
  with $|\psi| \ge h_{k,p}$. In this case, let $c$ be the core
  of~$\psi$ and let $\{e_1,\dots,e_{|\psi|}\} = \psi$ be the clauses 
  of~$\psi$ in some deterministic order. The rule \emph{yields} 
  \begin{align*}
    \operatorname{prune}_{\psi}(\phi) := (\phi
    \setminus \operatorname{link}_\phi(c)) \cup \{e_1,\dots,e_{h_{k,p}}\}.
  \end{align*}
\end{reduction rule}

\begin{figure}[htbp]
  \centering
  \begin{tikzpicture}[sunflower  figure]

    \scoped{
      \pictureexplain{$\operatorname{pluck}_{\psi}(\phi)$}
      \pictureliterals
      \scoped[]{\picturekernel}

      \scoped[green!50!black] {
        \pic {join 3 = l on 4 with a' on 4 with b on 4};
      }
    }

    \scoped[xshift=7cm] {
      \pictureexplain{$\operatorname{prune}_{\psi}(\phi)$}
      \pictureliterals
      \scoped[]{\picturekernel}

      \scoped[green!50!black] {
        \pic {join 3' = a' on 2 with b on 2 with c on 1};
        \pic {join 4' = a' on 3 with b on 3 with d' on 1 with j on 1};
        \pic {join 3' = a' on 4 with b on 4 with e on 1};
      }
    }
    
  \end{tikzpicture}
  \caption{Examples of the effects of applying the plucking and
    pruning rules to a formula. Recall $\phi \in \Lang{5cnfs}$ from
    Figure~\ref{fig-sunflower-only} on
    page~\pageref{fig-sunflower-only} and the sunflower $\psi
    \subseteq \phi$ (the
    solid  clauses in that figure) with core $\{x,\neg y,z\}$. Assuming that
    $h_{k,p} = 3$ holds (actually, the value will typically be
    \emph{much} larger), Rules \ref{rule-pluck} and~\ref{rule-prune}
    are applicable (the sunflower has size $|\psi| = 5 \ge 3$) and
    yield the depicted formulas. Note that the link of~$c$ is
    completely removed by the rules, so (unlike the formula~$\phi'$ on
    the right hand side of Figure~\ref{fig-sunflower-only}) the dotted
    lines also got removed, not just the sunflower itself.
  }
  \label{fig-sunflower}
\end{figure}

\begin{lemma}[Safety Lemma]\label{lemma-safety}\label{lem-sunflower-bounds}
  If Rule \ref{rule-pluck} or~\ref{rule-prune} is applicable to $\phi
  \in k\Lang{cnfs}$, the rule yields a formula that is gap-close
  to~$\phi$. Thus, both rules are safe by the No Tunneling Lemma.
\end{lemma}
\begin{proof}
  Let $\phi \in k\Lang{cnfs}$ be given and let $\psi \subseteq \phi$
  be a sunflower with core~$c$. By assumption, $|\psi| \ge
  h_{k,p}$ and let us just write $h$ for $h_{k,p}$ in the
  following. Let $\psi = \{e_1,\dots,e_{|\psi|}\}$.
  Observe that $\{c\} \FormImpl \operatorname{link}_\phi(c) \FormImpl \psi
  \FormImpl \{e_1,\dots,e_h\}$ holds and thus
  \begin{align}
    \operatorname{pluck}_\psi(\phi) \FormImpl \phi \FormImpl
    \operatorname{prune}_\psi(\phi), \label{eq-pluck-prune}
  \end{align}
  so it 
  suffices to show that $\operatorname{pluck}_\psi(\phi)$ and
  $\operatorname{prune}_\psi(\phi)$ are gap-close. For this, consider
  any assignment $\beta$ that satisfies
  $\operatorname{prune}_\psi(\phi)$, but not
  $\operatorname{pluck}_\psi(\phi)$. By definition, $\beta$ must
  then satisfy $\{e_1,\dots,e_h\}$ and cannot satisfy $\{c\}$. Thus,
  $\ProbBig{\operatorname{prune}_\psi(\phi)} -
  \ProbBig{\operatorname{pluck}_\psi(\phi)} \le
  \ProbBig{\{e_1,\dots,e_h\}} - \ProbBig{\{c\}} $ and  
  it suffices to show that $\{c\}$ and $\{e_1,\dots,e_h\}$ are
  gap-close. Since $\bigl\{e_1
  \setminus c, \dots, e_{h} \setminus c\}$ is a packing, the
  Packing Probability Lemma yields: 
  \begin{align*}
    \ProbBig{\{e_1,\dots,e_h\}} - \ProbBig{\{c\}} &=
    \overbrace{2^{-|c|}}^{\le1}\;  \prod\nolimits_{i=1}^h
    \overbrace{(1 - 2^{|e_i|-|c|})}^{\le 1-2^{-k}}\\& \le (1-2^{-k})^h 
    < (1-2^{-k})^{h-1} \overbrace{=}^{\smash{\text{by \eqref{eq-h}}}}
    \operatorname{spectral-gap}_{k\Lang{cnfs}}(p).
    \qedhere
  \end{align*}
\end{proof}

\begin{lstlisting}[
    style=pseudocode,
    backgroundcolor=,
    float=t,
    label={algo-sunflower},
    backgroundcolor=,
    caption={{
        Adaption of the sunflower-based kernel algorithm for hitting
        sets to our setting: On input of a formula $\phi \in
        k\Lang{cnfs}$, the algorithm simply applies a single
        $\mathit{rule}$ (either the pruning rule or the plucking rule)
        as long as possible and returns the final
        formula~$\phi^*$. The Safety Lemma together with the
        Sunflower Lemma imply that this is a kernel for
        $k\Lang{sat-pr}_{\ge p}$. In particular, we can check
        whether $\Prob{\phi} \ge p$ holds by checking for the
        constant-size formula $\phi^*_{\mathit{rule}}$ whether $\Prob{\phi^*_{\mathit{rule}}} \ge p$
        holds. 
    } }
  ]
algorithm $\Algo{kernel}(\phi,k,p,\mathit{rule})$ // $\commented{\mathit{rule} \in \{\mathrm{pluck},\mathrm{prune}\}}$, $\commented{\phi \in k\Lang{cnfs}}$
    while $\phi$ contains a sunflower $\psi$ of size at least $h_{k,p}+1$ do
        // The rule is applicable
        $\phi \gets \mathit{rule}_\psi(\phi)\label{line-col}$ 
    return $\phi$ // denoted $\commented{\phi^*_{\mathit{rule}}}$ later on
\end{lstlisting}

\subparagraph*{The Sunflower Kernel Algorithm and Its Properties.}
With the safety of the two reduction rules established, we get a kernel
algorithm (see Algorithm~\ref{algo-sunflower}): Simply apply a given
$\mathit{rule}$ as long as possible. The important properties of the
computed formulas are summarized in the Kernel Theorem from the
introduction, whose claim we restate here:

\begin{reclaim*}[of Theorem~\ref{thm-kernel}]
  For each $k$ and $p$, on input $\phi \in k\Lang{cnfs}$ we can
  compute formulas $\phi^*_{\mathrm{pluck}}, \phi^*_{\mathrm{prune}}
  \in k\Lang{cnfs}$ such that:
  \begin{enumerate}
  \item The computation takes linear time or is done by
    $\Class{AC}^0$ circuits. 
  \item $|\phi^*_{\mathrm{pluck}}| \le S_{k,p}$ and
    $|\phi^*_{\mathrm{prune}}| \le S_{k,p}$ for a constant~$S_{k,p}$
    depending only on $k$ and~$p$. 
  \item $\phi^*_{\mathrm{prune}} \subseteq \phi$.
  \item $\phi^*_{\mathrm{pluck}}(\phi) \FormImpl \phi \FormImpl
    \phi^*_{\mathrm{prune}}(\phi)$ and hence $\Prob{\phi^*_{\mathrm{pluck}}} \le \Prob{\phi} \le \Prob{\phi^*_{\mathrm{prune}}}$.
  \item $\Prob{\phi} \ge p$ iff $\Prob{\phi^*_{\mathrm{pluck}}} \ge p$ iff $\Prob{\phi^*_{\mathrm{prune}}} \ge p$.
  \end{enumerate}
\end{reclaim*}

Note that items 1, 2 and~5 together mean that both
$\phi^*_{\mathrm{pluck}}$ and $\phi^*_{\mathrm{prune}}$ are
kernels. Thus, we can check whether $\Prob{\phi} \ge p$ holds 
by checking whether either of $\Prob{\phi^*_{\mathrm{pluck}}} \ge p$
or  $\Prob{\phi^*_{\mathrm{prune}}} \ge p$ holds and
$\phi^*_{\mathrm{pluck}}$ and $\phi^*_{\mathrm{prune}}$ are a constant-size
formulas by item~2. 

\begin{proof}
  Set $S_{k,p}:=\bigl(2h_{k,p}\bigr){}^k\cdot k!$.
  Let $\phi^*_{\mathrm{pluck}}$ be the output of
  $\Algo{kernel}(\phi,k,p,\mathrm{pluck})$ from
  Algorithm~\ref{algo-sunflower} and let
  $\phi^*_{\mathrm{prune}}$ be the output of  
  $\Algo{kernel}(\phi,k,p,\mathrm{prune})$. The items now follow:
  \begin{enumerate}
  \item
    Each iteration of the while loop reduces
    the size of~$\phi$ by at least~$1$ as we search for a sunflower of
    size $h_{k,p}+1$ and replace it either by a single clause (the
    plucking rule) or by at most $h_{k,p}$ many clauses (the pruning
    rule). Note that finding large sunflowers is a bit of 
    an art and there is extensive literature on how to do this efficiently,
    see~\cite{BannachT20,FlumG06,vanBevern2014} for starting
    points, but the desired minimum size is 
    fixed in our case and we could even brute-force the search here.
  \item Corollary~\ref{cor-sunflower-lemma} states that as long as
    there are more than $(2h_{k,p})^k k!$ 
    clauses in~$\phi$, there is still a sunflower of size $h_{k,p}+1$ and,
    hence, the while loop will not have ended and the rule is still applicable. Thus,
    $|\phi^*| \le (2h_{k,p})^k k!$ as claimed.
  \item Since $\operatorname{prune}_\psi(\phi) \subseteq \phi$ holds,
    in each assignment $\phi \gets \operatorname{prune}_\psi(\phi)$ we  
    just remove clauses from~$\phi$. In particular, the final
    $\phi^*_{\mathrm{prune}}$ must have this property.
  \item By \eqref{eq-pluck-prune} from the proof of the Safety
    Lemma, 
    $\operatorname{pluck}_\psi(\phi) \FormImpl \phi \FormImpl
    \operatorname{prune}_\psi(\phi)$. By induction on the length of
    the while loop, the relations also hold for
    the final outputs. The inequalities follow trivially from this.
  \item This follows by induction once more, in conjunction with the
    Safety Lemma.\qedhere  
  \end{enumerate}
\end{proof}

Although we will see in a moment that the Kernel Theorem has a number
of interesting consequences, it is worthwhile to spell out the most
immediately corollary, namely that for ``fixed parameters $k$
and~$p$'' we can use kernels to decide $k\Lang{sat-pr}_{\ge
    p}$ very efficiently:

\begin{corollary}\label{cor-ac0-ge}
  $k\Lang{sat-pr}_{\ge
    p}$ lies in $\Class{AC}^0$ for all $k$
  and~$p \in [0,1]$.
\end{corollary}

\begin{proof}
  It is well-established~\cite{BannachST15,BannachT18} that kernels for
  hitting sets can be computed by
  $\Class{AC}^0$-circuits parameterized by the size of the hitting set
  and the size of the hyperedges. In particular, there exist
  $\Class{AC}^0$-circuits both for
  $\Algo{kernel}(\phi,k,p,\mathrm{prune})$ and
  $\Algo{kernel}(\phi,k,p,\mathrm{pluck})$, when $\phi$ is positive
  and $k$ and $p$ are the parameters. As the algorithms can easily be
  adapted to cope with the fact that sunflowers for formulas must take
  the ``signs'' of the literals in the cores into account, see the
  discussion prior to Corollary~\ref{cor-sunflower-lemma}, we get the
  claim.
\end{proof}

\subsubsection*{Application: The Spectra Are Topologically Closed}

The Kernel Theorem implies a purely combinatorial
statement: 

\begin{lemma}\label{lemma-kappa}
  For every $p \in [0,1]$ and $k$ there is a size~$S$ such that
  for all $\phi \in k\Lang{cnfs}$ with $\Prob{\phi} \ge
  p$ there is a formula $\phi^* \in k\Lang{cnfs}$ of size
  $|\phi^*| \le S$ with $\Prob{\phi} \ge
  \Prob{\phi^*} \ge p$. 
\end{lemma}
\begin{proof}
  Let $S = S_{k,p}$ from Theorem~\ref{thm-kernel} and let
  $\phi^* = \phi^*_{\operatorname{pluck}}$. The claim immediately follows from
  items~2, 4 and~5 of the theorem.
\end{proof}

This lemma provides us with an easy way of showing that
$k\CNFSSigmaSpectrum$ is \emph{topologically closed,} which means
that its complement is an open set. Note that this does not follow
from the fact that the spectra are well-ordered as the set
$\{1,\frac{1}{2},\frac{1}{4},\frac{1}{8},\dots\}$ is well-ordered, but
not closed (it misses~$0$).
\begin{lemma}\label{lemma-closed}
  Let $\Phi \subseteq k\Lang{cnfs}$. Then $\inf \{\Prob{\phi} \mid
  \phi \in \Phi\} \in k\CNFSSigmaSpectrum$. 
\end{lemma}
\begin{proof}
  Let $p = \inf \{\Prob{\phi} \mid \phi \in \Phi\}$. Then there
  must be a sequence $(\phi_0,\phi_1,\phi_2,\dots)$ with $\phi_i \in \Phi$
  and $\lim_{i\to\infty} \Prob{\phi_i} = p$. Consider the
  sequence $(\phi^*_0,\phi^*_1,\phi^*_2,\dots)$ where each $\phi^*_i$ is the
  formula from Lemma~\ref{lemma-kappa} for~$\phi_i$. Then, clearly,
  $\lim_{i \to \infty} \Prob{\phi^*_i} = p$. If necessary,
  rename the variables in each $\phi^*_i$ to that they are
  $\{v_1,\dots,v_q\}$ for $q = k\cdot S$, where $S$ is the constant
  from Lemma~\ref{lemma-kappa}, and note that this is always
  possible. Then $\Phi^* := \{\phi^*_i \mid i  \in \mathbb N\}$ is a
  \emph{finite} set as there are only finitely many different
  $\Lang{cnf}$ formulas over the variables $\{v_1,\dots,v_q\}$. This
  means that there is some $\mu \in \Phi^*$ with $\Prob{\mu} =
  \min \{\Prob{\rho} \mid \rho \in \Phi^*\}$. Then $\Prob{\mu} =
  p$ must hold and $\mu \in k\Lang{cnfs}$ witnesses $p
  \in k\CNFSSigmaSpectrum$.
\end{proof}

We get two interesting corollaries: First, the spectra are
topologically closed. Second, non-rational, non-dyadic thresholds
can always be ``replaced'' by dyadic rationals (recall $\mathbb D =
\{m/2^e \mid m,e \in \mathbb Z\}$).
\begin{corollary}\label{corollary-closed}
  For every $k$, the set $k\CNFSSigmaSpectrum$ is topologically closed.
\end{corollary}
\begin{corollary}
  For every $p \in [0,1] \setminus \mathbb D$ we have
  $p' := \inf\bigl\{\Prob{\phi} \mid \phi \in
  k\Lang{cnfs},\penalty0 \Prob{\phi} > p\bigr\} \in \mathbb D$ and
  $k\Lang{sat-pr}_{>p} = k\Lang{sat-pr}_{\ge p}
  =k\Lang{sat-pr}_{\ge p'}$.
\end{corollary}

\subsubsection*{Application: Locality Theorems}

Another way of seeing the Kernel Theorem as a purely
combinatorial statement is in the form of the locality
theorems for $k\Lang{cnfs}$, whose claims we repeat here:

\begin{reclaim*}[of Theorem~\ref{thm-constant-influence}]
  For every $k$ and~$p$ there is
  a number~$S$ so that for every $\phi \in k\Lang{cnfs}$ with 
  $\Prob{\phi} = p$ there is a $\psi \in k\Lang{cnfs}$ of
  size $|\psi| \le S$ with $\phi \equiv \psi$.
\end{reclaim*}

\begin{proof}[Proof]
  Let $S = S_{k,p}$ be the bound from item~2 of
  Theorem~\ref{thm-kernel}. For a given $\phi  \in k\Lang{cnfs}$ with 
  $\Prob{\phi} = p$, let $\psi := \phi^*_{\mathrm{pluck}}$. By item~2 we
  have $|\psi| \le S$. By first item~5, then item~4, and then the
  assumption, we  have $p \le \Prob{\psi} \le \Pr[\phi] = p$. Finally, once
  more by item~4, $\psi = \phi^*_{\mathrm{pluck}} \FormImpl \phi$.
  We conclude that the formulas $\psi$ and $\phi$ have the same
  number of satisfying assignments (since $\Prob{\psi}
  = \Pr[\phi]$) and every satisfying assignment of $\psi$
  also satisfies $\phi$ (since  $\psi \FormImpl
  \phi$). This is only possible when $\phi \equiv \psi$.
\end{proof}

\begin{reclaim*}[of Theorem~\ref{thm-local}]
  For every $k$ and~$p$ there is
  a size~$S$ so that for every $\phi \in k\Lang{cnfs}$ we have
  $\Prob{\phi} \ge p$ iff $\Prob{\phi'} \ge p$ holds for every $\phi'
  \subseteq \phi$ with $|\phi'| \le S$.
\end{reclaim*}

\begin{proof}[Proof]
  One direction is trivial:   Since $\phi' \subseteq \phi$ implies
  $\Prob{\phi'} \ge \Prob{\phi}$, 
  when $\Prob{\phi} \ge p$, then $\Prob{\phi'} \ge p$ holds for all $\phi'
  \subseteq \phi$. For the other direction, let $S = S_{k,p}$ once
  more be the  bound from item~2 of Theorem~\ref{thm-kernel} and assume
  $\Prob{\phi} < p$. Then by items~3 and 5, $\phi' =
  \phi^*_{\mathrm{prune}}$ has the properties $|\phi'| \le S$ and $\Prob{\phi'} < p$. 
\end{proof}

\subsubsection*{Application: Witnesses for Satisfaction Thresholds}

\!\emph{Witnesses} (sometimes also called
\emph{proofs,} depending on 
the context) are a powerful tool of complexity theory. The idea is
that for a language $A \subseteq \Sigma^*$ and an instance $x \in
\Sigma^*$, being shown a witness will ``immediately convince us'' that
$x \in A$ holds, while for $x \notin A$ ``no alleged witness could possibly
convince us.'' For a classical example, being shown a satisfying
assignment~$\beta$ for a formula~$\phi$ will ``immediately convince
us'' that $\phi \in \Lang{sat}$ holds. Similarly, being shown a
Hamiltonian cycle in a graph~$G$ will ``immediately convince us'' that
$G \in \Lang{hamiltonian}$ holds. Crucially, for unsatisfiable
formulas and non-Hamiltonian graphs, witnesses do not exist.
While witness-based arguments are commonly used to show that problems
lie in~$\Class{NP}$ or classes further up the polynomial hierarchy,
they can also help to show membership in classes as small as
$\Class{AC}^0$ -- namely, when the witnesses have logarithmic length. Formally:
\begin{definition}
  A \emph{small witness relation} for $A \subseteq
  \Sigma^*$ is a relation $W \subseteq \Sigma^* \times \Sigma^*$ with
  \begin{enumerate}
  \item for each $x \in A$ there is a $w \in \Sigma^*$ of length
    $O(\log |x|)$ with $(x,w) \in W$, 
  \item for each $x \notin A$ for all $w \in \Sigma^*$ we have $(x,w)
    \notin W$.
  \end{enumerate}
\end{definition}
\begin{lemma}\label{lem-comp-witness}
  If $A$ has a small witness relation in~$\Class{AC}^0$, then
  \begin{enumerate}
  \item
    there is a function in $\Class{AC}^0$ that maps each $x \in A$ to
    a witness~$w$, meaning $(x,w) \in W$, 
  \item
    and thus $A \in \Class{AC}^0$.
  \end{enumerate}
\end{lemma}
\begin{proof}
  For $x \in \Sigma^*$ there are at most $|x|^{O(1)}$ different
  possible witnesses, which can be checked in parallel. For the first
  item, just output the first of them for $x \in A$,  and for the
  second item, accept $x \in \Sigma^*$ if at least one of the possible
  witness is, indeed, a witness.
\end{proof}

Our previous algorithmic results for checking whether $\Prob{\phi} \ge
p$ holds (or not) for a given $\phi \in k\Lang{cnfs}$ can easily be
rephrased in terms of small witnesses. This will be helpful later on
when we study the complexity of $k\Lang{sat-pr}_{>p}$.  

\begin{lemma}[Small Witnesses for ``$\Prob{\phi}<p$'']\label{lem-witness-less}
  For $k$ and $p$, there is an $S \in \mathbb N$ such that 
  \begin{align*}
    \Lang{witness-}k\Lang{sat-pr}_{<p} := \{(\phi,\omega) \mid \phi \in k\Lang{cnfs},\,  \omega \subseteq \phi,\, |\omega| 
    \le S,\, \Prob{\omega} < p\}
  \end{align*}
  is a small witness relation in $\Class{AC}^0$ for
  $k\Lang{sat-pr}_{<p}$. 
\end{lemma}
\begin{proof}
  Let $S$ be the number from the Threshold Locality Theorem
  (Theorem~\ref{thm-local}). The two properties of a small witness
  relation hold since:
  \begin{enumerate}
  \item
    The theorem tells us that for each $\phi \in k\Lang{cnfs}$ with
    $\Pr[\phi] <p$ there
    is an $\omega \subseteq \phi$ of size $|\omega| \le S$
    with $\Pr[\omega] < p$, meaning that there is a size-$S$
    witness~$\omega$ for $\Pr[\phi] < p$. Crucially, the encoding
    of~$\omega$ needs only $S \cdot O(k\log
    \left|\operatorname{vars}(\phi)\right|) = O(\log n)$ bits (as
    $S$~is a constant).
  \item
    When $(\phi,\omega) \in \Lang{witness-}k\Lang{sat-pr}_{<p}$ holds,
    we have $\omega \subseteq \phi$ and hence $\Pr[\phi] \le \Pr[\omega]
    < p$. Thus, every $(\phi,\omega)$ in the relation does, indeed, witness
    $\Pr[\phi] < p$.
  \end{enumerate}
  In total, we get that $\Lang{witness-}k\Lang{sat-pr}_{<p}$ is a small 
  witness relation and membership in~$\Class{AC}^0$ is
  straightforward as $S$~is a constant. 
\end{proof}

\begin{lemma}[Small Witnesses for ``$\Prob{\phi} \ge p$'']\label{lem-witness-ge}
  For $k$ and $p$, there is an $S \in \mathbb N$ such that 
  \begin{align*}
    \Lang{witness-}k\Lang{sat-pr}_{\ge p} := \bigl\{(\phi,X) \bigm| {}& \phi \in k\Lang{cnfs},\, X \subseteq
    \operatorname{vars}(\phi),\, |X| 
    \le S,\,\\
    &\textstyle\Pr_{\beta:X \to \mathbb B}[\phi|_\beta = \emptyset] \ge p\bigr\}
  \end{align*}
  is a small witness relation in $\Class{AC}^0$ for $k\Lang{sat-pr}_{\ge p}$.
\end{lemma}
\begin{proof}
  Once more, let $S$ be the number $S_{k,p}$. from the Kernel
  Theorem (Theorem~\ref{thm-kernel}).
  \begin{enumerate}
  \item
    For $\phi \in k\Lang{cnfs}$ with $\Prob{\phi} \ge p$, consider
    $X = \operatorname{vars}(\phi^*_{\mathrm{pluck}})$. By the Kernel
    Theorem we have
    $|\phi^*_{\mathrm{pluck}}| \le S_{k,p}$ and, thus, $|X| \le S$. In
    particular, the binary encoding of $X$ has length $O(\log n)$,
    where $n$ is the number total of variables, as $S$ is a
    constant. Also by the Kernel Theorem,
    $\Prob{\phi^*_{\mathrm{pluck}}} \ge p$. Consider any $\beta \colon
    X \to \mathbb B$ with $\beta \models \phi^*_{\mathrm{pluck}}$. Then
    for each clause $c \in \phi^*_{\mathrm{pluck}}$, the assignment
    $\beta$ makes at least one literal true. However, this means that
    $\phi|_\beta = \emptyset$ as, by construction, every clause
    $d \in \phi$ contains a clause $c \in \phi^*_{\mathrm{pluck}}$ as
    a subclause, that is, $c \subseteq d$. In particular, each clause
    of $\phi$ contains a literal made true by~$\beta$ and, thus,
    $\phi|_\beta = \emptyset$. We conclude that the fraction of
    $\beta\colon X\to\mathbb B$ with $\phi|_\beta = \emptyset$ is at
    least the fraction of $\beta\colon X \to \mathbb B$ that satisfy
    $\phi^*_{\mathrm{pluck}}$. Since the latter fraction is at
    least~$p$, we get $(\phi,X) \in \Lang{witness-}k\Lang{sat-pr}_{\ge p}$ and $X$ has the allowed size. 
  \item
    For $\phi \in k\Lang{cnfs}$ with $\Prob{\phi} < p$, suppose there
    is an $X$ such that we have $(\phi,X) \in \Lang{witness-}k\Lang{sat-pr}_{\ge p}$. Consider those $\beta \colon X
    \to\mathbb B$ with $\phi|_\beta =\emptyset$. For them, 
    every clause of~$\phi$ contains a literal set to true
    by~$\beta$. In particular, for such a 
    $\beta\colon X\to\mathbb B$ every extension $\beta'\colon
    \operatorname{vars}(\phi) \to \mathbb B$ to the remaining variables
    of~$\phi$ satisfies~$\phi$. Thus, $\Prob{\phi} 
    \ge \Pr_{\beta:X\to\mathbb B}[\phi|_\beta = \emptyset]$ and
    $(\phi,X) \notin \Lang{witness-}k\Lang{sat-pr}_{\ge p}$, a contradiction. 
  \end{enumerate}
  Membership in $\Class{AC}^0$ follows once more from $|X|$ being constant.
\end{proof}

\subsubsection*{Application: Solving the Majority-of-Majority Problem}

The Threshold Locality Theorem also lies at the heart of an algorithm for
solving the majority-of-majority problem efficiently for
$k\Lang{cnfs}$. Recall that for this problem we are given a formula
$\phi \in k\Lang{cnfs}$ and a partition $X_1 \mathbin{\dot\cup} X_2$
of $\operatorname{vars}(\phi)$ and the 
question is whether for at least half of all assignments $\beta \colon
X_1 \to \mathbb B$ (the first majority) we have $\phi|_\beta \in
k\Lang{sat-pr}_{\ge\frac{1}{2}}$ (the second majority). We will solve this
problem by reducing it to $l\Lang{sat-pr}_{\ge\frac{1}{2}}$ (for
which we already have efficient algorithms) and the reduction will be
based on the Threshold Locality Theorem. It turns out that this
reduction is not only applicable to the majority-of-majority problem, but
also to probability thresholds other than $\frac{1}{2}$ and also to
``iterated'' majority-of-majority-$\dots$-of-majority
problems. Interestingly, we seem to get a clearer and easier-to-follow
proof if we describe and tackle the reduction in the more general
setting, so let us start by defining these kinds of problems rigorously: 
\begin{definition}\label{def-multi-thr}
  For $j\ge2$ and numbers $p_1,\dots,p_j \in [0,1]$, let
  $k\Lang{sat-pr}_{\ge p_1,\ge 
    p_2,\dots,\ge p_j}$ denote the set
  of all tuples $(\phi,X_1,\dots,X_j)$ such that $\phi \in
  k\Lang{cnfs}$, the $X_i$ form a partition of
  $\operatorname{vars}(\phi)$, and 
  \begin{align*}
    \Pr_{\beta_1:X_1\to\mathbb B} \biggl[
      \Pr_{\beta_2:X_2\to\mathbb B}\biggl[ \cdots\biggl[
        \Pr_{\beta_j:X_j\to\mathbb B}\Bigl[
          \phi|_{\beta_1}|_{\beta_2}|\dots|_{\beta_j} =
          \emptyset\Bigr]\ge p_j\biggr] \cdots
        \biggr]\ge p_2
      \biggr]\ge p_1.
  \end{align*}
\end{definition}
Of course, a few remarks concerning this rather intimidating
definition are in order: First, note that
$\phi|_{\beta_1}|_{\beta_2}|\dots|_{\beta_j}$ means
$\bigl(\cdots\bigl((\phi|_{\beta_1})|_{\beta_2}\bigr)|\dots\bigr)|_{\beta_j}$,
but the order of the ``$|_{\beta_i}$'' is actually not important. Second, 
observe that for any set $X \supseteq
\operatorname{vars}(\phi)$ and any $\beta \colon X \to \mathbb B$ we
have $\beta \models\phi$ iff $\phi|_\beta = \emptyset$. In particular,
the inner equality could also be written equivalently (but less 
uniformly) as $\beta_j \models
\phi|_{\beta_1}|_{\beta_2}|\dots|_{\beta_{j-1}}$. Third, note that the
above definition also makes sense for $j=1$, where it states that the
language contains all $(\phi,X_1)$ with $\phi \in k\Lang{sat}$ and
$X_1 = \operatorname{vars}(\phi)$ such that $\Pr_{\beta_1:
  X\to\mathbb B}[\phi|_{\beta_1} = \emptyset] \ge p_1$; and
as we just saw, the probability is the same as that of $\Pr_{\beta_1:
  \operatorname{vars}(\phi)\to\mathbb B}[\beta_1 \models \phi]=
\Prob{\phi}$. In other words, for $j=1$ the above definition yields
the language $k\Lang{sat-pr}_{\ge p}$ used throughout this paper,
only with formulas $\phi$ replaced by
$(\phi,\operatorname{vars}(\phi))$. Since this does not change the
complexity in any way, it is just a matter of convenience which
definition is used. Fourth, the problem $\Lang{maj-maj-}k\Lang{sat}$
studied by Akmal and Williams~\cite{AkmalW2022} is exactly
$k\Lang{sat-pr}_{\ge \frac{1}{2},\ge \frac{1}{2}}$. 

As mentioned earlier, our objective is to reduce the problems from
Definition~\ref{def-multi-thr} with their many thresholds to other
problems from Definition~\ref{def-multi-thr}, but now with fewer
thresholds. For this reduction, the following lemma will be the key
tool: 

\begin{lemma}[Threshold Encoding Lemma]\label{lemma-red}
  For each $k$ and $p \in [0,1]$ there are an~$l$ and a
  function $\omega_{k,p}$ in $\Class{AC}^0$ that maps any $(\phi,X)$
  with $\phi \in k\Lang{cnfs}$ and $X \subseteq
  \operatorname{vars}(\phi)$ to some $\omega := \omega_{k,p}(\phi,X)$ with
  \begin{enumerate}
  \item $\omega \in l\Lang{cnfs}$,
  \item $\operatorname{vars}(\omega) \subseteq X$, and
  \item for all $\beta \colon X \to \mathbb B$ we have $\Prob{\Restrict{\phi}{\beta}} \ge p \text{ iff } \beta \models \omega$.\label{eq-red}
  \end{enumerate}
\end{lemma}

Before we prove the lemma, two remarks are in order: First, note that
the size of the set~$X$ is \emph{not} assumed to be constant and
the $\Class{AC}^0$-circuit must be able to deal with arbitrarily
large~$X$. Second, note that for $X_j := \operatorname{vars}(\phi)
\setminus X$, the equivalence in item~\ref{eq-red} can be rewritten
equivalently as  
\begin{align}
  \textstyle\Pr_{\beta_j: X_j \to \mathbb B}[\phi|_\beta|_{\beta_j} =
      \emptyset]\ge p \text{ iff } \omega|_\beta =
  \emptyset, \label{eq-red2}
\end{align}
which clearly has a lot of similarity with the inner part of
Definition~\ref{def-multi-thr} already.

\begin{proof}
  In the following, we first describe the construction of $\omega$
  based on~$\phi$. For the description, we will argue what $\phi|_\beta$
  may or may not contain for a given~$\beta$, but note that the
  construction of~$\omega$ will not depend on any concrete~$\beta$:
  Rather, the final formula must have the property that for any
  concrete $\beta \colon X \to \mathbb B$ the equivalence from item~\ref{eq-red} holds.

  \begin{figure}[htpb]
    \centering
    \begin{tabular}{rll|rl}
      \emph{Literals~$l$ with $\operatorname{var}(l) \notin X$} &
      \emph{Literals~$l$ with $\operatorname{var}(l) \in X$}\\[.5em]
      $\phi = \bigl\{\;\{r,s,$ & $x,y\},$
      && \quad$\omega_{\{r,s\}\notin}= \bigl\{$\kern-3mm\hbox{}&$\{x,y\},$ \\
      $\{r,s,$ & $\neg x,y\},$ &&& $\{\neg x,y\},$ \\
      $\{r,s,$ & $x,y,z\},$ &&& $\{x,y,z\}\,\bigr\}$.\\
      $\{\neg r,t,$ & $x,y\},$ &&&\\
      $\{\neg r,t,$ & $z\},$ &&&\\
      $\{r,\neg s,$ & $y,\neg z\},$&&& \\
      $\{t,$ & $y,\neg z\},$
      && $\omega_{\{t\}\notin}= \bigl\{$\kern-3mm\hbox{}&$\{y,\neg z\},$ \\
      $\{t,$ & $\neg y,z\}\, \bigr\}$  &&& $\{\neg y,z\}\,\bigr\}$.
    \end{tabular}
    \caption{A formula $\phi$, where for each clause the literals~$l$
      with $\operatorname{var}(l) \notin X = \{x,y,z\}$ are shown left
      and the literals with $\operatorname{var}(l) \in X$ are shown
      right. The formula $\phi_{-X}$ contains all 
      clauses ``remaining on the left,'' that is, $\phi_{-X} =
      \bigl\{\{r,s\},\{\neg r,t\}, \{r,\neg s\}, \{t\}\bigr\}$. We have
      $\phi|_\beta \subseteq \phi_{-X}$ for all $\beta\colon
      X\to\mathbb B$, but suppose we are interested in those~$\beta$ for
      which $\Prob{\phi|_\beta} \ge \frac{1}{2}$ holds. If there is
      some $\psi \subseteq \phi|_\beta$ with $\Prob{\psi} <
      \frac{1}{2}$, then we also have $\Prob{\phi|_\beta}<
      \frac{1}{2}$ and $\beta$ must ``rule out'' such~$\psi$, meaning
      that at least one clause of $\psi$ is not a clause
      of~$\phi|_\beta$. For instance, to rule out $\psi =
      \bigl\{\{r,s\},\{t\}\bigl\}$ (with $\Prob{\psi} = \frac{3}{8} <
      \frac{1}{2}$), it suffices that
      $\beta\models\omega_{\{r,s\}\notin}$ (which ensures $\{r,s\}
      \notin \Restrict{\phi}{\beta}$) or $\beta\models\omega_{\{t\}\notin}$ (which ensures $\{t\} \notin
      \Restrict{\phi}{\beta}$). This
      disjunction is achieved by requiring $\beta \models
      \omega_{\psi\not\subseteq} = \bigl\{\{x,y\}\cup\{y,\neg z\},\penalty0\{\neg x,
      y\}\cup\{y,\neg z\},\penalty0\{x,y,z\}\cup\{y,\neg z\},\penalty0\{x,y\}\cup\{\neg y, z\},\penalty0\{\neg x,
      y\}\cup\{\neg y, z\},\penalty0\{x,y,z\}\cup\{\neg y, z\}\bigr\} \cap
      \Lang{non-\penalty0tautological-\penalty0clauses} =  \bigl\{\{x,y,\neg z\},\{\neg x,
      y,\neg z\}\bigr\}$. 
    }
    \label{fig-maj-maj}
  \end{figure}
  
  We start by considering the formula $\phi_{-X} :=
  \bigcup_{\beta':X\to\mathbb B} \phi|_{\beta'}$ that results
  from~$\phi$ if we simply remove all literals~$l$ with
  $\operatorname{var}(l) \in X$ from all clauses. This formula is, in a 
  sense, the ``worst case'' of what $\Restrict{\phi}{\beta}$ could look like
  regarding the satisfaction probability: $\phi_{-X}$ is the formula where
  \emph{no} clause is already satisfied by the assignment~$\beta$,
  leaving a maximum number of clauses that need to be
  satisfied. Note that $\Restrict{\phi}{\beta} \subseteq \phi_{-X}$ and, thus, if
  $\Prob{\phi_{-X}} \ge p$ happens to hold, we have
  $\Prob{\Restrict{\phi}{\beta}} \ge p$ for all~$\beta$ and could set
  $\omega$ to an arbitrary tautology. The interesting question is,
  thus, what happens when $\Prob{\phi_{-X}} <p$: For which $\beta$
  will $\phi|_\beta$ miss enough clauses from~$\phi_{-X}$ to raise the
  satisfaction probability above~$p$? 

  To answer this question (and to turn it into a formula $\omega$), we
  use the Threshold Locality Theorem. By this theorem, there is a
  constant~$S$ such that we have $\Prob{\Restrict{\phi}{\beta}} \ge p$ iff
  for every $\psi \subseteq \Restrict{\phi}{\beta}$ with $|\psi| \le S$ we have
  $\Prob{\psi} \ge p$. In particular, for every small $\psi
  \subseteq \phi_{-X}$ with $\Prob{\psi} < p$, we must have $\psi
  \not\subseteq \Restrict{\phi}{\beta}$ to have a chance that
  $\Prob{\Restrict{\phi}{\beta}} 
  \ge p$ holds, that is, the rather small set~$\psi$ must still
  contain a clause that $\beta$ rules out. Formally,
  $\psi \not\subseteq \Restrict{\phi}{\beta}$ means there is a clause
  $c\in \psi$ such that for all clauses $d \in \phi$ from which $c$ resulted,
  at least one $X$-literal in~$d$ is set to true by~$\beta$ (because,
  then, $c$ is not added to~$\Restrict{\phi}{\beta}$). 

  To summarize, we have two conditions to check:
  \begin{enumerate}
  \item For every $\psi \subseteq \phi_{-X}$ with $|\psi| \le S$ and
    $\Prob{\psi} < p$ 
    we must have $\psi \not\subseteq \Restrict{\phi}{\beta}$, which holds iff
  \item there is a clause $c\in \psi$ with $c \notin
    \Restrict{\phi}{\beta}$, meaning that for all $d 
    \in \phi$ from which $c$ resulted, at least one $X$-literal in~$d$ is
    set to true by~$\beta$.
  \end{enumerate}
  It turns out that we can express these conditions using a single formula
  $\omega \in l\Lang{cnfs}$ for a sufficiently large~$l$. Let us start
  with the second condition for a fixed $\psi \subseteq\phi_{-X}$ and
  let us try to find a single 
  formula $\omega_{\psi\not\subseteq} \in l\Lang{cnfs}$ expressing
  it. The condition is clearly a \emph{disjunction} (``there is a
  clause'') over all clauses $c \in   \psi$ of the following $k\Lang{cnf}$
  formulas $\omega_{c\notin}$ (see Figure~\ref{fig-maj-maj} for an example):
  \begin{align*}
    \omega_{c \notin} := \bigl\{ d \setminus c \bigm| d \in \phi,
    c\subseteq d, \operatorname{vars}(d) \setminus X =
    \operatorname{vars}(c) \bigr\}
  \end{align*}
  and observe that for every $\beta \colon X \to \mathbb B$ we have
  $\beta \models \omega_{c\notin}$ iff $c \notin \phi|_\beta$. 
  
  We can turn the disjunction of the at most~$S$ many
  $\omega_{c \notin} \in k\Lang{cnfs}$ for $c \in \psi$ into a single conjunction
  $\omega_{\psi\not\subseteq} \in l\Lang{cnfs}$ using the distributive
  law if we set $l := k \cdot S$: For $\psi =
  \{c_1,\dots,c_{|\psi|}\}$ define
  \begin{align*}
    \omega_{\psi\not\subseteq} := \bigl\{ e_1 \cup \cdots \cup
    e_{|\psi|} \bigm| e_1 \in \omega_{c_1\notin}, \dots, e_{|\psi|} \in
    \omega_{c_{|\psi|}\notin} \bigr\} \cap \Lang{non-tautological-clauses}
  \end{align*}
  where $\Lang{non-tautological-clauses}$ is the set of all clauses
  that do not contain both a variable and its negation (which we chose
  to forbid syntactically in $\Lang{cnf}$ formulas at the beginning of
  this paper). Observe that for
  every $\beta\colon X\to\mathbb B$ we now have
  \begin{align}
      \beta \models
      \omega_{\psi\not\subseteq}
      \iff
    \psi \not\subseteq \Restrict{\phi}{\beta}.\label{eq-not-sub}
  \end{align}

  It is now easy to express the first condition: Set
  \begin{align}
    \omega := \omega_{k,p}(\phi,X) := \textstyle\bigcup_{\psi \subseteq \phi_{-X},
      |\psi| \le S, \Prob{\psi} < p} \omega_{\psi\not\subseteq}\label{eq-omega}
  \end{align}
  and note that $\omega \in l\Lang{cnfs}$ holds. Note furthermore
  that an $\Class{AC}^0$-circuit can compute the function
  $\omega_{k,p}$ as $S$ is a constant and, hence, we can consider all
  size-$S$ subsets $\psi\subseteq \phi_{-X}$ in parallel and can
  hardwire the results of the tests $\Prob{\psi} <p$.

  For the correctness of the construction it remains to show that
  the third item from the lemma holds, which claimed that \emph{for
  all $\beta \colon X \to \mathbb B$ we have
  $\Prob{\Restrict{\phi}{\beta}} \ge p \text{ iff } \beta \models
  \omega$.} We show two
  directions. First, if $\beta \models \omega$, then, by~\eqref{eq-omega}, 
  \begin{enumerate}
  \item for all $\psi \subseteq \phi_{-X}$ \label{item-phi-minus}
  \item of size $|\psi| \le S$ \label{item-size}
  \item for which $\Prob{\psi} < p$ holds,  \label{item-pr}
  \end{enumerate}
  we have $\beta \models \omega_{\psi\not\subseteq}$ and
  by~\eqref{eq-not-sub} also $\psi
  \not\subseteq \Restrict{\phi}{\beta}$. Consider any $\psi \subseteq
  \Restrict{\phi}{\beta}$ of size $|\psi| \le S$. For such a~$\psi$,
  item~\ref{item-phi-minus} holds as we always have 
  $\Restrict{\phi}{\beta}\subseteq \phi_{-X}$ and item~\ref{item-size}
  holds by assumption, so item~\ref{item-pr} must be violated, meaning
  $\Prob{\psi} \ge p$. By the Threshold  Locality Theorem, we then
  have $\Prob{\Restrict{\phi}{\beta}} \ge p$. Second, 
  suppose $\Prob{\Restrict{\phi}{\beta}} \ge p$. Then $\Prob{\psi}
  \ge p$ trivially holds for every subset $\psi \subseteq
  \Restrict{\phi}{\beta}$. Thus, for every
  $\omega_{\psi\not\subseteq}$ considered in~$\omega$, 
  we have $\psi \not\subseteq \Restrict{\phi}{\beta}$. But then, again
  by~\eqref{eq-not-sub}, $\beta \models
  \omega_{\psi\not\subseteq}$. Thus, $\beta \models \omega$.
\end{proof}

Let us now use the lemma to reduce satisfaction probability problems
with $j$ thresholds to just $j-1$ thresholds:

\begin{theorem}
  Let $p_1,\dots,p_j \in [0,1]$ and let $k$ be a number. Then
  there is a number $l$ such that $k\Lang{sat-pr}_{\ge p_1,\dots,\ge
    p_j}$ reduces to $l\Lang{sat-pr}_{\ge p_1,\dots,\ge p_{j-1}}$. 
\end{theorem}

\begin{proof}
  Applying Lemma~\ref{lemma-red} to $k$ and $p_j$ we get a
  function~$\omega_{k,p_j}$. Let functions $\beta_1\colon
  X_1\to\mathbb B$ to $\beta_{j-1}\colon X_{j-1} \to \mathbb B$ be given. Setting $X = X_1
  \cup \dots \cup X_{j-1}$ and letting $\beta\colon X \to \mathbb B$ be
  the ``union assignment'' of $\beta_1$ to $\beta_{j-1}$ (so $\beta(x) = \beta_i(x)$
  for $x \in X_i$), equivalence~\eqref{eq-red2} tells us that we have
  $\Pr_{\beta_j: X_j \to \mathbb B}[\phi|_\beta|_{\beta_j} =
    \emptyset]\ge p_j$ iff $\omega_{k,p_j}(\phi,X)|_\beta = \emptyset$. Spelled
  out, this means:
  \begin{align*}
    \Pr_{\beta_j: X_j \to
      \mathbb B}\bigl[\phi|_{\beta_1}|\dots|_{\beta_{j-1}}|_{\beta_j} = 
      \emptyset\bigr]\ge p_j \text{ iff } \omega_{k,p_j}(\phi,X_1\cup \dots
    \cup X_{j-1})|_{\beta_1}|\dots|_{\beta_{j-1}} =
    \emptyset.
  \end{align*}
  Plugging this into inner part of the main formula from
  Definition~\ref{def-multi-thr}, we immediately get that
  $(\phi,X_1,\dots,X_j) \mapsto \bigl(\omega_{k,p_j}(\phi,X_1 \cup
  \cdots \cup X_{j-1}),X_1,\dots,X_{j-1}\bigr)$ is the desired
  reduction.  
\end{proof}

\begin{corollary}
  $k\Lang{sat-pr}_{\ge p_1,\dots,\ge
    p_j} \in \Class{AC}^0$ for all $p_1,\dots,p_j \in [0,1]$ and~$k$.
\end{corollary}

\begin{corollary}\label{cor-maj-maj}
  $\Lang{maj-maj-}k\Lang{sat} \in \Class{AC}^0$ for all~$k$.
\end{corollary}

An interesting remark concerning to theorems  of Akmal and Williams
\cite{AkmalW2021} is in order here: They study
$k\Lang{sat-pr}_{>0,\ge p}$ (under the name $\Lang{e-maj-}k\Lang{sat}$
for $p=1/2$) and
show $2\Lang{sat-pr}_{>0,\ge p} \in \Class P$ in their Theorem~6.1,  while
$3\Lang{sat-pr}_{>0,\ge 1/2}$ is $\Class{NP}$-complete by their
Theorem~6.2. For the latter result, the crucial containment in
$\Class{NP}$ easily follows from the Threshold Encoding Lemma (just
``chip away'' that final majority-of to reduce the problem to 
$l\Lang{sat-pr}_{>0}$ for some~$l$). In contrast,
$2\Lang{sat-pr}_{>0,\ge p} \in \Class P$ does \emph{not} seem to
follow from the results presented in this section and it seems that
the dedicated algorithm presented in \cite{AkmalW2021} for the problem
is really needed to decide it efficiently.

\subsection{The Gap Size Oblivious Algorithm}
\label{section-algo-intervals}
\label{sec-oblivious}

The algorithms presented up to now all need to have (at least a lower
bound on) the size of the spectral gap hardwired into their code. Of
course, we \emph{do} have such bounds by the results of
Section~\ref{sec-bounds}, so instead of hardwiring 
the real spectral gap, we could compute a lower bound at the start of
the algorithms and use that. Unfortunately, even if the bound were
very good (and it is not clear whether we have already achieved this),
spectral gaps may be very small, meaning that for instance the Kernel
Algorithm will essentially ``do nothing'' 
unless there are really \emph{huge} sunflowers. To get a feeling for
the scope of the problem, consider $k=2$ and $p= \frac{1}{2} +
\frac{1}{2^{1000}}$. Then $\operatorname{spectral-gap}_{2\Lang{cnfs}}(p) =
\frac{1}{2^{1001}}$ and the kernel algorithm \emph{will do nothing unless
there are at least $h_{2,p} = 1+\log_{3/4}\frac{1}{2^{1001}}
> 2{,}412$ clauses.} Even if it does something, the \emph{returned
kernel may contain up to $S_{2,p} = (2h_{2,p})^2\cdot 2! >
46{,}541{,}952$ clauses,} meaning that this is the size of formulas up to
which we have to compute satisfaction probabilities by
brute force (using recent improved
bounds~\cite{10.4007/annals.2021.194.3.5,10.19086/da.11887} for the
sizes of graphs that must contain sunflowers, one could lower these numbers
quite a bit, but not in a fundamental way). The same troubling effect
also renders the Akmal--Williams 
algorithm~\cite{AkmalW2022} (at least for $k>2$), the randomized
folklore approximation algorithm (see the introduction), and also
Trevisan's derandomization thereof~\cite{Trevisan2004} useless from
any practical perspective.

All of this seems particularly unfortunate as, in reality, it is quite 
easy to decide whether $\Prob{\phi} \ge p = \frac{1}{2} +
\frac{1}{2^{1000}}$ holds for $\phi \in 2\Lang{cnfs}$: First check
whether $\left|\operatorname{pack}(\phi)\right| \ge 3$ holds (recall that
$\operatorname{pack}(\phi) \subseteq \phi$ is a maximal packing) and,
if so, output ``$\Prob{\phi} \le (\frac{3}{4})^3 < 1/2 <
p$.'' Otherwise, check whether $\sum_{\beta:
  \operatorname{vars}(\operatorname{pack}(\phi))\to\mathbb B} 
\Prob{\RestrictAdd{\phi}{\beta}} \ge p$ holds and note that the sum
equals $\Prob{\phi}$, that it consists of \emph{at most 16 summands of the
form $\Prob{\psi}$ for some $\psi \in \Lang{1cnfs}$} (and at most nine
of them are not trivially~$0$) and that it is thus quite trivial to compute. 

Astute readers may have noticed that the just-sketched algorithm for
$2\Lang{sat-pr}_{\ge p}$ is somewhat reminiscent 
of a concept introduced in Section~\ref{sec-bounds}, namely 
\emph{expansion sequences.} It turns out that we can turn expansions
sequences, which were introduced as a technical tool to bound the
sizes of spectral gaps, into a tool for algorithmic
purposes. A bit ironically, this will allow us to construct a gap size
\emph{oblivious} algorithm even though we introduced expansions sequences to
bound the sizes of spectral gaps in the first place. (Readers who have
followed the advice given earlier to skip Section~\ref{sec-bounds}
during the first reading might wish to consult the introduction of that
section now, though the following presentation will be self-contained.) 

Definition~\ref{def-exp} states that the expansion sequence of a
formula $\phi \in k\Lang{cnfs}$ is a sequence of sets $\Phi_i$ of
formulas, starting with $\Phi_1 := \{\phi\}$, such that $\ProbOr{\Phi_i}
:= \sum_{\psi \in \Phi_i} \Prob{\psi}$ stays constant. The next
element in the sequence is determined by picking $\psi \in \Phi_i
\setminus 1\Lang{cnfs}$ such that $\operatorname{pack}(\psi)$ has
maximum satisfaction probability and setting $\Phi_{i+1} =
(\Phi_{i} \setminus \{\psi\}) \cup \penalty-100 \bigl\{\RestrictAdd{\psi}{\beta}
\bigm| \penalty100 \beta\colon
\operatorname{vars}(\operatorname{pack}(\psi))\to\mathbb B\bigr\}$. As shown in
Theorem~\ref{thm-spec-quant}, the sequence will always end ``at or
above~$p$'' or ``well below~$p$'' and this will happen after a number
of steps that depends only on $k$ and~$p$.
The exact definitions of ``at or above~$p$'' and especially
``well below~$p$'' are somewhat complex (inequality~\eqref{eq-tc-a}
is not idly called the \emph{technical} condition), but this is partly
due to the fact that in 
Section~\ref{sec-bounds} our objective was to establish bounds on the
sizes of spectral gaps. When one is only interested in deciding
whether $\Prob{\phi} < p$ holds, it turns out that \eqref{eq-tc-a}
can be replaced by a simpler check, namely the one in
line~\ref{line-less} of Algorithm~\ref{algo-expand-largest}.

\begin{lstlisting}[
    style=pseudocode,
    backgroundcolor=,
    float=htpb,
    label={algo-expand-largest},
    backgroundcolor=,
    caption={
      A gap size oblivious algorithm for deciding whether $\Prob{\phi}
      \ge p$ holds based on computing expansion sequences. In the
      algorithm, %$\Lang{packings} \subseteq \Lang{cnfs}$ is the set of
      %all packings and
      $\operatorname{maxpack}(\Psi \setminus \Lang{1cnfs})$
      is the formula $\psi \in \Psi \setminus \Lang{1cnfs}$ for which
      $\Prob{\operatorname{pack}(\psi)}$ is maximum (break ties
      arbitrarily). 
    }
  ]
algorithm $\Algo{gap-size-oblivious}(\phi,k,p)$ // $\commented{\phi \in k\Lang{cnfs}}$ must hold
    $\Psi \gets \{\phi\}$
    loop forever
        if $\sum_{\psi \in \Psi \cap \Lang{1cnfs}} \Prob{\psi} \ge p$ then output ``$\Prob{\phi} \ge p$'' and stop$\label{line-more}$
        if $\sum_{\psi \in \Psi} \Prob{\operatorname{pack}(\psi)}  < p$ then output ``$\Prob{\phi} < p$'' and stop$\label{line-less}$
        $\psi \gets \operatorname{maxpack}(\Psi \setminus \Lang{1cnfs})$
        $\Psi \gets (\Psi \setminus \{\psi\}) \cup \bigl\{\RestrictAdd{\psi}{\beta} \bigm| \beta \colon \operatorname{vars}(\operatorname{pack}(\psi)) \to \mathbb B\bigr\}$
\end{lstlisting}

\begin{theorem}
  For each~$k$ and $p$ there is a number~$r_{k,p}$ such that
  for every $\phi \in k\Lang{cnfs}$
  Algorithm~\ref{algo-expand-largest} stops after at most 
  $r_{k,p}$ iterations and produces a correct output.
\end{theorem}
\begin{proof}
  Let $\Psi_i$ denote the value of~$\Psi$ at the beginning of the
  $i$th iteration. We claim that the sequence
  $(\Psi_1,\Psi_2,\dots)$ is a prefix of the expansion sequence
  $\operatorname{exp-seq}_{k,p}(\{\phi\})$: First note
  that the update rule for~$\Psi$ from the last line of the 
  algorithm exactly matches the update rule from
  Definition~\ref{def-exp}. Second, we claim that before the 
  expansion sequence ends, one of the two termination conditions
  from the algorithm (lines \ref{line-more} 
  and~\ref{line-less}) is met.
  There are three reasons why an expansion sequence can end (see
  Definition~\ref{def-exp}): First,  $\sum_{\psi \in \Psi \cap
    \Lang{1cnfs}} \Prob{\psi} \ge p$ may hold -- but this is
  \emph{exactly} what we check in line~\ref{line-more}. Second, 
  $\sum_{\psi \in \Psi_i} \Prob{\operatorname{pack}(\psi)}  \le
  p_{\le|\Psi_i|+1}$ may hold, but then $\sum_{\psi \in \Psi_i}
  \Prob{\operatorname{pack}(\psi)} < p$, which is what we check in
  line~\ref{line-less}. Finally, the expansion sequence may end
  ``exhausted,'' which means that $\Psi \subseteq \Lang{1cnfs}$ holds
  and, then, in both termination conditions we actually compare
  $\Pr[\phi]$ to~$p$. In particular, exactly one condition will be
  met. In conclusion, the number of iterations done 
  by the algorithm is at most the length of the sequence (since, then,
  one of the two conditions in lines \ref{line-more}
  or~\ref{line-less} will be met) -- and this length is at most
  $s_{k,p}(1)$ for the function $s_{k,p}$ from
  Lemma~\ref{lemma-fourth} (actually, the lemma bounds the size 
  of the last~$\Psi_r$, but this also bounds the sequence's~length).

  Concerning the correctness of the output, note that $\Prob{\phi} =
  \ProbOr{\Psi_i} \ge \ProbOr{(\Psi_i \cap \Lang{1cnfs})}$ and also
  $\Prob{\phi} = \ProbOr{\Psi_i} = \sum_{\psi \in \Psi_i}
  \Prob{\psi} \le \sum_{\psi \in \Psi_i}
  \Prob{\operatorname{pack}(\psi)}$. 
\end{proof}

Actual implementations may wish to incorporate some
optimizations: Instead of  checking whether we have $\sum_{\psi \in
  \Psi \cap \Lang{1cnfs}} \Prob{\psi} 
\ge p$, we could also check whether $\sum_{\psi \in \Psi \cap
  \Lang{easy}} \Prob{\psi} \ge p$ holds, where $\Lang{easy}$ is any
superset of $\Lang{1cnfs}$ for which it is ``easy'' to compute
$\Pr[\psi]$ for $\psi \in \Lang{easy}$. For instance, for $\Lang{easy}
= \{\psi \in \Lang{cnfs} \mid \psi = \operatorname{pack}(\psi)\}$, the
set of formulas that are packings,
we can easily compute the satisfaction probability. Indeed, by
equation~\eqref{eq-pr-sunflower}, 
we can even compute $\Pr[\psi]$ easily whenever $\psi$ is a
\emph{sunflower}. By replacing the first check in the algorithm by
$\sum_{\psi \in \Psi \cap \Lang{sunflowers}}
\Prob{\psi} \ge p$, we may stop the loop earlier -- but the output
will, of course, still be correct. A bit less obviously, we can then
also modify which formula we pick for expansion: Instead of the
assignment $\psi  \gets \operatorname{maxpack}(\Psi \setminus
\Lang{1cnfs})$, we can also use $\psi \gets
\operatorname{maxpack}\bigl(\Psi \setminus \Lang{sunflowers})$,
that is, we never expand those formulas for which we can compute the
exact satisfaction probability easily.

\section{Structural Complexity Results}
\label{section-complexity}

With the groundwork laid in the previous sections, we can proceed to 
the main structural complexity result of this paper,
Theorem~\ref{thm-main}. Here is the claim once more:
\begin{reclaim*}[of Theorem~\ref{thm-main}] 
  For each $k$ and $p$:
  \begin{enumerate}
  \item If $k\Lang{cnfs}$ have room for $3\Lang{sat}$ at~$p$, then
    $k\Lang{sat-pr}_{> p}$ is $\Class{NP}$-complete.
  \item If $k\Lang{cnfs}$ have room for $2\Lang{sat}$ at~$p$, but not for
    $3\Lang{sat}$, then $k\Lang{sat-pr}_{> 
    p}$  is $\Class{NL}$-complete. 
  \item In all other cases, $k\Lang{sat-pr}_{> p}$ lies in
    $\Class{AC}^0$.
  \end{enumerate}
\end{reclaim*}

Similar to the previous sections, for the proof we will develop some
technical tools and it will be helpful to start with the overall
picture -- the details will be presented later. A closer look at the
to-be-proved claim shows that we have to prove some \emph{hardness
results} and some \emph{membership results} for the classes $\Class{AC}^0$,
$\Class{NL}$, and~$\Class{NP}$. It is, of course, no coincidence that
the problems $\Lang{1sat}$, $\Lang{2sat}$, and $\Lang{3sat}$ happen to
be complete for these classes. We introduce two
technical conditions on numbers~$t$ (which can be
arbitrary, but it is best to think of~$t$ as $1$, $2$,~or~$3$) and
then prove three lemmas that, taken together, imply the above
claim. As always, $k$ and $p$ are considered to be fixed numbers in
the following. 

The first condition just gives a shorter name to the notion of
``having room for $\dots$ at $\dots$'' from Definition~\ref{def-open}:

\begin{definition}[Hardness Condition]\label{tc-b}
  A number $t \in \mathbb N$ \emph{meets the
   hardness condition} if $k\Lang{cnfs}$ have room for
  $t\Lang{sat}$ at~$p$.
\end{definition}

The second condition is (much) more complex and we will devote most of
Section~\ref{sec-upper} to explaining the rationale behind the
definition. Fortunately, these details are not yet important as the
claims of the three lemmas do not refer to them at all.

\begin{definition}[Membership Condition]\label{tc-c}
  A number $t \in \mathbb N$ \emph{meets the
   membership condition} if for
  all $\phi \in k\Lang{cnfs}$ and $X \subseteq
  \operatorname{vars}(\phi)$ with $\Prob{\phi} > p$ and
  $\Pr_{\beta:X\to\mathbb B}[\phi|_\beta=\emptyset] = p$ there is a
  $\beta\colon X \to \mathbb B$ such that $\phi|_\beta \in t\Lang{sat}
  \setminus 0\Lang{cnfs}$.
\end{definition}

\begin{lemma}\label{lem-lower}
  If $t$ meets the hardness condition, then $t\Lang{sat}
  \le^{\mathrm{fo}}_{\mathrm m} k\Lang{sat-pr}_{>p}$.
\end{lemma}
\begin{lemma}\label{lem-upper}
  If $t$ meets the membership condition, then $k\Lang{sat-pr}_{>p}
  \le^{\mathrm{fo}}_{\mathrm{dtt}} t\Lang{sat}$.
\end{lemma}
\begin{lemma}\label{lem-link}
  $t+1$ meets the hardness condition or $t$ meets the membership
  condition.
\end{lemma}

The three subsections of the present section each explain and then
prove one of the three lemmas. In particular, readers will find
details on the used reductions (``$\le^{\mathrm{fo}}_{\mathrm m}$'' and
``$\le^{\mathrm{fo}}_{\mathrm{dtt}}$'') there, while right now it is only of
importance that the classes $\Class{AC}^0$, $\Class{NL}$, and
$\Class{NP}$ are all closed under these reductions. Once we have these
three lemmas, we get Theorem~\ref{thm-main} via the following proof:
\begin{proof}[Proof of Theorem~\ref{thm-main}]
  Let $k$ and $p$ be fixed. Let us look at the three items of the claim.

  For the first item, the assumption is
  that $k\Lang{cnfs}$ have room for $\Lang{3sat}$ at~$p$. In other
  words, $t=3$ meets the hardness condition. By Lemma~\ref{lem-lower},
  $\Lang{3sat}$ then reduces to $k\Lang{sat-pr}_{>p}$, which is, 
  thus, $\Class{NP}$-hard. On the other hand, $k\Lang{cnfs}$ never
  have room for $(k+1)\Lang{sat}$ at any~$p$ (clauses~$c_*$ of size
  $k-(k+1) = -1$ do not exist). Thus, $k+1$ does not meet the
  hardness condition and so by Lemma~\ref{lem-link}, the number $t=k$ meets the
  membership condition. By Lemma~\ref{lem-upper} we conclude that
  $k\Lang{sat-pr}_{>p}$ reduces to $t\Lang{sat} \in \Class{NP}$ and,
  thus, it lies in $\Class{NP}$. All told, when $k\Lang{cnfs}$ have
  room for $3\Lang{sat}$ at~$p$, then $k\Lang{sat-pr}_{>p}$ is
  $\Class{NP}$-complete. 

  The theorem's second item follows by nearly the same argument:
  Assuming that $k\Lang{cnfs}$ have room for $\Lang{2cnfs}$ at~$p$,
  but not for   $\Lang{3cnfs}$, 
  Lemma~\ref{lem-lower} tells us that $\Lang{2sat}$ reduces to
  $k\Lang{sat-pr}_{>p}$, while Lemma~\ref{lem-link} tells us that
  when $3$ does not meet the hardness condition, then $2$ meets the
  membership condition and, thus,   $k\Lang{sat-pr}_{>p}$ reduces to
  $\Lang{2sat}$ by Lemma~\ref{lem-upper}. All told, 
  $k\Lang{sat-pr}_{>p}$ is $\Class{NL}$-complete.

  For the theorem's third item, just observe that ``in all other
  cases'' means, in particular, that $k\Lang{cnfs}$ have no room for
  $\Lang{2sat}$ at~$p$. Thus, $2=:t+1$ does not meet the hardness condition and
  by Lemma~\ref{lem-link} once more, $t=1$ must meet the membership
  condition. Lemma~\ref{lem-upper} tells us that
  $k\Lang{sat-pr}_{>p}$ reduces to $\Lang{1sat}$ and, thus, lies in
  $\Class{AC}^0$.
\end{proof}

\subsection{The Hardness Condition Implies Hardness}

\label{sec-lower}

Our objective in the present subsection is to prove
Lemma~\ref{lem-lower}, which stated:

\begin{reclaim*}[of Lemma~\ref{lem-lower}]
  If $t$ meets the hardness condition, then $t\Lang{sat}
  \le^{\mathrm{fo}}_{\mathrm m} k\Lang{sat-pr}_{>p}$.
\end{reclaim*}

The claim refers, firstly, to the \emph{hardness condition,} by which
$k\Lang{cnfs}$ must have room for $t\Lang{sat}$ at~$p$. Recall from
Definition~\ref{def-open} that this means that there is an irredundant
$\omega \in k\Lang{cnfs}$ with $p = \Prob{\omega}$ containing a clause
$c_*$ of size at most $k-t$. An example for $t=2$, $k=3$, and $p = 7/32$, was
the irredundant $\omega = \bigl\{\{a\},\{b\},\penalty0\{c_1,c_2,c_3\}\bigr\}
\in \Lang{3cnfs}$ that contains a clause $c_* = \{a\}$ of size $3-2=1$
and has $\Pr[\omega] = 7/32$. As was already sketched in the
introduction, $\omega$ will serve as a ``target for a
reduction from $t\Lang{sat}$ as we can add clauses of size at most~$2$ 
to~$c_*$ and this will increase the satisfaction probability when the
clauses are satisfiable'' (this will be made precise in a moment,
namely in the proof below). 

The claim refers, secondly, to the reduction type
``$\le^{\mathrm{fo}}_{\mathrm{m}}$.''\label{def-many-one} This stands for      
\emph{first-order many-to-one reducible,} which was already mentioned
in the introduction.  Here, \emph{first-order}
means that the reduction is first-order definable in the sense of
descriptive complexity (see \cite{Immerman1998} for an introduction)
and this is known to be the same as being computable by  
$\Class{AC}^0$-circuits (see \cite{Immerman1998} once
more). \emph{Many-to-one} means that each instance for the
to-be-reduced problem is mapped to an instance for the to-be-reduced-to
problem (and \emph{many}-to-one just emphasizes that many source
instances can be mapped to the same target instance). All told, the
reduction is rather weak (which is a desirable property for hardness
proofs), but still robust and it is one of the standard
reductions commonly used for hardness proofs. 

\begin{proof}[Proof of Lemma~\ref{lem-lower}]
  Let $t$ meet the hardness condition. Then there is an irredundant $\omega 
  \in k\Lang{cnfs}$ with $\Pr[\omega] = p$ and a clause $c_* \in
  \omega$ with $|c_*| \le k-t$. Let $\beta_*\colon
  \operatorname{vars}(\omega) \to \mathbb B$ witness $c_*$'s
  irredundancy in~$\omega$, that is,  $\beta_* \models
  \omega\setminus\{c_*\}$ but $\beta_* \not\models \{c_*\}$. To
  many-one-reduce $t\Lang{sat}$ to
  $k\Lang{sat-pr}_{>p}$, let $\psi \in t\Lang{cnfs}$ be a given
  input. If necessary, rename the variables in~$\psi$ to ensure 
  $\operatorname{vars}(\psi) \cap \operatorname{vars}(\omega) =
  \emptyset$. Map $\psi$ to $\rho :=  (\omega \setminus \{c_*\})
  \cup \{ c_* \cup d  \mid d \in \psi \}$. Observe that $\omega
  \FormImpl \rho$ (every clause of $\rho$ is a superclause of some
  clause of~$\omega$) and thus $\Pr[\rho] \ge \Pr[\omega] = p$. Thus,
  the to-be-proved equivalence ``$\Prob{\rho} > p$ iff $\psi \in
  t\Lang{sat}$'' is the same as ``$\rho$ has an \emph{additional}
  satisfying assignment compared to~$\omega$ iff $\psi$ is
  satisfiable.'' We prove this latter equivalence.  

  For the first direction, let there be an assignment~$\gamma$ with
  $\gamma \models \rho$ and $\gamma 
  \not\models \omega$. Since all clauses of $\omega$ except for
  $c_*$ are also present in $\rho$, we conclude $\gamma
  \not\models \{c^*\}$. On the other hand, $\gamma \models \rho
  \supseteq \{c^* \cup d \mid d \in \psi\}$. Thus, each clause of the
  form $c^* \cup d$ for $d  
  \in \psi$ is satisfied by~$\gamma$, but $c^*$ is not, implying that
  $\gamma$ satisfies each $d \in \psi$. In other words, $\gamma$
  witnesses $\psi \in t\Lang{sat}$. 
  Second, assume that $\psi \in t\Lang{sat}$ is witnessed by some
  satisfying assignment $\alpha \colon
  \operatorname{vars}(\psi) \to \mathbb B$. Consider the 
  assignment $\gamma \colon \operatorname{vars}(\psi) \cup
  \operatorname{vars}(\omega) \to \mathbb B$ defined by $\gamma(v) =
  \alpha(v)$ for $v \in \operatorname{vars}(\psi)$ and $\gamma(v) =
  \beta_*(v)$ for $v \in \operatorname{vars}(\omega)$. Trivially, $\gamma
  \not\models \omega$ as $\beta_* \not\models \{c_*\} \subseteq
  \omega$. However, $\gamma \models \rho$: Each clause $c \in \omega
  \setminus \{c_*\}$ is satisfied by $\beta_*$ and each clause of the
  form $c_* \cup d$ with $d \in \psi$ is satisfied as $\alpha
  \models \psi$.  
\end{proof}

\subsection{The Membership Condition Implies Membership}

\label{sec-upper}

We move on to the second lemma, Lemma~\ref{lem-upper}, which stated:

\begin{reclaim*}[of Lemma~\ref{lem-upper}]
  If $t$ meets the membership condition, then $k\Lang{sat-pr}_{>p}
  \le^{\mathrm{fo}}_{\mathrm{dtt}} t\Lang{sat}$.
\end{reclaim*}

Just as in the previous section, we begin by explaining the two
central concepts of the claim (the \emph{membership condition} and the
reduction type ``$\le^{\mathrm{fo}}_{\mathrm{dtt}}$''). We start with
the reduction, which is probably less familiar to some readers. 

\subparagraph*{Disjunctive Truth-Table Reductions.} The abbreviations
in  ``$\le^{\mathrm{fo}}_{\mathrm{dtt}}$'' stand for 
\emph{first-order disjunctively truth-table reducible.} The first-order
part is the same as before (so the reduction is computable by
$\Class{AC}^0$-circuits). What is new is
that in a \emph{disjunctive truth-table reduction} an input instance 
is not mapped to a single output instance, but to a whole set of
output instances (in principle, the size of this set could be
polynomial, but we will only need sets of constant size). The crucial
condition is that the input instance must be a member of the
to-be-reduced problem iff \emph{at least one} of the
output instances is a member of the to-be-reduced-to
problem. Formally, $A \le^{\mathrm{fo}}_{\mathrm{dtt}} B$ means that
when $x$ is mapped to $\{y_1,\dots,y_z\}$, then we must have $x \in A$
iff $\bigvee_{i=1}^z (y_i \in B)$ (hence the name
``disjunctive''). Note that many-to-one reductions are special cases
of disjunctive reductions (the set of output instances is just a
singleton set), so disjunctive reductions ``just'' offer more
flexibility and may be easier to construct in certain cases (such as
in our proof of Lemma~\ref{lem-upper}). Most importantly, the
classes $\Class{AC}^0$, $\Class{NL}$, and $\Class{NP}$ are all closed
under these reductions and, thus, showing that $k\Lang{sat-pr}_{>p}$
reduces to $t\Lang{sat}$ via a disjunctive truth-table reduction
suffices to show that the problem lies in these classes. 

\subparagraph*{The Membership Condition.} Let us now turn our
attention to the (rather complex) Definition~\ref{tc-c} of the
\emph{membership condition,} which is met by a number~$t$ if:
\begin{quote}
  \emph{For all $\phi \in k\Lang{cnfs}$ and $X \subseteq
  \operatorname{vars}(\phi)$ with $\Prob{\phi} > p$ and
  $\Pr_{\beta:X\to\mathbb B}[\phi|_\beta=\emptyset] = p$ there is a
  $\beta\colon X \to \mathbb B$ such that $\phi|_\beta \in t\Lang{sat}
  \setminus 0\Lang{cnfs}$.}
\end{quote}
To get a handle on this, first observe that
``$\Pr_{\beta:X\to\mathbb B}[\phi|_\beta=\emptyset] = p$'' is almost the
same as the notion of a \emph{small witness} from
Lemma~\ref{lem-witness-ge}, the only difference is that in the lemma
we had the requirement ``$\ge p$'' instead of ``$=p$.'' Nevertheless,
$X$ is a witness is the sense of Lemma~\ref{lem-witness-ge} and its
very existence proves that $\Prob{\phi} \ge p$ holds (the probability
on the left hand side of the requirement is a lower bound for
$\Prob{\phi}$). Thus, the membership condition starts with the
\emph{assumption} that $\Prob{\phi} > p$ holds and with a \emph{given
witness} that $\Prob{\phi} \ge p$ holds (namely the witness~$X$, which
has the additional property that
$\Pr_{\beta:X\to\mathbb B}[\phi|_\beta=\emptyset]$ is even equal to $p$
and not just bounded from below by~$p$).

Following all these assumptions and witnesses, we then have the main
condition, namely the existence of a $\beta$ with a certain
property. The rough idea is that this $\beta$ will be an
``easy-to-find witness for $\Prob{\phi} > p$ rather than just
$\Prob{\phi} \ge 
p$.'' To better explain what is meant by this, we need to review a
classical concept from \textsc{fpt} theory: \emph{Backdoor sets}.

In general, \emph{backdoor sets} are powerful
tools~\cite{WilliamsGS2003} for deciding satisfiability problems,
with a rich theory around them. While our objective is not quite the 
same (we are interested in satisfaction probability thresholds rather
than ``just'' satisfiability), we can adapt the ideas underlying
backdoor sets for our purposes. Towards this aim, let us start by
rephrasing the basic properties of backdoor sets ``in terms of
satisfaction probabilities'': Given a formula $\phi \in k\Lang{cnfs}$,
consider any set $X \subseteq \operatorname{vars}(\phi)$ of
variables. As we have observed repeatedly in this paper, we have 
\begin{align}
  \Prob{\phi} = \sum_{\beta:X \to \mathbb B}
  \frac{\Prob{\Restrict{\phi}{\beta}}}{2^{|X|}}. \label{eq-simple-pr}
\end{align}
By this equation, in order to decide whether $\Prob{\phi} > 0$ holds
(that is, whether $\phi$ is satisfiable), it suffices to check whether
$\Prob{\Restrict{\phi}{\beta}} > 0$ holds for at least one of the
$2^{|X|}$ many $\beta\colon X \to \mathbb B$. The key observation is
that for \emph{a cleverly chosen~$X$,} the formulas
$\Restrict{\phi}{\beta}$ may be \emph{syntactically 
simple} and deciding $\Prob{\Restrict{\phi}{\beta}} >
0$ may be easy for them. For instance, $\Restrict{\phi}{\beta} \in
t\Lang{cnfs}$ might hold for all~$\beta$ for $t=1$ or $t=2$ (other
``simple'' set such as $\Lang{horn-cnfs}$ are also possible, but let
us focus on $\Lang{1cnfs}$ and $\Lang{2cnfs}$). A set~$X$ such that
all $\Restrict{\phi}{\beta}$ lie in~$t\Lang{cnfs}$ is
called a \emph{strong backdoor set into 
$t\Lang{cnfs}$}. Being able to compute, for a given formula, a
constant-size strong backdoor set into $t\Lang{cnfs}$ for $t\le 2$ 
means that we are able to decide satisfiability for the formula
efficiently.

As the name suggests, \emph{strong} backdoor sets are strong
tools. Unfortunately, this also means that a small given~$X$ is only
rarely a strong backdoor set: Typically, there will be some $\beta
\colon X \to \mathbb B$ such that $\Restrict{\phi}{\beta} \notin
\Lang{2cnfs}$. A closer look at equation~\eqref{eq-simple-pr}
reveals, however, that this is not always a problem: For the sum to be
positive, it suffices that \emph{for at least one syntactically simple
$\Restrict{\phi}{\beta}$ we have $\Prob{\Restrict{\phi}{\beta}} > 0$.}
In this case, it does not matter whether the syntactically
``difficult'' $\Restrict{\phi}{\beta}$ are satisfiable or not. To
formalize this idea, we rewrite~\eqref{eq-simple-pr}:

\begin{align}
  \Prob{\phi} = 
  \sum_{\substack{\beta:X\to\mathbb B,\\ \Restrict{\phi}{\beta}
      \in t\Lang{cnfs}}}
  \frac{\Prob{\Restrict{\phi}{\beta}}}{2^{|X|}} + \sum_{\substack{\beta:X\to\mathbb B,\\ \Restrict{\phi}{\beta}
      \notin t\Lang{cnfs}}}
  \frac{\Prob{\Restrict{\phi}{\beta}}}{2^{|X|}}. \label{eq-two-pr}
\end{align}
A \emph{weak backdoor set into $t\Lang{cnfs}$} for a formula~$\phi$
is a set~$X$ such that whenever $\Prob{\phi}$ is positive, so is the
\emph{first} sum of equation~\eqref{eq-two-pr} (the reverse is trivial: if any
of the sums are positive, so is $\Prob{\phi}$). Note that there is no
obvious way of checking whether a given set~$X$ is a weak backdoor set
(short of just deciding satisfiability for all
$\Restrict{\phi}{\beta}$): Some other argument must ensure that when
the first sum of equation~\eqref{eq-two-pr} is zero, so is the second one --
we cannot check this ourselves.

In order for backdoor sets to be useful in the context of the present
paper, we must lift them to the situation where the question
is not ``Does $\Prob{\phi} > 0$ hold?,'' but ``Does
$\Prob{\phi} > p$ hold?'' for some arbitrary $p \in [0,1]$. The key
observation is that we can rewrite equation~\eqref{eq-two-pr} using
\emph{three} sums by splitting up the first sum once more:
\begin{align}
  \Prob{\phi} = 
    \sum_{\substack{\beta:X\to\mathbb B,\\ \Restrict{\phi}{\beta}
      \in 0\Lang{cnfs}}}
    \frac{\Prob{\Restrict{\phi}{\beta}}}{2^{|X|}}
  +
  \sum_{\substack{\beta:X\to\mathbb B,\\ 
      \Restrict{\phi}{\beta} \notin 0\Lang{cnfs},\\
      \Restrict{\phi}{\beta} \in t\Lang{cnfs}\phantom{,}
  }}
  \frac{\Prob{\Restrict{\phi}{\beta}}}{2^{|X|}}
  +
  \sum_{\substack{\beta:X\to\mathbb B,\\ \Restrict{\phi}{\beta}
      \notin t\Lang{cnfs}}}
  \frac{\Prob{\Restrict{\phi}{\beta}}}{2^{|X|}}
  . \label{eq-three-pr}
\end{align}

A \emph{weak backdoor set into $t\Lang{cnfs}$ for threshold~$p$
for~$\phi$} is then a set $X$ such that the first sum equals~$p$,
while the second sum is positive iff $\Prob{\phi} > p$ holds. In other
words, the third sum (for which a positivity check is difficult) is
\emph{irrelevant} to distinguish between the cases $\Prob{\phi} = p$ and
$\Prob{\phi} > p$.

The membership condition is directly related to the sums of
equation~\eqref{eq-three-pr}. Recall that there are two requirements
in the membership condition:
\begin{enumerate}
\item
  $\Pr_{\beta:X\to\mathbb B}[\phi|_\beta=\emptyset] = p$.
\item
  There is a $\beta\colon X \to \mathbb B$ such that $\phi|_\beta \in
  t\Lang{sat} \setminus 0\Lang{cnfs}$.
\end{enumerate}
Observe that the probability in the first item is \emph{exactly
the first sum in equation~\eqref{eq-three-pr}:} In $\sum_{\beta:X\to\mathbb B,
  \Restrict{\phi}{\beta} \in 0\Lang{cnfs}}
\Prob{\Restrict{\phi}{\beta}}/2^{|X|}$ we always have
$\Prob{\Restrict{\phi}{\beta}} \in \{0,1\}$ and, thus, the sum is
exactly the fraction of $\beta \colon X \to \mathbb B$ for which
$\Restrict{\phi}{\beta} = \emptyset$ holds. Next, observe that the
second item holds \emph{iff the second sum is positive} as this means
that for some $\beta\colon X \to \mathbb B$ we have that $\phi|_\beta
\in t\Lang{cnfs}$, but also $\phi|_\beta \notin 0\Lang{cnfs}$ and
$\Prob{\phi|_\beta} > 0$. In other words, $\phi|_\beta \in
t\Lang{sat}$ must hold, but $\phi|_\beta$ should not lie in
$0\Lang{cnfs}$ (since the only satisfiable formula in $0\Lang{cnfs} =
\{ \emptyset, \{\emptyset\}\}$ is the trivial tautology~$\emptyset$,
this just means that $\phi|_\beta \neq \emptyset$ holds).

\subparagraph*{Putting it All Together.}
The just-made observations on the membership condition allow us to
``decipher'' it as follows: It requires that for $\Prob{\phi} >
p$, witnesses $X$ with $\Pr_{\beta:X\to\mathbb B}[\phi|_\beta=\emptyset]
= p$ must already be weak backdoor sets into $t\Lang{cnfs}$ for
threshold~$p$. Since computing witnesses is easy (by
Lemmas~\ref{lem-comp-witness} and \ref{lem-witness-ge}), and since the
test ``is there some $\beta\colon X \to \mathbb B$ with $\phi|_\beta \in
t\Lang{sat} \setminus 0\Lang{cnfs}$'' can clearly be answered using a
disjunctive truth-table reduction to $t\Lang{sat}$, we get
Lemma~\ref{lem-upper}, see the following proof for details: 

\begin{proof}[Proof of Lemma~\ref{lem-upper}]
  Let $k$ and $p$ be given and let $t$ meet the membership condition
  from Definition~\ref{tc-c}. We wish to reduce
  $k\Lang{sat-pr}_{>p}$ to $t\Lang{sat}$ via a disjunctive
  truth-table reduction. In other words, on input $\phi \in
  k\Lang{cnfs}$ we must come up with a set of $t\Lang{cnf}$ formulas
  such that $\Prob{\phi} > p$ holds iff at least one of the formulas
  is satisfiable. 

  On input $\phi \in k\Lang{cnfs}$, the reduction first computes a
  witness set~$X$ for the small witness relation
  $\Lang{witness-}k\Lang{sat-pr}_{\ge p}$ from
  Lemma~\ref{lem-witness-ge} using the $\Class{AC}^0$-circuit from
  Lemma~\ref{lem-comp-witness}. This  set $X$ will have constant size
  and the following properties: 
  \begin{enumerate}
  \item If $\Pr_{\beta:X \to \mathbb B}[\phi|_\beta = \emptyset] < p$,
    then $\Prob{\phi} < p$ will hold.  
  \item If $\Pr_{\beta:X \to  \mathbb B}[\phi|_\beta = \emptyset] > p$,
    then $\Prob{\phi} > p$ will hold.
  \item If $\Pr_{\beta:X \to  \mathbb B}[\phi|_\beta = \emptyset] = p$,
    then $\Prob{\phi} > p$ will hold iff there is a
    $\beta\colon X \to \mathbb B$ such that $\phi|_\beta \in t\Lang{sat}
    \setminus 0\Lang{cnfs}$.
  \end{enumerate}
  The first item holds because of the witness property
  (Lemma~\ref{lem-comp-witness} would compute a witness, if this were
  possible). The second item holds as  $\Prob{\phi} \ge \Pr_{\beta:X \to
    \mathbb B}[\phi|_\beta = \emptyset]$. For the third
  item, the left-to-right-direction of the ``iff'' is exactly the membership
  condition (which holds by assumption). For the
  right-to-left-direction, the assumption is that there is a
  $\beta^*\colon X \to \mathbb B$ such that $\Pr[\phi|_{\beta^*}] > 0$, but
  $\phi|_{\beta^*} \neq \emptyset$. Once more, we split the
  sum in equation~\eqref{eq-simple-pr}, but now according to whether
  or not $\phi|_\beta=\emptyset$ holds, to obtain
  \begin{align*}
    \Prob{\phi} &\overbrace{=}^{\text{by \eqref{eq-simple-pr}}}
    \sum_{\beta:X \to \mathbb B}
    \frac{\Prob{\Restrict{\phi}{\beta}}}{2^{|X|}}
    =
    \overbrace{
    \sum_{\substack{\beta:X\to\mathbb B,\\ \Restrict{\phi}{\beta} = \emptyset}}
    \frac{\Prob{\Restrict{\phi}{\beta}}}{2^{|X|}}
    }^{ = p} +
    \overbrace{
    \sum_{\substack{\beta:X\to\mathbb B,\\ \Restrict{\phi}{\beta} \neq \emptyset}}
    \frac{\Prob{\Restrict{\phi}{\beta}}}{2^{|X|}}}^{>0} > p.
  \end{align*}
  The ``${=}p$'' is due to the fact that in the first sum in all
  summands $\Prob{\Restrict{\phi}{\beta}} = \Prob{\emptyset} =
  1$, so the sum equals $\Pr_{\beta:X \to \mathbb
    B}[\Restrict{\phi}{\beta} = \emptyset]$, which is~$p$ by
  assumption. The ``${>}0$'' is witnessed by~$\beta^*$.

  By the first two items, if $\Pr_{\beta:X \to
    \mathbb B}[\phi|_\beta = \emptyset]$ is strictly above or strictly
  below~$p$, we immediately know that the same holds
  for~$\Prob{\phi}$. Thus, the reduction can directly accept or reject
  the 
  input (meaning that it formally outputs the singleton set containing a
  tautology or a contradiction). The remaining case is the third
  item. However, clearly, we can check whether such a $\beta$ exists
  by asking whether at least one formula in the set $\Psi := \bigl\{\phi|_\beta
  \bigm| \penalty0 \beta \colon X \to\penalty100 \mathbb B\bigr\} \cap
  (t\Lang{cnfs} \setminus   0\Lang{cnfs})$ lies in
  $t\Lang{sat}$. Since $X$ has constant size,   the size of $\Psi$ is
  constant. 
\end{proof}

\subsection{Linking the Hardness and Membership Conditions}

\label{sec-link}

The final claim left to prove is Lemma~\ref{lem-link}:

\begin{reclaim*}[of Lemma~\ref{lem-link}]
  For every $t$ we have: $t+1$ meets the hardness condition or $t$
  meets the membership condition.
\end{reclaim*}

\begin{proof}
  Assume that $t$ does not meet the membership condition, which means by
  Definition~\ref{tc-c} that there are a
  formula $\phi \in k\Lang{cnfs}$ and a set $X \subseteq
  \operatorname{vars}(\phi)$ with
  \begin{align}
    &\text{$\Prob{\phi} > p$ and $\textstyle\Pr_{\beta:X\to\mathbb
      B}[\phi|_\beta=\emptyset] =p$ and}\label{eq-prop1} \\
    &\text{for all $\beta\colon X \to
      \mathbb B$ with $\phi|_\beta \in t\Lang{cnfs} \setminus
      0\Lang{cnfs}$ the formula $\phi|_\beta$ is not
      satisfiable.}\label{eq-prop2}
  \end{align}
  We will show that $t+1$ meets the hardness condition. By
  Definition~\ref{tc-b}, we must show that, there is a formula
  $\omega \in  k\Lang{cnfs}$ that
  \begin{enumerate}
  \item
    has $\Prob{\omega} = p$, 
  \item
    contains a clause~$c_*$ of size at most $k-(t+1)$ and
  \item
    is irredundant.
  \end{enumerate}
  We first construct an~$\omega$ having the first two properties, but
  which still needs to be ``made irredundant'' by removing redundant
  clauses. Crucially, we will show that we never need to remove the
  last clause of size at most $k-(t+1)$ during this process, which
  then yields all the three items.   
  For the construction, for a clause~$c$ let $c_X := \{ l \in c \mid
  \operatorname{var}(l) \in X\}$ and $c_{\overline X} := \{ l \in c
  \mid \operatorname{var}(l) \notin X\}$. For example,
  $\{x, \neg y, a, \neg b\}_{\{x,y,z\}} = \{x,\neg y\}$ and $\{x,
  \neg y,\penalty0 a, \neg b\}_{\overline{\{x,y,z\}}} = \{a,\neg b\}$. Define: 
  \begin{align*}
    \omega := \bigl\{c_X \bigm| c\in \phi,
    |c_{\overline X}| > t\bigr\} \cup  \bigl\{ c \bigm| c \in \phi,
    |c_{\overline X}| \le t\bigr\}
  \end{align*}
  and note that $\omega\FormImpl\phi$ holds by construction.
  Spelled out, $\omega$ results from $\phi$ by leaving the clauses of
  $\phi$ unchanged when $|c_{\overline X}|$ is ``small, meaning at most $t$,''
  but ``cutting off'' or ``restricting'' $c$ to the variables in $X$
  when $|c_{\overline X}|$ is ``large, meaning larger than~$t$.'' For an
  example of how this works, see Figure~\ref{fig-omega-last}.

  \begin{figure}[htpb]
    \centering
    \begin{tabular}{rll|rl}
      \emph{Clause part~$c_{\overline X}$} &
      \emph{Clause part~$c_X$ \rlap{for $X = \{x,y,z\}$}}\\[.5em]
      $\phi = \bigl\{\quad\{\textcolor{red}{r,s,t,}$ & $\textcolor{blue}{\neg x,y}\},$ &&
      \quad$\omega= \bigl\{\quad\{$&$\textcolor{blue}{\neg x,y}\},$\\      
      $\{\textcolor{red}{r,\neg s, t,}$ & $\textcolor{blue}{\neg x}\},$ &&
      $\{$ & $\textcolor{blue}{\neg x}\},$\\
      $\{\textcolor{red}{u,v,w,}$ & $\textcolor{blue}{\neg x}\},$ &&
      $\{$ & $\textcolor{blue}{\neg x}\},$\\
      $\{\textcolor{red}{t,u,v,}$ & $\textcolor{blue}{x,y}\},$ &&
      $\{$ & $\textcolor{blue}{x,y}\},$\\
      $\{\textcolor{red}{r,t, u,\neg v,}$ & $\textcolor{blue}{y}\},$ &&
      $\{$ & $\textcolor{blue}{y}\},$\\[.25em]
      $\{\textcolor{green!50!black}{r,s,}$ & $x,y,z\},$ &&
      $\{\textcolor{green!50!black}{r,s,}$ & $x,y,z\},$\\
      $\{\textcolor{green!50!black}{t,u,}$ & $x,y,z\},$ &&
      $\{\textcolor{green!50!black}{t,u,}$ & $x,y,z\},$\\
      $\{\textcolor{green!50!black}{v,w,}$ & $x,y,z\},$ &&
      $\{\textcolor{green!50!black}{v,w,}$ & $x,y,z\},$\\
      $\{\textcolor{green!50!black}{t,}$ & $\neg y,z\},$ &&
      $\{\textcolor{green!50!black}{t,}$ & $\neg y,z\},$\\
      $\{\textcolor{green!50!black}{s,\neg u,}$ & $\neg y,z\},$ &&
      $\{\textcolor{green!50!black}{s,\neg u,}$ & $\neg y,z\},$\\
      $\{\textcolor{green!50!black}{s,\neg u,}$ & $x,y,\neg z\}\quad\bigr\}$ &&
      $\{\textcolor{green!50!black}{s,\neg u,}$ & $x,y,\neg z\}\quad\bigr\}$
    \end{tabular}
    \caption{
      Example of the construction of~$\omega$ from $\phi$ in the proof
      of Lemma~\ref{lem-link} for the depicted $\phi \in 5\Lang{cnfs}$
      (so $k=5$), $t=2$, and $X = \{x,y,z\}$. By definition, $\omega
      =\bigl\{\textcolor{blue}{c_X} \bigm| c\in \phi,\,
      \textcolor{red}{|\smash{c_{\overline X}}| > t}\smash{\bigr\}} \cup 
      \bigl\{ c \bigm| c \in\penalty2000 \phi,\penalty0\,
      \textcolor{green!50!black}{|\smash{c_{\overline X}}| \le t}\smash{\bigr\}}
      $, meaning that the clauses $c = \textcolor{red}{c_{\overline
          X}} \cup \textcolor{blue}{c_X} \in \phi$ in the upper
      part, where the size of~$\textcolor{red}{c_{\overline X}}$ is at
      least $t+1=3$, are ``cut off'' and only
      $\textcolor{blue}{c_X}$ of size $\left|\textcolor{blue}{c_X}\right| \le k - (t+1) = 5-3 = 
      2$ is added to~$\omega$; while for the
      clauses $c = \textcolor{green!50!black}{c_{\overline
          X}} \cup \textcolor{blue}{c_X}  \in \phi$ where
      $\textcolor{green!50!black}{c_{\overline X}}$ is at most~$t=2$,
      the clauses are added to~$\omega$ unchanged. Note that
      $\omega\FormImpl\phi$ holds and that any assignment $\alpha$
      with $\alpha \models \phi$ but $\alpha \not\models \omega$ (like
      for instance the assignment mapping all variables to~$1$) must satisfy all clauses
      of the lower part (they are present in both $\phi$
      and~$\omega$), must not satisfy at least one blue clause
      $\textcolor{blue}{c_*} \in \omega$ in the upper part (like
      $\textcolor{blue}{c_*} = \{\textcolor{blue}{\neg x}\}$  and note
      $|\textcolor{blue}{c_*}| \le 
      k-(t+1)$), and must satisfy the red part of all clauses in
      $\phi$ from which $\textcolor{blue}{c_*}$ resulted (so $\alpha
      \models \bigl\{\{\textcolor{red}{r,\neg s,t}\},
      \{\textcolor{red}{u,v,w}\}\bigr\}$).  
    }
    \label{fig-omega-last}
  \end{figure}
  
  To prove that $\omega$ has the three claimed properties, let us
  start by showing that $\Prob{\omega} = p$ holds. Recall that by
  assumption~\eqref{eq-prop1} we have $\Pr_{\beta:X\to\mathbb
    B}[\phi|_\beta=\emptyset] = p$, so it suffices to show
  $\Prob{\omega} = \Pr_{\beta:X\to\mathbb
    B}[\phi|_\beta=\emptyset]$. This clearly follows from the
  claim below, where for any given assignment $\alpha
  \colon \operatorname{vars}(\phi) \to \mathbb B$ we let $\beta
  \colon X \to \mathbb B$ denote the restriction of $\alpha$ to the
  set~$X$:
  \begin{claim}\label{claim-omega-iff}
    $\alpha \models \omega$ holds iff $\phi|_\beta = \emptyset$.
  \end{claim}
  \begin{proof}
    For the direction from right to left, assume $\phi|_\beta
    =\emptyset$. Then all clauses in  
    $\phi$ contain a literal whose variable lies in~$X$ and is made
    true by~$\beta$ (``$\beta$ hits the blue or black part of each
    clause of~$\phi$ in Figure~\ref{fig-omega-last}''). In
    particular, every clause of~$\omega$ is also made 
    true by~$\alpha$ and, thus, $\alpha\models \omega$ holds whenever
    $\phi|_\beta = \emptyset$. For the other direction, assume $\phi|_\beta \neq
    \emptyset$ (``$\beta$~does not hit the blue or black parts of
    some clauses of~$\phi$, and $\phi|_\beta$ contains the red or green
    parts of the clauses not hit'').

    First consider the case $\phi|_\beta \in
    t\Lang{cnfs}$. If $\phi|_\beta\in 0\Lang{cnfs} =
    \{\emptyset,\{\emptyset\}\}$, then $\phi|_\beta$ must equal the
    contradiction~$\{\emptyset\}$. Otherwise, by
    assumption~\eqref{eq-prop2} we have $\phi|_\beta \notin
    t\Lang{sat}$. In either case, there must be a
    clause $c_{\overline X} \in \phi|_\beta$, resulting from some $c
    \in \phi$, not made true by the restriction of~$\alpha$ to
    $\operatorname{vars}(\phi) \setminus X$. Since we also have $\beta
    \not\models \{c\}$ (we would not have added $c_{\overline X}$
    to~$\phi|_\beta$ otherwise), we get $\alpha \not\models
    \{c\}$. Since $\phi|_\beta \in t\Lang{cnfs}$, we 
    conclude $|c_{\overline X}| \le t$ (so ``$c_{\overline X}$ is
    green'') and, thus, $c \in \omega$ holds. In particular, $\alpha
    \not\models \omega$. 

    Now consider the case $\phi|_\beta \notin 
    t\Lang{cnfs}$. Then there is a clause $c \in \phi$ that is not made 
    true already by~$\beta$ (meaning that all literals in~$c_X$ are set
    to false by~$\beta$) for which $|c_{\overline X}| > t$
    (``$\beta$~does not hit the blue part of some clause''). However,
    this means  
    that in $\omega$ we have a clause $c_X$ in which all literals are
    set false by~$\beta$ and hence also by~$\alpha$. In particular,
    $\alpha \not\models \omega$.
  \end{proof}

  To establish the second and third desired properties
  of~$\omega$ (the existence of a small clause~$c_*$ and the
  irredundancy), we first prove the following claim:
  \begin{claim}\label{claim-alpha-star}
    There is an assignment $\alpha_* \colon
    \operatorname{vars}(\omega) \to \mathbb B$ with  
    \begin{enumerate}
    \item $\alpha_* \models \{d\}$ for all $d \in \omega$ of size $|d| > k - (t+1)$, but
    \item  $\alpha_* \not\models \{c_*\}$ for at least one $c_* \in
      \omega$ of size $|c_*| \le k -(t+1)$.
    \end{enumerate}
  \end{claim}
  (``Some $\alpha_*$ hits all large green--black clauses
  of~$\omega$, but not all small blue clauses.'')   
  
  \begin{proof} 
    Since $\Pr[\phi] > \Pr[\omega]$ and $\omega \FormImpl \phi$, there
    must exist an assignment $\alpha_*\colon \operatorname{vars}(\phi)
    \to \mathbb B$ with $\alpha_* \models \phi$ but $\alpha_*
    \not\models \omega$. 
    
    For the first item, let $d \in \omega$ be given of size $|d| > k -
    (t+1)$. If we even have $d \in \phi$, we trivially have $\alpha_*
    \models \{d\}$ as $\alpha_* \models \phi$. Otherwise (``$d \in
    \omega\setminus \phi$ means that $d$ is blue''), by
    construction of~$\omega$ there must be a clause $c \in \phi$ with
    $d = c_X$ and $|c_{\overline X}| > t$ (``$c_{\overline X}$ must be
    red''), so $|d| < k - t$ (``blue clauses are small''). This contradicts $|d| > k - (t+1)$. 
    
    For the second item, since $\alpha_* \not\models \omega$, by
    Claim~\ref{claim-omega-iff} we have 
    $\phi|_{\beta_*} \neq \emptyset$, where $\beta_*\colon X \to
    \mathbb B$ is the restriction of $\alpha_*$ to~$X$.
    Since $\beta_*$ is the restriction of an $\alpha_*$ with $\alpha_*
    \models \phi$, we know that $\phi|_{\beta_*}$ is satisfiable. By
    assumption~\eqref{eq-prop2}, $\phi|_{\beta_*} \notin 
    t\Lang{sat} \setminus 0\Lang{cnfs}$ must hold. As $\phi|_{\beta_*}$ is
    satisfiable and not equal to~$\emptyset$, we even know
    $\phi|_{\beta_*} \notin t\Lang{cnfs}$ and there must exist a clause $d \in
    \phi|_{\beta_*}$ of size $|d| > t$ (a ``red clause''). Since all
    clauses in $\phi|_{\beta_*}$ are of the form $c_{\overline X}$ for
    some $c \in \phi$, we conclude that $|d| = 
    |c_{\overline X}|$ for some $c \in \phi$. By the definition of~$\omega$, we
    conclude that for (``the blue'') $c_* := c_X$ we have $c_* \in
    \omega$ and note that $|c_*| < k-t$. Most importantly, $\beta_*
    \not\models \{c_*\}$ since, otherwise, $d=c_{\overline X}$ would
    not have been included in~$\phi|_{\beta_*}$. Finally, since
    $\operatorname{vars}(c_*) \subseteq X$, we also have $\alpha_*
    \not\models \{c_*\}$. 
  \end{proof} 
  Claim~\ref{claim-alpha-star} clearly implies that $\omega$ contains a clause~$c_*$ of 
  size at most $k-(t+1)$, but we still need to ``make $\omega$
  irredundant.'' This is achieved by removing redundant clauses as
  long as possible, that is, by executing $\omega \gets \omega
  \setminus \{c\}$ as long as possible, 
  where $c$ is any clause with $\omega \setminus \{c\} \FormImpl \{c\}$. Clearly, this
  will eventually result in an irredundant $\omega' \subseteq \omega$
  with $\omega' \equiv \omega$ and hence $\Pr[\omega'] =
  \Pr[\omega]$. Now, this process may remove both small and large
  clauses and, in particular, $c_*$ may be removed at some
  point. However, the final $\omega'$ will still contain at least 
  one small clause~$c_*'$: Suppose $\omega'$ only contained clauses~$d$ of size
  $|d| > k-(t+1)$. Then $\alpha_*\models \omega'$ would hold by the
  first item of Claim~\ref{claim-alpha-star}; contradicting
  $\alpha_*\not\models \{c_*\} \subseteq \omega \equiv \omega'$.
\end{proof}

\section{Conclusion and Outlook}

The results of the present paper settle the complexity of
$k\Lang{sat-pr}_{>p}$ from a structural complexity
view: The problem is either $\Class{NP}$-complete or
$\Class{NL}$-complete or lies in $\Class{AC}^0$ -- and which of these
is the case depends on whether or not $p = \Prob{\omega}$ holds
for some irredundant formula $\omega\in k\Lang{cnfs}$ containing a
clause of size $k-3$ or of size $k-2$. The proof is based on the insight that
the spectra $k\CNFSSigmaSpectrum$ are 
well-ordered with respect to~$>$. We saw that being well-ordered (or,
equivalently, having gaps below all values) is a key property of the
spectra with numerous algorithmic consequences. In particular, the 
standard sunflower-based kernel algorithm for hitting sets
allows us to compute kernels for $k\Lang{sat-pr}_{\ge p}$. By
extending the notion of weak backdoor sets to the threshold setting,
we also saw that the computed kernels will always form ``weak backdoor
sets for threshold~$p$ into $\Lang{2cnfs}$ (or $\Lang{1cnfs}$)''
when $k\Lang{cnfs}$ do \emph{not} have room for $\Lang{3sat}$
(or~$\Lang{2sat}$). 

An attempt to visualize the ``landscape'' of the complexity of
$k\Lang{sat-pr}_{>p}$ for $k \le 4$ can be found
in Figure~\ref{fig-spec} on page~\pageref{fig-spec}. For $k=4$, two 
values of special interest are $p_1 = \frac{15}{32} = \frac{1}{2} -
\frac{1}{32}$ and $p_2 = \frac{63}{128} = \frac{1}{2} -
\frac{1}{128}$.

There is a ``red triangle'' (signaling $\Class{NP}$-completeness) in
the figure at~$p_1 = \frac{15}{32}$, meaning that
$\smash{4\Lang{sat-pr}_{>\frac{15}{32}}}$ is
$\Class{NP}$-complete. The reason is that the first item of
Theorem~\ref{thm-main} applies to $ \frac{15}{32} =
\ProbBig{\bigr\{\{a\},\{x, y, z, w\}\bigr\}}$ as $\bigr\{\{a\},\{x, y,
z, w\}\bigr\}$ is clearly irredundant and contains a clause of size
$4-3 =1$ and hence has room for $\Lang{3sat}$. 

In contrast, there is a ``green triangle'' (signaling
$\Class{NL}$-completeness) at $p_2= \frac{63}{128}$ (as well as at 
many, many other positions in $\bigl(\frac{15}{32},\frac{1}{2}\bigr)$,
but still only at a nowhere dense subset despite the ``solid 
line'' in the visualization) as $\smash{4\Lang{sat-pr}_{>\frac{63}{128}}}$ is 
$\Class{NL}$-complete. This is because, on the one hand, $\omega =
\bigl\{\{a,b\},\penalty0\{c, d\},\penalty0\{ e, f, g\}\bigr\}$ is
irredundant, contains a clause of size $4-2=2$, and $\Pr[\omega] =
\frac{63}{128}$, proving that $4\Lang{cnfs}$ have room for
$2\Lang{sat}$ at~$\frac{63}{128}$. On the other hand, no
$\Lang{4cnf}$ formula~$\omega$ with $\Prob{\omega} = \frac{63}{128}$ can
have room for~$\Lang{3sat}$ as this would mean that $\omega$ contains
a size-1 clause and, hence, $\Pr[\omega] \le 1/2$. Since $\omega$
would need to contain at least one more clause, $\Pr[\omega] \le
\frac{1}{2}\cdot \frac{15}{16} = \frac{15}{32} < \frac{63}{128}$. 

For larger values of~$k$, observe that, on the one hand,
$k\Lang{sat-pr}_{>1-2^{-(k-2)}}$ is  $\Class{NL}$-complete for
all~$k$ (since $\ProbBig{\bigr\{\{a_1,\dots,\penalty0a_{k-2}\}\bigr\}} =
1-2^{-(k-2)}$, but $\Prob{\omega} \neq 1-2^{-(k-2)}$ for all $\omega
\in k\Lang{cnfs}$ containing a clause of size $k-3$ as this clause
already lowers the satisfaction probability to at most
$1-2^{-(k-3)}< 1-2^{-(k-2)}$); while on the other hand,
$k\Lang{sat-pr}_{>2^{-i}}$ is $\Class{NP}$-complete for all $k\ge 4$
and $i \ge 1$.

Since the arguments presented in this paper depend so heavily on the
size of spectral gaps, it is of interest to determine these
sizes precisely. We established such bounds in
Section~\ref{sec-bounds}, but it is unclear whether these
hyperexponential bounds are even remotely tight and it would also be of
interest to determine explicit values: A close look at Figure~\ref{fig-spec} reveals 
$\operatorname{spectral-gap}_{\Lang{2cnfs}}(1/2) = 1/32$, but what is
the value of $\operatorname{spectral-gap}_{\Lang{3cnfs}}(1/2)$?

A bit frustratingly, it is not clear how difficult it is to decide on
input of numbers $k$ and~$p$ (with $p$~encoded as, say, two
integers~$m$ and~$e$ with $p = m/2^e$) which of the three cases in the
Spectral Trichotomy Theorem applies. The obvious difficulty lies in
determining, for $t\in \{2,3\}$, whether there is a formula $\omega
\in k\Lang{cnfs}$ that has room for $t\Lang{sat}$ at~$p$. It is
clearly trivial to check whether a given formula~$\omega$ contains
a clause of size $k-t$ and still easy to check whether $\omega$ is
irredundant (in the context of decidability, ``easy'' liberally
includes ``in exponential time'', but see
\cite{DBLP:conf/ecai/Liberatore02} for better complexity
bounds). However, the search space is  
the infinite set of $k\Lang{cnf}$ formulas with $\Pr[\omega] = p$. We
know by Theorem~\ref{thm-constant-influence} that all such
$\omega$ are \emph{equivalent to some k$\Lang{cnf}$ formula~$\psi$ of
size~$S_{k,p}$} and that \emph{$\psi$ is obtainable from~$\omega$ just
through removing literals from the clauses.} Using the explicit
lower bounds on the sizes of spectral gaps from
Theorem~\ref{thm-spec-quant}, we can compute an 
upper bound on the number~$S_{k,p}$. All of this means that if
$k\Lang{cnfs}$ have room for $t\Lang{sat}$ at some~$p$, 
we can algorithmically search for and find a ``small'' formula~$\psi
\in k\Lang{cnfs}$ (of a size computable from $k$ and~$p$) with
$\Pr[\psi] = p$ that contains a clause~$c_*$ of size
$k-t$. The trouble is, however, that $c_*$ might have become a
redundant clause! This problem would disappear, and
the cases of Theorem~\ref{thm-main} would become decidable, if one
could prove the following conjecture:
\begin{conjecture}
  For every $\psi \in k\Lang{cnfs}$ there is an equivalent,
  irredundant $\psi' \in k\Lang{cnfs}$ whose smallest clause is no
  larger than the smallest clause of~$\psi$.
\end{conjecture}

\subsection*{Outlook: Other Versions}

The focus of this paper was on the question of whether $\Prob{\phi}
\ge p$ or $\Prob{\phi} > p$ holds for formulas $\phi \in
k\Lang{cnfs}$, and we looked at this question from different
angles. However, there are numerous further ``versions'' or
``variants'' that are also of interest for theoretical
or practical reasons and some of these versions are sketched in the
following.

\subparagraph*{The Equal-To Version.}
In this paper, we focused on the ``strictly greater than'' 
problem $k\Lang{sat-pr}_{>p}$, since we ``boringly'' always have
$k\Lang{sat-pr}_{\ge p} \in 
\Class{AC}^0$. However, by combining this with the Spectral Trichotomy
Theorem, we get an interesting corollary for the ``equal to'' version:
\begin{corollary}
  For  the same $k$ and $p$ as in Theorem~\ref{thm-main}, the
  problem $k\Lang{sat-pr}_{= p}$ is $\Class{coNP}$-complete,
  $\Class{NL}$-complete, or lies in $\Class{AC}^0$.
\end{corollary}
Spelled out, we get results like the following: ``It is $\Class{NL}$-complete
to decide on input of a $\Lang{3cnf}$ formula whether
exactly half of the assignments are satisfying'' and ``it is
$\Class{coNP}$-complete to decide on input of a $\Lang{4cnf}$
formula whether exactly half of the assignments are
satisfying,'' but also stranger ones like ``it is $\Class{NL}$-complete
to decide on input of a $\Lang{4cnf}$ formula whether
the fraction of satisfying assignments is exactly $\frac{1}{2}-\frac{1}{128}$'' while
``it is $\Class{coNP}$-complete 
to decide on input of a $\Lang{4cnf}$ formula whether
the fraction of satisfying assignments is exactly $\frac{1}{2}-\frac{1}{32}$.'' 

\subparagraph*{The CSP Version.} Many of the results in the present
paper appear to generalize to more (or less) general versions of
constraint satisfaction problems. The reason is that it seems possible
to prove the Spectral Well-Ordering Theorem also for more general
\textsc{csp}s as long as a version of the Packing Probability Lemma
also holds for them. A bit more formally, a \emph{constraint language
for a domain~$D$} is a 
set~$\Gamma$ of sets of relations over~$D$. The set
$\Lang{csps}(\Gamma)$ of instances for $\Gamma$ contains pairs $I=
(V,\{C_1,\dots,C_m\})$ of sets $V$ of variables and sets of
constraints. Each $C_i$ consists of a tuple~$t$ of 
variables from~$V$ together with an element $R \in \Gamma$ whose
arity equals $t$'s length. A \emph{solution for $I$} is
an assignment $\alpha \colon V \to D$ such that for each constraint
$C_i = (t,R)$ with $t = (v_1,\dots,v_k)$ we have
$(\alpha(v_1),\dots,\alpha(v_k)) \in R$. Writing $\Prob{I}$ for the
probability that a random $\alpha\colon V \to D$ is a
solution of~$I$, let $\Lang{csp($\Gamma$)-pr}_{\ge
  p}$ be the set $\{I \mid \Prob{I} \ge p, I \in
\Lang{csps}(\Gamma)\}$. Then $k\Lang{sat-pr}_{\ge k}$ is exactly
$\Lang{csp($\Gamma$)-pr}_{\ge p}$ where $\Gamma = \{ B  \subseteq
\mathbb B^k \mid \left|B\right| = 2^k-1 \}$ is the set of all relations
that exclude one possibility for $(\alpha(v_1),\dots,\alpha(v_k))$.

It seems that the Spectral Well-Ordering Theorem holds for all finite
constraint languages~$\Gamma$. It also seems that
$\Lang{csp($\Gamma$)-pr}_{\ge p} \in \Class{AC}^0$ holds. However,
neither of these results are (at least trivial) consequences of the
work done in the present paper: For the proof of the Spectral Well-Ordering
Theorem, the start of the induction gets more involved; for the
membership in $\Class{AC}^0$ it is no longer quite clear which of the
sunflower-based algorithms still work (indeed, it is no longer clear
how, exactly, sunflowers should be defined for \textsc{csp}s).

In another direction, we can also consider constraint satisfaction
problems $\Lang{csps}(\Gamma)$ that are \emph{easier} than
$k\Lang{sat}$. For instance, we can consider positive $k\Lang{cnf}$
formulas, which corresponds to $\Gamma^+ =
\bigl\{\mathbb B^k\setminus\{0^k\},
\mathbb B^{k-1}\setminus\{0^{k-1}\},\dots, \{01,10,11\}, \{1\} \bigr\}$.
Of course, the satisfiability problem for positive formulas is not
particularly interesting (assigning~$1$ to all variables is always a
solution). However, determining the parity of the number
of satisfying assignments of positive $k\Lang{cnfs}$ is as hard as for
$k\Lang{cnfs}$ by the results of \cite{Cyganetal2021}; and the
complexity of computing $\Prob{\cdot}$ for positive
$k\Lang{cnfs}$ seems to be an open problem. This makes
$\Lang{csp}(\Gamma^+)\Lang{-pr}_{\ge p}$ an interesting
problem. Even more interesting is the fact that
$\Lang{csp}(\Gamma^+)\Lang{-pr}_{> p}$ might have the same
complexity: Intuitively, we should be able to use 
equation~\eqref{eq-three-pr} and note that deciding whether the second and
third sums are positive is trivial as $\Restrict{\phi}{\beta}$ is
\emph{always} satisfiable unless $\emptyset \in \Restrict{\phi}{\beta}$. 
All told, it seems that for every $p \in [0,1]$ and every~$d$, an
$\Class{AC}^0$-circuit can decide on input of a $d$-hypergraph (every
hyperedge contains at most $d$ vertices), whether the fraction of
vertex subsets that are hitting sets is strictly less than~$p$, equal
to~$p$, or larger than~$p$.

\subparagraph*{The Algebraic Version.}
A natural generalization of satisfaction probability threshold
problems like $\Lang{sat-pr}_{\ge p}$ or $k\Lang{sat-pr}_{\ge
  p}$ are ``weighted'' or ``algebraic'' versions. Given $p
\in [0,1]$ and a polynomial $P \in \mathbb Q[x_1,\dots,x_n]$ as input
(encoded in some appropriate way), we are
asked to decide whether the following holds:
\begin{align}
  \Prob{P} := \frac{1}{2^n}
  \sum_{a_1,\dots,a_n \in \{0,1\}} P(a_1,\dots,a_n) \ge p. \label{eq-algebra}
\end{align}
The connection to propositional logic is simple (known as an
\emph{algebraization} of the formulas): For a propositional formula
$\phi$ we define $\operatorname{algebraic}(\phi)$ as follows: If $\phi$
is a propositional variable~$x_i$, the algebraization is just the
rational variable~$x_i$ with $x_i =0$ meaning false and
$x_i = 1$ meaning true; for negations we have
$\operatorname{algebraic}(\neg \phi) =
1-\operatorname{algebraic}(\phi)$; and for conjunctions we have
$\operatorname{algebraic}(\phi_1 \land \phi_2) =
\operatorname{algebraic}(\phi_1) \cdot
\operatorname{algebraic}(\phi_2)$. In this way, we have $\beta \models
\phi$ iff $\operatorname{algebraic}(\phi)(\beta(x_1),\dots,\beta(x_n))
= 1$, we have that $\operatorname{algebraic}(\phi)(a_1,\dots,a_n) \in \{0,1\}$ holds 
whenever $a_i \in \{0,1\}$ (we call such $a_i$ \emph{binary}), and we have
$\Prob{\phi} = \Prob{\operatorname{algebraic}(\phi)}$.

For $\phi \in k\Lang{cnfs}$, the polynomial 
$\operatorname{algebraic}(\phi)$ is a product $\prod_{i=1}^m p_i$ of
``clause'' polynomials~$p_i$ that depend on at most~$k$ variables and
have degree at most~$k$.
Akmal and Williams~\cite{AkmalW2022} ask about the computational
complexity of deciding~\eqref{eq-algebra} when $P = \prod_{i=1}^m p_i$
for arbitrary degree-$k$ polynomials $p_i \in \mathbb
Q[x_1,\dots,x_n]$ that are given as input (instead of the
exponentially many factors of~$P$); and also for special cases like the 
restriction that $p_i(a_1,\dots,a_n) \in [0,1]$ holds for binary $a_i$.
Given the importance of algebraization techniques in computational
complexity theory, this seems like a natural and important question to
ask -- and the results of the present paper shed some light on it
(although they do not completely solve it) as we can use the same
techniques to show that whenever the range of
values that the polynomials~$p_i$ may have for binary~$a_i$ is a
well-ordered subset of~$[0,1]$, then deciding \eqref{eq-algebra} is
easy (lies in $\Class{AC}^0$), while otherwise it is hard (is
$\Class{NP}$- or even $\Class{PP}$-hard).

\subparagraph*{The Descriptive Versions.}
Since practically all results of computational complexity can be
rephrased in terms of descriptive complexity (see \cite{Immerman1998}
for an introduction), it is no coincidence
that our results on the complexity of $k\Lang{sat-pr}_{> p}$ can
be recast in the descriptive framework. In short,
``for all $k$ and $p$ we have $k\Lang{sat-pr}_{\ge p} \in
\Class{AC}^0$'' translates to ``second-order threshold quantifiers
followed by universal 
first-order quantifiers can be replaced by only universal first-order
quantifiers.'' As another example, ``$3\Lang{sat-pr}_{> 1/2} \in
\Class{NL}$'' translates to ``a second-order strict majority
quantifier followed by three universal first-order quantifiers can be
replaced by first-order formulas with transitive closure.''

An intriguing  question is whether these
translations need the $\Lang{bit}$ predicate or not. This predicate is
normally used ubiquitously in descriptive complexity theory (see
\cite{Immerman1998} for a detailed discussion), but it seems that we
do not even need an \emph{ordering} of the universe for the above
results.

\subparagraph*{The Infinite Version.}

A final question, which brings us far beyond the realm of
computational complexity and finite model theory, is whether any of
the results of the present paper also hold in an infinite
setting. We can
simply drop the requirement that a $\Lang{cnf}$ formula must be a
\emph{finite} set of clauses. This is quite natural; indeed, nothing
needs to be changed concerning the semantics of when an assignment
$\beta \colon X \to \mathbb B$ satisfies an infinite formula~$\phi$
and formulas like
$\bigl\{\{x_0,x_1\},\{x_0,x_2\},\{x_0,x_3\},\dots\bigr\}$ express
sensible properties of infinite sets like $X = \{x_0,x_1,x_2,\dots\}$ of
propositional variables (indeed, even for uncountable~$X$ such
formulas make sense).

Writing $\phi \in k\Lang{cnfs-infinite}$ if there is a
set~$X$ of variables such that $\phi$ is a (possibly) infinite set of
clauses over~$X$ such that for all $c \in \phi$ we have $|c| \le k$
and never have both $x \in c$ and also $\neg x \in c$ for any~$x \in
X$, we can still define the \emph{satisfaction probability}
of~$\phi$ as $\Prob{\phi} = \Pr_{\beta: X \to \mathbb B}[\beta
  \models \phi]$. (Note that it is not immediately clear whether this 
probability is well-defined as the set $\{\beta \colon X \to \mathbb B \mid
\beta \models \phi\}$ might not even be measurable. But since $\phi
\in k\Lang{cnfs-infinite}$, the set can be expressed as an infinite
intersection of measurable sets. Exploring these measure-theoretic
questions further is far beyond the scope of this paper, however.) Observe
that $\Prob{\phi} > 0$ 
is no longer the same as saying that $\phi$ is satisfiable: The
$1\Lang{cnfs-infinite}$ formula $\bigl\{\{x_2\}, \{x_4\},\{x_6\},
\dots\bigr\}$ is clearly satisfiable (it has infinitely many
satisfying assignments as
we can set the odd-numbered variables arbitrarily as long as we set
all even-numbered variables to true), but the satisfaction probability
is~$0$.

Even though the satisfaction probabilities of infinite formulas have
properties different from those in the finite settings, we can still
ask what values $\Prob{\phi}$ can take for $\phi \in
k\Lang{cnfs-infinite}$, that is, we can ask what
$k\Lang{cnfs-infinite-pr-spectrum}$ looks like. Curiously, it seems that
the added power of infinite formulas does not change the spectrum,
that is, that $k\Lang{cnfs-pr-spectrum} = k\Lang{cnfs-infinite-pr-spectrum}$
holds for all~$k$. 
In the infinite setting it makes less sense to consider
``algorithms'' that work on these infinite inputs. However, more
``abstract'' results still appear to
hold, including the Threshold Locality Lemma. In particular, it seems
that the following holds: \emph{For each $k$ and $p \in [0,1]$
there is an~$S \in 
\mathbb N$ so that for all $\phi \in k\Lang{cnfs-infinite}$ we have
$\Prob{\phi} \ge p$, iff $\Prob{\psi} \ge p$ for
all $\psi \subseteq \phi$ with~$|\psi| \le S$.}

\bibliography{2021-spectrum}

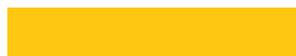
\end{document}